\documentclass[a4paper,11pt]{article} 
\usepackage{url}
\usepackage{fullpage}
\usepackage{amsmath,amsthm,amssymb}
\usepackage{xcolor}
\usepackage{graphicx}
\usepackage{textcomp} 
\usepackage{siunitx}
\usepackage{bbm}
\usepackage{mathabx} 
\usepackage{authblk}

\renewcommand{\vec}[1]{\boldsymbol{#1}}
\newcommand{\tensor}[1]{\tilde{\boldsymbol{#1}}}
\newcommand{\mat}[1]{\boldsymbol{#1}}
\newcommand{\vnabla}{\vec{\nabla}}
\def\Ms{M_\text{s}}
\def\gammae{\gamma_\text{e}}
\def\mub{\mu_\text{B}}

\def\Di{D_\text{i}}
\def\Db{D_\text{b}}

\def\heff{\vec{H}^\text{eff}}
\def\hdemag{\vec{H}^\text{dem}}
\def\hdemags{H^\text{dem}}
\def\hex{\vec{H}^\text{ex}}

\def\hzee{\vec{H}^\text{zee}}
\def\haniso{\vec{H}^\text{ani}}
\def\htorque{\vec{H}^\text{T}}

\def\Edemag{E^\text{dem}}
\def\Eex{E^\text{ex}}
\def\Ezee{E^\text{zee}}
\def\Eaniso{E^\text{ani}}
\def\Eanisou{E^\text{aniu}}
\def\Eanisoc{E^\text{anic}}
\def\Edmi{E^\text{dmi}}
\def\Edmii{E^\text{dmii}}
\def\Edmib{E^\text{dmib}}
\def\Eiex{E^\text{iex}}
\def\lambdasf{\lambda_\text{sf}}
\def\lambdaj{\lambda_\text{J}}
\def\tausf{\tau_\text{sf}}
\def\tauj{\tau_\text{J}}
\def\tauphi{\tau_\phi}
\def\je{\vec{j}_\text{e}}
\def\js{\tilde{\vec{j}}_\text{s}}
\def\thetash{\theta_\text{SH}}
\def\betavf{\beta''}
\def\lambdavf{\lambda}

\def\rhos{\rho^\ast}
\def\gammavf{\gamma_\text{sf}}
\def\jevf{\vec{J}_\text{e}}
\def\jssvf{\vec{J}_\text{s}}
\def\jsvf{\tilde{\vec{J}}_\text{s}}
\def\uvf{\phi}
\def\svf{\vec{\phi}_\text{s}}
\def\ssvf{\phi_\text{s}}
\def\rsb{r_\text{b}^\ast}
\def\gup{g_\updownarrows}
\def\n{n}

\DeclareSymbolFont{extraup}{U}{zavm}{m}{n}
\DeclareMathSymbol{\vardiamond}{\mathalpha}{extraup}{87}

\newcommand{\ddiff}[3]{\mathchoice{\frac{#1#2}{#1#3}}{#1#2 / #1#3}{#1#2 / #1#3}{#1#2 / #1#3}}
\newcommand{\diff}[2]{\ddiff{\text{d}}{#1}{#2}}
\newcommand{\pdiff}[2]{\ddiff{\partial}{#1}{#2}}
\newcommand{\pdiffs}[2]{\partial_{#2} #1}
\newcommand{\vdiff}[2]{\ddiff{\delta}{#1}{#2}}

\newcommand{\dx}{\,\text{d}\vec{x}}
\newcommand{\ds}{\,\text{d}\vec{s}}
\newcommand{\dt}{\,\text{d}t}

\title{
  Micromagnetics and spintronics:\\
  Models and numerical methods
}

\author[1]{Claas Abert}
\affil[1]{Christian Doppler Laboratory for Advanced Magnetic Sensing and Materials, Faculty of Physics, University of Vienna, 1090 Vienna, Austria}

\begin{document}

\maketitle

\begin{abstract}
  Computational micromagnetics has become an indispensable tool for the theoretical investigation of magnetic structures. 
  Classical micromagnetics has been successfully applied to a wide range of applications including magnetic storage media, magnetic sensors, permanent magnets and more.
  The recent advent of spintronics devices has lead to various extensions to the micromagnetic model in order to account for spin-transport effects.
  This article aims to give an overview over the analytical micromagnetic model as well as its numerical implementation.
  The main focus is put on the integration of spin-transport effects with classical micromagnetics.
\end{abstract}
  
\newpage

\tableofcontents
\newpage

\section{Introduction}\label{sec:intro}
The micromagnetic model has proven to be a reliable tool for the theoretical description of magnetization processes on the micron scale.
In contrast to purely quantum mechanical theories, such as density functional theory, micromagnetics does not account for distinct magnetic spins nor nondeterministic effects due to collapse of the wave function.
However, micromagnetics integrates quantum mechanical effects that are essential to ferromagnetism, like the exchange interaction, with a classical continuous field description of the magnetization in the sense of expectation values.
The main assumption of this model is that the organizing forces in the magnetic material are strong enough to keep the magnetization in parallel on a characteristic length scale $\lambda$ well above the lattice constant $a$
\begin{equation}
  \vec{S}_i \approx \vec{S}_j
  \quad \text{for} \quad
  |\vec{x}_i - \vec{x}_j|
  < \lambda
  \gg a
  \label{eq:neighbors_equal}
\end{equation}
where $\vec{S}_{i/j}$ and $\vec{x}_{i/j}$ are distinct spins and their positions respectively.
For a homogeneous density of spins, the discrete distribution of magnetic moments $\vec{S}_i$ is well approximated by a continuous vector density $\vec{M}(\vec{x})$ such that
\begin{equation}
  \int_\Omega \vec{M}(\vec{x}) \dx
  \approx
  \sum_i \mathbbm{1}_\Omega (\vec{x}_i) \vec{S}_i
  \label{eq:discrete_to_continuum}
\end{equation}
renders approximately true for arbitrary volumes $\Omega$ of the size $\lambda \times \lambda \times \lambda$ and larger with $\mathbbm{1}_\Omega$ being the indicator function of $\Omega$.
The continuous magnetization $\vec{M}(\vec{x})$ has a constant norm due to the homogeneous density of spins and can thus be written in terms of a unit vector field $\vec{m}(\vec{x})$
\begin{equation}
  \vec{M}(\vec{x}) = \Ms \, \vec{m}(\vec{x})
  \quad \text{with} \quad
  | \vec{m}(\vec{x}) | = 1.
  \label{eq:normalized_magnetization}
\end{equation}
where $\Ms$ is the spontaneous magnetization.
In the case of zero temperature, which is often considered for micromagnetic modeling, $\Ms$ is the saturation magnetization which is a material constant.
While $\vec{m}$ and $\vec{M}$ have to be strictly distinguished, both are referred to as magnetization throughout this work for the sake of simplicity.

Due to the combination of classical field theory with quantum mechanical effects, micromagnetics is often referred to as semiclassical theory.
Opposed to the macroscopic Maxwell equations, the micromagnetic model resolves the structure of magnetic domains and domain walls.
This enables accurate hysteresis computations of macroscopic magnets, since hysteresis itself is the direct result of field-induced domain generation and annihilation.
While static hysteresis computations are very important for the development of novel permanent magnets \cite{schrefl2007numerical}, another application area for micromagnetics is the description of magnetization dynamics on the micron scale.
The time and space-resolved description of magnetic switching processes and domain-wall movements is essential for the development of novel storage and sensing technologies such as magnetoresistive random-access memory (MRAM) \cite{huai2008spin}, sensors for read heads \cite{parkin2004giant}, and angle sensors \cite{granig2006integrated}.
Besides the manipulation of the magnetization with external fields, the interaction of spin polarized electric currents with the magnetization plays an increasing role for novel devices.
Several extensions to the classical micromagnetic theory have been proposed in order to account for these spintronics effects.

A lot of articles and books have been published on analytical \cite{brown_1963,doring1948tragheit,kronmuller2007general} as well as numerical \cite{schrefl2007numerical,miltat2007numerical,Leliaert_2018} micromagnetics.
This review article is supposed to serve two purposes.
First, it is meant to give a comprehensive yet compact overview over the micromagnetic theory, both on an analytical as well as a numerical level. 
The article describes the most commonly used discretization strategies, namely the finite-difference method and the finite-element method, and discusses advantages, disadvantages and pitfalls in their implementation.
The second and main purpose of this article, however, is to give an overview over existing models for spin transport in the context of micromagnetics.
This article reviews the applicability and the limits of these models and discusses their discretization.

\section{Energetics of a ferromagnet}\label{sec:energetics}
The total energy of a ferromagnetic system with respect to its magnetization is composed by a number of contributions depending on the properties of the respective material.
While some of these contributions, like the demagnetization energy and the Zeeman energy can be described by classical magnetostatics, other contributions like the exchange energy and the magnetocrystallin anisotropy energy have a quantum mechanical origin.
This section aims to give an overview over typical energy contributions and their representation in the micromagnetic model.

\subsection{Zeeman energy}\label{sec:energetics_zeeman}
The energy of a ferromagnetic body highly depends on the external field $\hzee$.
The corresponding energy contribution is often referred to as Zeeman energy.
According to classical electromagnetics the Zeeman energy of a magnetic body $\Omega_m$ is given by
\begin{equation}
  \Ezee = - \mu_0 \int_{\Omega_m} \Ms \vec{m} \cdot \hzee \dx
  \label{eq:energetics_zeeman}
\end{equation}
with $\mu_0$ being the vacuum permeability.

\subsection{Exchange energy}\label{sec:energetics_exchange}
The characteristic property of ferromagnetic materials is its remanent magnetization, i.e. even for a vanishing external field, a ferromagnetic system can have a nonvanishing macroscopic magnetization.
In a system where the spins are coupled by their dipole--dipole interaction only, the net magnetization always vanishes for a vanishing external field as known from classical electrodynamics \cite{jackson_1999}.

However, in ferromagnetic materials, the spins are subject to the so-called exchange interaction.
For two localized spins, this quantum mechanical effect energetically favors a parallel over an antiparallel spin alignment.
The origin of this energy contribution can be attributed to the Coulomb energy of the respective two-particle system, typically consisting of two electrons.
Depending on the spin-alignment, the two-particle wave function is either symmetric or anti-symmetric leading to higher expectation value of the distance, and thus a lower expectation value of the Coulomb energy, in case of an parallel alignment.
The classical description of the exchange interaction is given by the Heisenberg model.
Details on its derivation can be found in any textbook on quantummechanics, e.g. \cite{griffiths_1994}.
The Heisenberg formulation of the exchange energy of two unit spins $\vec{s}_i$ and $\vec{s}_j$ is defined as
\begin{equation}
  \Eex_{ij} = - J \, \vec{s}_i \cdot \vec{s}_j
  \label{eq:energetics_heisenberg}
\end{equation}
with $J$ being the so-called exchange integral.
With respect to the continuous magnetization field $\vec{m}(\vec{x})$, the exchange energy associated with all couplings of a single spin site $\vec{x}$ is given by
\begin{align}
  \Eex_{\vec{x}}
  &= - \sum_i \frac{J_i}{2} \, \vec{m}(\vec{x}) \cdot \vec{m}(\vec{x} + \Delta\vec{x}_i) \label{eq:energetics_m_heisenberg}\\
  &= - \sum_i \frac{J_i}{2} \left[ 1 - \frac{1}{2} ( \vnabla \vec{m}^T \cdot \Delta\vec{x}_i )^2 \right] + \mathcal{O}(\Delta\vec{x}_i^3) \label{eq:energetics_m_heisenberg_taylor}
\end{align}
where the index $i$ runs over all exchange coupled spin sites at positions $\vec{x} + \Delta\vec{x}_i$, and $J_i$ denotes the exchange integral with the respective spin.
Expression \eqref{eq:energetics_m_heisenberg_taylor} is obtained by application of the unit-vector identity $(\vec{n}_1 - \vec{n}_2)^2 = 2 - 2 \vec{n}_1 \cdot \vec{n}_2$ and performing Taylor expansion of lowest order.
The exchange integral $J_i$ highly depends on the distance of the spin sites.
Hence, significant contributions to the exchange energy are only provided by nearby spins, usually next neighbors.
The transition from the discrete Heisenberg model to a continuous expression for the total exchange energy is done by integrating \eqref{eq:energetics_m_heisenberg_taylor} while considering a regular spin lattice, i.e. a regular spacing of the spin sites $\vec{x}$ as well as identical $J_i$ and $\Delta\vec{x}_i$ for each site.
In the most general form this procedure yields
\begin{equation}
  \Eex  = C + \int_{\Omega_m} \sum_{i,j,k} A_{jk} \pdiff{m_i}{x_j} \pdiff{m_i}{x_k} \dx
  \label{eq:energetics_exchange_continuous_3}
\end{equation}
where the coefficients of the matrix $A_{jk}$ depend on the crystal structure and the resulting exchange couplings of the spins in the magnetic body.
The term $C$ results from the integration of the constant part of \eqref{eq:energetics_m_heisenberg_taylor} and is usually neglected since it does not depend on $\vec{m}$ and thus only gives a constant offset to the energy without changing the physics of the system.
The matrix $A$ can always be diagonalized by a proper choice of coordinate system \cite{doering_1966} which yields
\begin{equation}
  \Eex = \int_{\Omega_m} \sum_{i,j} A_{j} \left( \pdiff{m_i}{x_j'} \right)^2 \dx'.
\end{equation}
For cubic and isotropic lattice structures, the exchange coupling constants $A_j$ simplify further to the scalar exchange constant $A$ which results in the typical micromagnetic expression for the exchange energy
\begin{equation}
  \Eex
  = \int_{\Omega_m} A \sum_{i,j} \left( \pdiff{m_i}{x_j} \right)^2 \dx
  = \int_{\Omega_m} A (\vnabla \vec{m})^2 \dx
  \label{eq:energetics_exchange}
\end{equation}
where $(\vnabla \vec{m})^2 = \sum_{i,j} (\pdiff{m_i}{x_j})^2$ is to be understood as a Frobenius inner product.
Although derived for localized spins and isotropic lattice structures, this energy expression turns out to accurately describe a large number of materials including band magnets and anisotropic materials \cite{hubert_1998}.
This is explained by the fact, that \eqref{eq:energetics_exchange} exactly represents the lowest order phenomenological energy expression that penalizes inhomogeneous magnetization configurations.

\subsection{Demagnetization energy}\label{sec:energetics_demag}
The demagnetization energy accounts for the dipole--dipole interaction of a magnetic system.
This energy contribution, that is also referred to as magnetostatic energy or stray-field energy, owes its name to the fact that magnetic systems energetically favor macroscopically demagnetized states if they are subject to dipole--dipole interaction only.
For a continuous magnetization $\vec{M} = \Ms \vec{m}$ the demagnetization energy can be derived from classical electromagnetics.
Assuming a vanishing electric current $\je = 0$, Maxwell's macroscopic equations reduce to
\begin{align}
  \vnabla \cdot  \vec{B} &= 0   \label{eq:energetics_demag_zero_div} \\
  \vnabla \times \hdemag & = 0  \label{eq:energetics_demag_zero_curl}
\end{align}
where the magnetic flux $\vec{B}$ can be written in terms of the magnetic field $\hdemag$ and the magnetization $\vec{M}$
\begin{equation}
  \vec{B} = \mu_0 (\hdemag + \vec{M}).
\end{equation}
According to \eqref{eq:energetics_demag_zero_curl}, the magnetic field $\hdemag$ is conservative and thus has a scalar potential $\hdemag = - \vnabla u$.
With these definitions \eqref{eq:energetics_demag_zero_div} -- \eqref{eq:energetics_demag_zero_curl} can be reduced to the single equation
\begin{equation}
  \vnabla \cdot ( - \vnabla u + \vec{M} ) = 0
  \label{eq:energetics_demag_poisson_raw}
\end{equation}
which is solved in the whole space $\mathbb{R}^3$.
Assuming a localized magnetization configuration, the boundary conditions are given in an asymptotical fashion by
\begin{equation}
  u(\vec{x}) = \mathcal{O}(1 / |\vec{x}|) \text{ for } |\vec{x}| \rightarrow \infty
  \label{eq:energetics_demag_open_boundary}
\end{equation}
which is referred to as open boundary condition since the potential is required to drop to zero at infinity.
The defining equation \eqref{eq:energetics_demag_poisson_raw} is often transformed in Poisson's equation
\begin{equation}
  \Delta u = \vnabla \cdot \vec{M}.
  \label{eq:energetics_demag_poisson}
\end{equation}
However, in contrast to the original equation \eqref{eq:energetics_demag_poisson_raw}, the divergence on the right-hand side of \eqref{eq:energetics_demag_poisson} may become singular at the boundary of the magnetic material in the case of a localized magnetization.
In this case, \eqref{eq:energetics_demag_poisson} is well defined only in a distributional sense.

\begin{figure}
  \centering
  \includegraphics{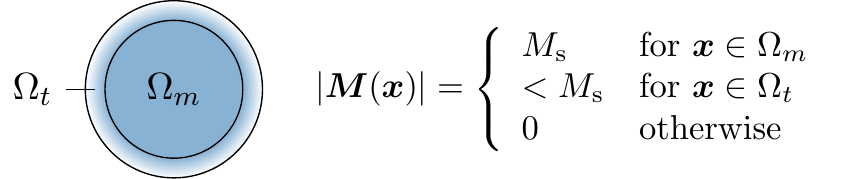}
	\caption{
    Limiting procedure for the demagnetization-field calculation of a finite magnetic body.
    The magnetization $\vec{M}$ is defined in the magnetic region $\Omega_m$ and continuously decreases in the transitional shell region $\Omega_t$.
    The overall continuous and differentiable definition of $\vec{M}$ allows for the application of Green's theorem.
	}
	\label{fig:energetics_demag_limit}
\end{figure}

The potential $u$ can be expressed in terms of an integral equation by considering the well-known fundamental solution to the Laplacian that naturally fulfills the required open boundary conditions \cite{jackson_1999}
\begin{equation}
  u(\vec{x}) = - \frac{1}{4 \pi} \int_{\mathbb{R}^3} \frac{\vnabla' \cdot \vec{M}(\vec{x}')}{|\vec{x} - \vec{x}'|} \dx'.
  \label{eq:energetics_demag_greens_r3}
\end{equation}
For a localized magnetization with sharp boundaries, this solution, like \eqref{eq:energetics_demag_poisson}, suffers from a singular divergence at the boundary of the magnetic body.
However, this singularity is integrable as demonstrated in the following.
Consider a finite magnet with $|\vec{M}(\vec{x})| = \Ms$ for $\vec{x} \in \Omega_m$ surrounded by a thin transition shell $\Omega_t$ where the magnetization continuously decays to zero, see Fig.~\ref{fig:energetics_demag_limit}.
In this case, the integration over $\mathbb{R}^3$ in \eqref{eq:energetics_demag_greens_r3} can be reduced to an integration over $\Omega_m \cup \Omega_t$ since the magnetization, and thus the integrand of \eqref{eq:energetics_demag_greens_r3}, vanishes outside the magnet.
Further, the integral is split into integration over $\Omega_m$ and $\Omega_t$, and the integral over $\Omega_t$ is transformed with Green's theorem
\begin{equation}
  \int_{\Omega_t} \frac{\vnabla' \cdot \vec{M}(\vec{x}')}{|\vec{x} - \vec{x}'|} \dx'
  =
  \int_{\partial \Omega_t} \frac{\vec{M}(\vec{x}') \cdot \vec{n}}{|\vec{x} - \vec{x}'|} \ds'
  - \int_{\Omega_t} \vec{M}(\vec{x}') \cdot \vnabla' \frac{1}{|\vec{x} - \vec{x}'|} \dx'
  \label{eq:energetics_demag_green}
\end{equation}
where $\ds'$ denotes the areal measure to $\vec{x}'$ and $\vec{n}$ is an outward-pointing normal vector.
In order to obtain the potential for an ideal magnet with a sharp transition of the magnetic region $\Omega_m$ to the air region, we consider the limit of a vanishing transition region $\Omega_t \rightarrow 0$.
In this case, the right-hand side of \eqref{eq:energetics_demag_green} reduces to the boundary integral.
Furthermore, the boundary integral vanishes for the outer boundary of $\Omega_t$, because of a vanishing magnetization $\vec{M}$.
The inner boundary, however, coincides with with the boundary of the magnetic region $\partial \Omega_m$ except for its orientation.
A complete integral form for the magnetic scalar potential of an ideal localized magnet in region $\Omega_m$ accordingly reads
\begin{equation}
  u(\vec{x}) =
  - \frac{1}{4 \pi}  \left[
    \int_{\Omega_m} \frac{\vnabla' \cdot \vec{M}(\vec{x}')}{|\vec{x} - \vec{x}'|} \dx'
  - \int_{\partial \Omega_m} \frac{\vec{M}(\vec{x}') \cdot \vec{n}}{|\vec{x} - \vec{x}'|} \ds'
  \right].
  \label{eq:energetics_demag_jackson}
\end{equation}
In analogy to the integral equation for the electric field, the terms $\rho = - \vnabla \cdot \vec{M}$ and $\sigma = \vec{M} \cdot \vec{n}$ are often referred to as magnetic volume charges and magnetic surfaces charges respectively.
An alternative integral expression for the potential is obtained by applying Green's theorem to \eqref{eq:energetics_demag_jackson}
\begin{equation}
  u(\vec{x}) =
  \frac{1}{4 \pi} 
  \int_{\Omega_m} \vec{M}(\vec{x}') \cdot \vnabla' \frac{1}{|\vec{x} - \vec{x}'|} \dx'.
  \label{eq:energetics_demag_potential_convolution}
\end{equation}
Starting from this formulation, the demagnetization field $\hdemag$ can be expressed as a convolution
\begin{equation}
  \hdemag(\vec{x}) =
  - \vnabla u(\vec{x}) =
  \int_{\Omega_m} \tensor{N}(\vec{x} - \vec{x}') \vec{M}(\vec{x}') \dx'
	\label{eq:energetics_demag_field}
\end{equation}
with the so-called demagnetization tensor $\tensor{N}$ given by
\begin{equation}
  \tensor{N}(\vec{x} - \vec{x}') = - \frac{1}{4 \pi} \vnabla \vnabla' \frac{1}{|\vec{x} - \vec{x}'|}.
\end{equation}
According to classical electrodynamics, the energy connected to the demagnetization field is given by
\begin{equation}
  \Edemag = - \frac{\mu_0}{2} \int_{\Omega_t} \vec{M} \cdot \hdemag \dx
  \label{eq:energetics_demag}
\end{equation}
where the factor $1/2$ accounts for the quadratic dependence of the energy on the magnetization $\vec{M}$.

\begin{figure}
  \centering
  \includegraphics{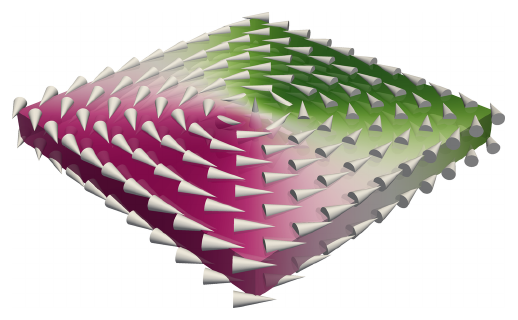}

  \caption{
    Magnetic vortex configuration in a square-shaped thin film.
    The magnetization can be roughly divided into four triangular domains, each aligned with one of the film edges.
    In the vortex core the magnetization points out of plane.
  }
  \label{fig:vortex}
\end{figure}
The competition of the demagnetization energy and the exchange energy leads to the formation of magnetic domains.
A graphic example for this effect is the magnetic vortex structure depicted in Fig.~\ref{fig:vortex}.
In order to minimize surface charges $\sigma$, that contribute to the demagnetization energy, the magnetization field aligns parallel with edges and surfaces, which exlains the curl-like configuration.
The exchange energy favors a parallel alignment of the magnetization which leads to the creation of four distinct, almost homogeneously magnetized, triangular domains, each aligned with one of the square's edges.
A perfect in-plane curl configuration, which completely avoids surface charges, is very unfavorable with respect to the exchange energy because it leads to a singularity in the center of the curl.
In order to reduce the exchange energy, the magnetization rotates out-of-plane in a distinct area around the center of the vortex, called the vortex core.

\subsection{Crystalline anisotropy energy}\label{sec:energetics_anisotropy}
Another important contribution to the total free energy of a magnet is the anisotropy energy that favors the parallel alignment of the magnetization to certain axes referred to as easy axes.
The origin of this energy lies in the spin-orbit coupling either due to an anisotropic crystal structure or due to lattice deformation at material interfaces \cite{hubert_1998}.
Depending on the symmetry of these anisotropies, the respective material will exhibit one or multiple easy axes.
These axes are undirected and thus the energy does not depend on the sign of the magnetization
\begin{equation}
	\Eaniso(\vec{m}) = \Eaniso(-\vec{m}).
	\label{eq:energetics_aniso_symmetry}
\end{equation}
For the simplest case of a single easy axis, the anisotropy energy is given by
\begin{equation}
  \Eanisou = - \int_{\Omega_m} [
    K_\text{u1} ( \vec{m} \cdot \vec{e}_\text{u} )^2
  + K_\text{u2} ( \vec{m} \cdot \vec{e}_\text{u} )^4
  + \mathcal{O}(\vec{m}^6)
  ] \dx
  \label{eq:energetics_aniso_uniaxial}
\end{equation}
where $\vec{e}_\text{u}$ is a unit vector parallel to the easy axis and $K_\text{u1}$ and $K_\text{u2}$ are the scalar anisotropy constants.
This phenomenological expression is obtained by symmetry considerations.
For a uniaxial anisotropy, the energy may depend only on the angle between the magnetization and easy axis $\vec{m}\cdot\vec{e}_u$.
Furthermore only even powers in $\vec{m}\cdot\vec{e}_u$ are considered in order to fulfill condition \eqref{eq:energetics_aniso_symmetry}.
Uniaxial anisotropy typically occurs in materials with a hexagonal or tetragonal crystal structure, e.g. cobalt.

Materials with cubic lattice symmetry such as iron, which has a body-centered cubic structure, exhibit three easy axes $\vec{e}_i$ which are pairwise orthogonal
\begin{equation}
  \vec{e}_i \cdot \vec{e}_j = \delta_{ij}.
\end{equation}
Like for the uniaxial anisotropy, the expression for the cubic anisotropy energy is developed as series in magnetization components along the easy axes up to sixth order
\begin{equation}
  \Eanisoc = \int_\Omega [
    K_\text{c1} (m_1^2 m_2^2 + m_2^2 m_3^2 + m_3^2 m_1^2)
  + K_\text{c2} m_1^2 m_2^2 m_3^2
  ] \dx
  \label{eq:energetics_aniso_cubic}
\end{equation}
where $m_i = \vec{e}_i\cdot\vec{m}$ is the projection of the magnetization $\vec{m}$ on the anisotropy axis $\vec{e}_i$.
Only contributions compatible with the symmetry condition \eqref{eq:energetics_aniso_symmetry} are considered.
Moreover, the resulting expression is required to be constant under permutation of magnetization components $m_i$ in order to have a cubic symmetry.

While the magnetization prefers to align parallel to the respective axes in the case of positive anisotropy constants $K_\text{u1}$, $K_\text{u2}$, $K_\text{c1}$, $K_\text{c2}$, the magnetization avoids a parallel configuration for negative anisotropy constants.
In the case of uniaxial anisotropy, this leads to an easy-plane anisotropy.
In the case of cubic anisotropy, this leads to four easy axes as is the case for nickel which has a face-centered cubic structure.

Both, equations \eqref{eq:energetics_aniso_uniaxial} and \eqref{eq:energetics_aniso_cubic} hold for magnetic anisotropies in bulk material.
If magnetic anisotropy is caused at material interfaces, either due to lattice deformation or due to the electric band structure, the energy depends on the magnetization configuration $\vec{m}$ at this interface only.
The energy for such a surface anisotropy is obtained by similar expressions as \eqref{eq:energetics_aniso_uniaxial} and \eqref{eq:energetics_aniso_cubic}.
However, instead of integrating over the magnetic volume $\Omega_m$ the integration in this case has to be carried out over the respective interface $\partial \Omega_m$ only.

Although being derived only phenomenologically, the expressions \eqref{eq:energetics_aniso_uniaxial} and \eqref{eq:energetics_aniso_cubic} have proven to describe anisotropy effects with high accuracy.
For many application it is even sufficient to consider the lowest order contributions only and setting the higher order constants $K_\text{u2}$ and $K_\text{c2}$ to zero.

\subsection{Antisymmetric exchange energy}
As discovered by Dzyaloshinskii \cite{dzyaloshinsky1958thermodynamic} and Moriya \cite{moriya1960anisotropic}, neighboring spins can be subject to an antisymmetric exchange interaction in addition to the regular exchange interaction discussed in Sec.\ref{sec:energetics_exchange}.
This effect, that is often referred to as Dzyaloshinskii-Moriya interaction (DMI), is caused by the spin-orbit coupling in certain material systems.
The general antisymmetric exchange energy of two spins $\vec{s}_i$ and $\vec{s}_j$ is given as
\begin{equation}
  \Edmi_{ij} = \vec{d}_{ij} \cdot (\vec{s}_i \times \vec{s}_j)
\end{equation}
where the vector $\vec{d}_{ij}$ depends on the symmetry of the system.
A typical system that gives rise to DMI is a magnetic layer with an interface to a heavy-metal layer.
In this case, the antisymmetric exchange between two neighboring magnetic spins near the interface is mediated by a single atom in the heavy metal layer and the vector $\vec{d}_{ij}$ is given as 
\begin{equation}
  \vec{d}_{ij} = d (\Delta\hat{\vec{x}} \times \vec{e}_\text{d})
\end{equation}
where $d$ is a scalar coupling constant, $\Delta\hat{\vec{x}} = \Delta\vec{x} / |\Delta\vec{x}|$ is a unit vector pointing from spin site $i$ to spin site $j$, and $\vec{e}_\text{d}$ is the interface normal.
The transition to continuum theory is done similar to the exchange interaction.
Namely, in a first step, the energy of the couplings for a single spin site is expressed in terms of the continuous magnetization field $\vec{m}$ and the magnetization at the neighboring site $\vec{m}(\vec{x} + \Delta \vec{x})$ is expanded in powers of $\Delta\vec{x}$ to the lowest order
\begin{align}
  \Edmii_{\vec{x}}
	=& \sum_i
  \frac{d_i}{2} \big[ \Delta\hat{\vec{x}}_i \times \vec{e}_\text{d} \big] \cdot \big[ \vec{m}(\vec{x}) \times \vec{m}(\vec{x} + \Delta\vec{x}_i) \big]\\
	=& \sum_i
  \frac{d_i}{2} \big[ \Delta\hat{\vec{x}}_i \cdot \vec{m} \big] \big[\vec{e}_\text{d} \cdot (\vec{m} + \vnabla\vec{m}^T\Delta\vec{x}_i) \big]
  - \frac{d_i}{2} \big[ \Delta\hat{\vec{x}}_i \cdot (\vec{m} + \vnabla\vec{m}^T\Delta\vec{x}_i)\big] \big[\vec{e}_\text{d} \cdot \vec{m}\big] + \mathcal{O}(\Delta\vec{x}_i^2)\\
	=& \sum_i
  \frac{d_i}{2} \big[ \Delta\hat{\vec{x}}_i \cdot \vec{m} \big] \big[\vec{e}_\text{d} \cdot (\vnabla\vec{m}^T\Delta\vec{x}_i) \big]
  - \frac{d_i}{2} \big[ \Delta\hat{\vec{x}}_i \cdot (\vnabla\vec{m}^T\Delta\vec{x}_i) \big] \big[\vec{e}_\text{d} \cdot \vec{m}\big] + \mathcal{O}(\Delta\vec{x}_i^2)
\end{align}
where the vector identity $(\vec{a}\times\vec{b})\cdot(\vec{c}\times\vec{d}) = (\vec{a}\cdot\vec{c})(\vec{b}\cdot\vec{d}) - (\vec{a}\cdot\vec{d})(\vec{b}\cdot\vec{c})$ was used and the summation is carried out over the coupled neighboring spins.
Performing integration and assuming isotropic coupling $d_i$ as well as an isotropic lattice spacing $\Delta\vec{x}_i$, similar to the exchange interaction in Sec.~\ref{sec:energetics_exchange}, yields the continuous expression
\begin{equation}
  \Edmii = \int_{\Omega_m} \Di \big[ \vec{m} \cdot \vnabla (\vec{e}_\text{d} \cdot \vec{m}) - (\vnabla \cdot \vec{m}) (\vec{e}_\text{d} \cdot \vec{m}) \big] \dx
  \label{eq:energetics_dmi_interface}
\end{equation}
for the total antisymmetric exchange energy for interface DMI.
The scalar coupling constant $\Di$ depends on the coupling constants $d_i$ as well as the relative positions $\Delta\vec{x}_i$.

Another class of materials exhibiting DMI, are magnetic bulk materials lacking inversion symmetry \cite{yu2010real,yu2012magnetic}.
For these materials, the coupling vector $\vec{d}_{ij}$ is given as 
\begin{equation}
  \vec{d}_{ij} = - d \Delta\hat{\vec{x}}
\end{equation}
which results in the following energy $E_{\vec{x}}$ for the couplings of a single spin site $\vec{x}$
\begin{align}
  \Edmib_{\vec{x}}
	&= - \sum_i
    \frac{d_i}{2} \Delta\hat{\vec{x}}_i \big[\vec{m}(\vec{x}) \times \vec{m}(\vec{x} + \Delta\vec{x}_i) \big]\\
	&= - \sum_i
    \frac{d_i}{2} \Delta\hat{\vec{x}}_i \big[\vec{m}(\vec{x}) \times (\vnabla\vec{m}^T \Delta\vec{x}_i) \big].
\end{align}
Again, assuming isotropic coupling $d_i$ and lattice spacing $\Delta\vec{x}_i$ results in the continuous formulation for the energy
\begin{equation}
  \Edmib = \int_{\Omega_m} \Db \vec{m} \cdot (\vnabla \times \vec{m}) \dx
  \label{eq:energetics_dmi_bulk}
\end{equation}
with the coupling constant $\Db$ depending on the atomistic coupling constants $d_i$ and the lattice spacing $\Delta\vec{x}_i$.
Besides the prominent interface and bulk DMI, further antisymmetric exchange couplings are defined by Lifshitz invariants \cite{bogdanov2001chiral,cortes2013influence}.

\begin{figure}
  \centering
  \includegraphics{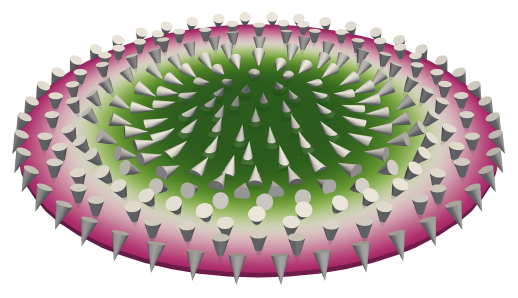}
  \caption{
    Magnetic skyrmion configuration in a circular thin film.
    This configuration is characterized by a continuous rotation of the magnetization $\vec{m}$ across the center.
  }
  \label{fig:skyrmion}
\end{figure}
The antisymmetric exchange counteracts the regular exchange energy that favors homogeneous magnetization configurations and penalizes domain walls.
It gives rise to a magnetization configuration called skyrmion, see Fig.~\ref{fig:skyrmion}.
A skyrmion is characterized by a continuous rotation of the magnetization field on any lateral axis crossing the center.
It has topological charge meaning that the skyrmion configuration is not continuously transformable into the homogeneous ferromagnetic state.

\subsection{Interlayer-exchange energy}
The magnetic layers of a multilayer structure may be exchange coupled even when separated by a nonmagnetic spacer layer.
This coupling, which was first proposed by Ruderman and Kittel \cite{ruderman1954indirect}, is mediated by the conducting electrons of the nonmagnetic layer.
The coupling constant $A$ shows oscillatory behavior with respect to the thickness of the spacer layer, i.e. depending on its thickness, the coupling of the magnetic layers may be either ferromagnetic or antiferromagnetic.
This effect was described in a more generalized theory by Kasuya \cite{kasuya1956theory} and Yosida \cite{yosida1957magnetic} and is referred to as Ruderman-Kittel-Kasuya-Yosida (RKKY) interaction.

In the continuous approximation, the interaction is assumed to couple the interface between one magnetic layer and the spacer layer $\Gamma_1$ and the interface between the other magnetic layer and the spacer layer $\Gamma_2$.
These interfaces are assumed to have equal size.
Integration of the Heisenberg interaction \eqref{eq:energetics_heisenberg} yields the continuous expression for the interlayer-exchange energy
\begin{equation}
  \Eiex = - \int_{\Gamma_1} A \, \vec{m}(\vec{x}) \cdot \vec{m}[P(\vec{x})] \ds
  \label{eq:energetics_interlayer_exchange_energy}
\end{equation}
with $A$ being the exchange constant whose sign and strength depend on the thickness of the spacer layer and $P: \Gamma_1 \rightarrow \Gamma_2$ being an isomorphism that maps any point on $\Gamma_1$ to its nearest point on $\Gamma_2$.

\begin{figure}
  \centering
  \includegraphics{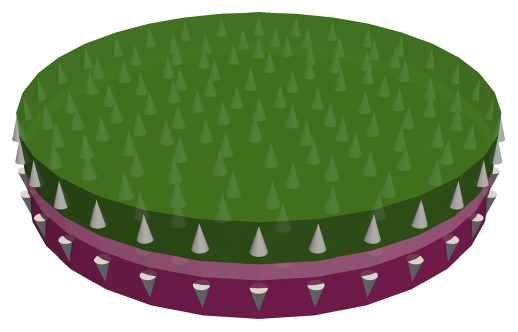}
  \caption{
    Synthetic antiferromagnet.
    Two magnetic layers with perpendicular crystalline anisotropy are antiferromagnetically coupled through a nonmagnetic layer by the RKKY interaction.
  }
  \label{fig:saf}
\end{figure}
The RKKY interaction is often exploited in order to build so-called synthetic antiferromagnets, see Fig.~\ref{fig:saf}.
For this purpose, the thickness of the spacer layer is chosen such that $A$ is negative which results in an antiferromagnetic coupling of the magnetic layers.
Synthetic antiferromagnets are important for applications due to their stability and lack of strayfield.

\subsection{Other energy contributions and effects}
While this work will focus on the energy contributions introduced above, there are numerous additional energy contributions and other effects that may play important roles in certain systems \cite{brown_1963}.
For instance, the effect of magnetostriction coupled the mechanical properties of magnetic materials to its magnetization configuration\cite{fabian1996include,shu2004micromagnetic}.
If magnetic systems are subject to charge currents, eddy currents \cite{torres2003micromagnetic,hrkac2005three} and the Oersted field \cite{hertel2014hybrid} need to be considered.

Another important area of research is finite-temperature micromagnetics.
Various approaches have been proposed in order account for temperature effects in micromagnetics, among them Langevin dynamics \cite{scholz2003scalable,berkov2002fast,chubykalo2002langevin} and the Landau-Lifshitz-Bloch equation \cite{garanin1997fokker,atxitia2007micromagnetic,evans2012stochastic}.
However, this comprehensive topic is out of the scope of this article.

\section{Static micromagnetics}\label{sec:static}
Static micromagnetics is the theory of stable magnetization configurations and hence a valuable model for the investigation of material properties such as hysteresis.
The prerequisite for a stable magnetization configuration is a minimum of the total free energy $E$ of the system with respect to its magnetization $\vec{m}$.
In order to be a valid micromagnetic solution, the solution $\vec{m}$ is further required to fulfill the micromagnetic unit-sphere constraint
\begin{equation}
  \min E(\vec{m}) \quad\text{subject to}\quad |\vec{m}(\vec{x})| = 1.
	\label{eq:static_min_e}
\end{equation}
Since the solution variable $\vec{m}$ is a continuous vector field, variational calculus is applied in order to solve for an energetic minimum.
A necessary condition for a minimum is a functional differential $\delta E$ that vanishes for arbitrary test functions $\vec{v} \in V_m$ with $V_m$ being the function space of the magnetization $\vec{m}$
\begin{equation}
  \delta E(\vec{m}, \vec{v}) =
	\diff{}{\epsilon} E(\vec{m} + \epsilon \vec{v}) =
	\lim_{\epsilon \rightarrow 0} \frac{E(\vec{m} + \epsilon\vec{v}) - E(\vec{m})}{\epsilon} =
	0 \quad\forall\quad \vec{v} \in V_m.
	\label{eq:static_functional_differential}
\end{equation}
An alternative formulation for this condition can be stated in terms of the functional derivative $\vdiff{E}{\vec{m}}$ that is defined as
\begin{equation}
  \int_{\Omega_m} \vdiff{E}{\vec{m}} \cdot \vec{v} \dx =
	\delta E(\vec{m}, \vec{v})
	\quad\forall\quad
	\vec{v} \in V_m^0
	\label{eq:static_functional_derivative}
\end{equation}
where the function space $V_m^0 \subset V_m$ includes only functions of $V_m$ that vanish on the boundary $v(\partial \Omega_m) = 0$.
That said, depending on the considered energy $E$, the differential $\delta E$ as defined in \eqref{eq:static_functional_differential} in general differs from the left-hand side of \eqref{eq:static_functional_derivative} by a boundary integral, i.e.
\begin{equation}
	\delta E(\vec{m}, \vec{v}) =
  \int_{\Omega_m} \vdiff{E}{\vec{m}} \cdot \vec{v} \dx +
  \int_{\partial \Omega_m} \vec{B}(\vec{m}) \cdot \vec{v} \ds
	\quad\forall\quad
	\vec{v} \in V_m.
\end{equation}
This means, that the knowledge of the functional derivative \eqref{eq:static_functional_derivative} is not sufficient in order to solve the minimization problem \eqref{eq:static_min_e}.
In general, additional boundary conditions defined by $\vec{B}(\vec{m})$ have to be considered.
All variational considerations so far do not account for the unit-sphere constraint $|\vec{m}|=1$.
This constraint can be incorporated by a Lagrange multiplier technique where the modified functional $E_\lambda(\vec{m}, \lambda, \mu)$, given by
\begin{equation}
	E_\lambda(\vec{m}, \lambda, \mu) = E(\vec{m}) + \int_{\Omega_m} \lambda(\vec{x}) \big( |\vec{m}|^2 - 1 \big) \dx
    + \int_{\partial \Omega_m} \mu(\vec{x}) \big( |\vec{m}|^2 - 1 \big) \ds,
	\label{eq:static_elambda}
\end{equation}
is minimized with respect to both the magnetization $\vec{m}$ and the Lagrange multiplier fields $\lambda$ and $\mu$ implementing the constraint on the volume and surface respectively.
The solution to this minimization problem is again obtained by variational calculus where the variations of the solution variables $\vec{v}_m$, $v_\lambda$ and $v_\mu$ can be treated separately
\begin{align}
  \delta E_\lambda(\{\vec{m}, \lambda, \mu\}, \vec{v}_m) &= 0 \quad\forall\quad \vec{v}_m \in V_m \label{eq:static_lagrange_1}\\
  \delta E_\lambda(\{\vec{m}, \lambda, \mu\}, v_\lambda) &= 0 \quad\forall\quad v_\lambda \in V_\lambda \label{eq:static_lagrange_2}\\
  \delta E_\lambda(\{\vec{m}, \lambda, \mu\}, v_\mu)     &= 0 \quad\forall\quad v_\mu\in V_\mu \label{eq:static_lagrange_3}
\end{align}
where $V_\lambda$ and $V_\mu$ are appropriate function spaces for the variation of $\lambda$ and $\mu$ respectively.
Expanding \eqref{eq:static_lagrange_1} considering the definition \eqref{eq:static_elambda} yields
\begin{align}
    \delta E_\lambda(\{\vec{m}, \lambda, \mu\}, \vec{v}_m)
	=&
	\delta E(\vec{m}, \vec{v}_m) \nonumber \\
    &+
	\diff{}{\epsilon}\left[
	\int_{\Omega_m} \lambda
	\big( |\vec{m} + \epsilon \vec{v}_m|^2 - 1\big) \dx
	\right]_{\epsilon = 0} \nonumber \\
    &+
	\diff{}{\epsilon}\left[
	\int_{\partial \Omega_m} \mu
	\big( |\vec{m} + \epsilon \vec{v}_m|^2 - 1\big) \ds
	\right]_{\epsilon = 0}\\
	=&
    \int_{\Omega_m} \vdiff{E}{\vec{m}} \cdot \vec{v}_m \dx +
    \int_{\partial \Omega_m} \vec{B} \cdot \vec{v}_m \ds \nonumber \\
    &
	+ 2 \int_{\Omega_m} \lambda \vec{m} \cdot \vec{v}_m \dx
	+ 2 \int_{\partial \Omega_m} \mu \vec{m} \cdot \vec{v}_m \ds \label{eq:static_lagrange_m_1}.
\end{align}
Since \eqref{eq:static_lagrange_m_1} has to vanish for arbitrary $\vec{v}_m \in V_m$, it also vanishes for test functions with vanishing boundary values $\vec{v}_m \in V_m^0$.
Hence the functional derivative of the energy has to fulfill 
\begin{equation}
  \vdiff{E}{\vec{m}} = -2 \lambda\vec{m}
\end{equation}
in $\Omega_m$  which is required to hold for arbitrary $\lambda \in V_\lambda$.
This condition is satisfied if and only if $\vdiff{E}{\vec{m}}$ is parallel to $\vec{m}$ and hence
\begin{equation}
    \vec{m} \times \vdiff{E}{\vec{m}} = 0
	\label{eq:static_brown_condition}
\end{equation}
which is exactly Brown's condition \cite{brown_1963}.
Testing \eqref{eq:static_lagrange_m_1} with functions that are defined on the boundary only $\vec{v}(\Omega_m \setminus \partial \Omega_m) = 0$ and considering the surface Lagrange multiplier $\mu$ in the same manner as above yields the additional boundary condition
\begin{equation}
    \vec{m} \times \vec{B} = 0.
    \label{eq:static_boundary_condition}
\end{equation}
Moreover, inserting \eqref{eq:static_elambda} into \eqref{eq:static_lagrange_2} yields
\begin{align}
  \delta E_\lambda(\{\vec{m}, \lambda, \mu\}, v_\lambda)
	&= 
  \diff{}{\epsilon}\left[
  \int_{\Omega_m} (\lambda + \epsilon v_\lambda) 
  \big( |\vec{m}|^2 - 1\big) \dx
  \right]_{\epsilon = 0}\\
  &=
  \int_{\Omega_m} v_\lambda \big( |\vec{m}|^2 - 1 \big) \dx\\
  &= 0
\end{align}
which is required to hold for arbitrary $v_\lambda \in V_\lambda$ and thus represents the micromagnetic constraint $|\vec{m}|^2 = 1$.
Due to the interface Lagrange multiplier $\mu$, this constraint is further specifically enforced on the boundary by \eqref{eq:static_lagrange_3}.
In the following, the functional derivatives and boundary conditions for the energy contributions introduced in Sec.~\ref{sec:energetics} will be discussed in detail.

\subsection{Zeeman energy}
The energy differential $\vdiff{E}{\vec{m}}$ for the Zeeman energy is obtained by variation of \eqref{eq:energetics_zeeman} which yields
\begin{align}
  \delta \Ezee(\vec{m}, \vec{v}_m)
	&=
  \diff{}{\epsilon}\left[
	  - \mu_0 \int_{\Omega_m} \Ms (\vec{m} + \epsilon \vec{v}_m) \cdot \hzee \dx
  \right]_{\epsilon = 0}\\
	&=
	- \int_{\Omega_m} \mu_0 \Ms
  \hzee \cdot \vec{v}_m \dx.
\end{align}
The variation does not give rise to any additional boundary integral.
Hence, the derivative and boundary term $\vec{B}$ for the Zeeman energy are given by
\begin{align}
  \vdiff{\Ezee}{\vec{m}} &= - \mu_0 \Ms \hzee\\
  \vec{B} &= 0.
\end{align}

\subsection{Exchange energy}\label{sec:energetics_static_exchange}
The differential $\delta \Eex$ for the exchange energy is derived from \eqref{eq:energetics_exchange} resulting in
\begin{align}
  \delta \Eex(\vec{m}, \vec{v}_m)
	&=
  \diff{}{\epsilon}\left[
		\int_{\Omega_m} A \big[ \vnabla (\vec{m} + \epsilon \vec{v}_m) \big]^2 \dx
  \right]_{\epsilon = 0}\\
	&=
  2 \int_{\Omega_m} A \vnabla \vec{m} : \vnabla \vec{v}_m \dx \label{eq:static_exchange_variation}\\
	&=
  -2 \int_{\Omega_m} \left[ \vnabla \cdot (A \vnabla \vec{m}) \right] \cdot \vec{v}_m \dx
	+ 2 \int_{\partial \Omega_m} A \pdiff{\vec{m}}{\vec{n}} \cdot \vec{v}_m \ds.
\end{align}
Here, integration by parts is performed in order to eliminate spatial derivatives of the test functions $\vec{v}_m$.
The resulting volume integral is of the same form as the integral in \eqref{eq:static_functional_derivative} which enables the identification of the functional derivative $\vdiff{\Eex}{\vec{m}}$.
However, this necessary step also gives rise to a surface integral and thus to a boundary term $\vec{B}$.
The resulting derivative and boundary term for the exchange energy read
\begin{align}
  \vdiff{\Eex}{\vec{m}} &= - 2 \vnabla \cdot (A \vnabla \vec{m}) \label{eq:static_exchange_field}\\
  \vec{B} &= 2 A \pdiff{\vec{m}}{\vec{n}} \label{eq:static_exchange_bc}
\end{align}
where \eqref{eq:static_exchange_field} can be simplified to $\vdiff{\Eex}{\vec{m}} = - 2 A \Delta \vec{m}$ if $A$ is assumed constant throughout the magnetic region $\Omega_m$.

\subsection{Demagnetization energy}
The differential for the demagnetization energy is obtained similarly to the differential for the Zeeman energy.
A decisive difference to the Zeeman energy, however, is the linear relation of the demagnetization field $\hdemag(\vec{M})$ to the magnetization $\vec{M}$.
The variation of the magnetization therefore leads to an additional factor of 2 which results in the differential
\begin{align}
  \delta \Edemag(\vec{m}, \vec{v}_m)
	&=
  \diff{}{\epsilon}\left[
	  - \frac{\mu_0}{2} \int_{\Omega_m} \Ms (\vec{m} + \epsilon \vec{v}_m) \cdot \hdemag(\vec{m} + \epsilon \vec{v}_m) \dx
  \right]_{\epsilon = 0}\\
	&=
	- \int_{\Omega_m} \mu_0 \Ms
  \hdemag \cdot \vec{v}_m \dx.
\end{align}
Consequently the derivative and boundary term for the demagnetization energy are given by
\begin{align}
  \vdiff{\Edemag}{\vec{m}} &= - \mu_0 \Ms \hdemag\\
  \vec{B} &= 0.
\end{align}
\subsection{Anisotropy energy}\label{sec:static_aniso}
For the uniaxial anisotropy \eqref{eq:energetics_aniso_uniaxial} the derivative and boundary terms are given by
\begin{align}
  \vdiff{\Eanisou}{\vec{m}} &= - 2 K_\text{u1} \vec{e}_u (\vec{e}_u \cdot \vec{m}) - 4 K_\text{u2} \vec{e}_u (\vec{e}_u \cdot \vec{m})^3 \label{eq:energetics_aniso_uniaxial_field}\\
	\vec{B} &= 0
\end{align}
and for the cubic anisotropy \eqref{eq:energetics_aniso_cubic} the respective terms read
\begin{align}
	\vdiff{\Eanisoc}{\vec{m}}
  &= 2 K_\text{c1}
	\begin{pmatrix} m_1 m_2^2 + m_1 m_3^2 \\ m_2 m_3^2 + m_2 m_1^2 \\ m_3 m_1^2 + m_3 m_2^2 \end{pmatrix}
	+ 2 K_\text{c2}
  \begin{pmatrix} m_1   m_2^2 m_3^2 \\ m_1^2 m_2   m_3^2 \\ m_1^2 m_2^2 m_3 \end{pmatrix} \label{eq:energetics_aniso_cubic_field}\\
  \vec{B} &= 0.
\end{align}
For interface anisotropy contributions, the variational derivative $\vdiff{E}{\vec{m}}$ obviously vanishes and the influence of the energy contribution reduces to the boundary term
\begin{equation}
    \vec{B} = \int_{\partial \Omega_m} \epsilon(\vec{x}) \dx
\end{equation}
with $\epsilon(\vec{x})$ being the respective areal energy density.

\subsection{Antisymmetric exchange energy}
Similar to the exchange energy, the variation of the antisymmetric exchange energy \eqref{eq:energetics_dmi_interface} needs to be transformed by partial integration in order to eliminate spatial derivatives of the test functions $\vec{v}_m$
\begin{align}
  \delta \Edmii(\vec{m}, \vec{v}_m)
	=& 
  \diff{}{\epsilon}\bigg[
	  \int_{\Omega_m}
    \Di \big[ (\vec{m} + \epsilon\vec{v}_m) \cdot \vnabla \big(\vec{e}_\text{d} \cdot (\vec{m} + \epsilon \vec{v}_m)\big) \nonumber\\
	&- \vnabla \cdot (\vec{m} + \epsilon\vec{v}_m) \big(\vec{e}_\text{d} \cdot (\vec{m} + \epsilon\vec{v}_m) \big) \big] \dx
  \bigg]_{\epsilon = 0}\\
	=&
	\int_{\Omega_m} \Di \bigg[
	  \vec{v}_m \cdot \vnabla(\vec{e}_\text{d} \cdot \vec{m}) + \vec{m} \cdot \vnabla(\vec{e}_\text{d} \cdot \vec{v}_m)\nonumber\\
	&- \vnabla \cdot \vec{v}_m (\vec{e}_\text{d} \cdot \vec{m}) - \vnabla \cdot \vec{m} (\vec{e}_\text{d} \cdot \vec{v}_m)
	\bigg] \dx\\
	=&
	2 \int_{\Omega_m} \Di \bigg[
	  \vnabla(\vec{e}_\text{d} \cdot \vec{m}) - (\vnabla \cdot \vec{m}) \vec{e}_\text{d}
	\bigg] \vec{v}_m \dx\nonumber\\
	&- \int_{\partial\Omega_m} \Di \big[
	  ( \vec{e}_\text{d} \times \vec{n}) \times \vec{m}
  \big] \cdot \vec{v}_m \ds.
\end{align}
The resulting variational derivative and the boundary term for the interface DMI energy read
\begin{align}
  \vdiff{\Edmii}{\vec{m}} &= 2 \Di \big[
	  \vnabla(\vec{e}_\text{d} \cdot \vec{m}) - (\vnabla \cdot \vec{m}) \vec{e}_\text{d}
	\big]\\
  \vec{B} &= - \Di ( \vec{e}_\text{d} \times \vec{n}) \times \vec{m}.
\end{align}
For the antisymmetric bulk exchange \eqref{eq:energetics_dmi_bulk} the differential is given by
\begin{align}
  \delta \Edmib(\vec{m}, \vec{v}_m)
	=& 
  \diff{}{\epsilon}\bigg[
	  \int_{\Omega_m}
    \Db (\vec{m} + \epsilon \vec{v}_m)  \cdot \vnabla \times (\vec{m} + \epsilon \vec{v}_m)
	\dx \bigg]_{\epsilon=0} \\
	=&
  \int_{\Omega_m} \Db \bigg[
	  \vec{m} \cdot \vnabla \times \vec{v}_m +
	  \vec{v}_m \cdot \vnabla \times \vec{m}
	\bigg] \dx\\
	=&
	2 \int_{\Omega_m} \Db (\vnabla \times \vec{m}) \cdot \vec{v}_m \dx
  - \int_{\partial\Omega_m} \Db (\vec{n} \times \vec{m}) \cdot \vec{v}_m \ds,
\end{align}
which leads to the following variational derivative and boundary term
\begin{align}
  \vdiff{\Edmib}{\vec{m}} &= 2 \Db \vnabla \times \vec{m}\\
  \vec{B} &= - \Db \vec{n} \times \vec{m}.
\end{align}
\subsection{Energy minimization with multiple contributions}\label{sec:static_multiple_bc}
In order to minimize the total energy of a system subject to multiple energy contributions both Brown's condition \eqref{eq:static_brown_condition} and the boundary condition \eqref{eq:static_boundary_condition} have to be fulfilled for the composite energy functional.
Namely, if a system is subject to the exchange energy \eqref{eq:energetics_exchange} and the demagnetization energy \eqref{eq:energetics_demag}, the respective conditions for an energy minimum read
\begin{align}
    \vec{m} \times \vdiff{E}{\vec{m}} &= \vec{m} \times (- 2 A \Delta \vec{m} - \mu_0 \Ms \hdemag) = 0 \label{eq:static_specific_brown}\\
    \vec{m} \times \vec{B} &= \vec{m} \times \left(2 A \pdiff{\vec{m}}{\vec{n}}\right) = 0 \label{eq:static_specific_boundary}
\end{align}
with \eqref{eq:static_specific_boundary} being the ``classical'' micromagnetic boundary condition.
Spatial derivatives of the magnetization $\vec{m}$ are always orthogonal to the magnetization due to the micromagnetic unit-sphere constraint.
Hence, the boundary condition \eqref{eq:static_specific_boundary} is usually simplified to $\pdiff{\vec{m}}{\vec{n}} = 0$.
If the system is additionally subject to the antisymmetric exchange \eqref{eq:energetics_dmi_interface}, both Brown's condition \eqref{eq:static_specific_brown} and the boundary condition \eqref{eq:static_specific_boundary} are supplemented with the respective contributions.
The resulting boundary condition reads
\begin{equation}
  2 A \pdiff{\vec{m}}{\vec{n}} - \Di (\vec{e}_\text{d} \times \vec{n}) \times \vec{m} = 0
\end{equation}
where the cross product $\vec{m} \times \vec{B}$ was again neglected by orthogonality arguments.
Depending on the considered energy contributions, this boundary condition is supplemented by additional terms.
Hence, adding energy contributions does not add additional boundary conditions, but changes the single boundary condition instead.

\section{Dynamic micromagnetics}\label{sec:llg}
\begin{figure}
  \centering
  \includegraphics{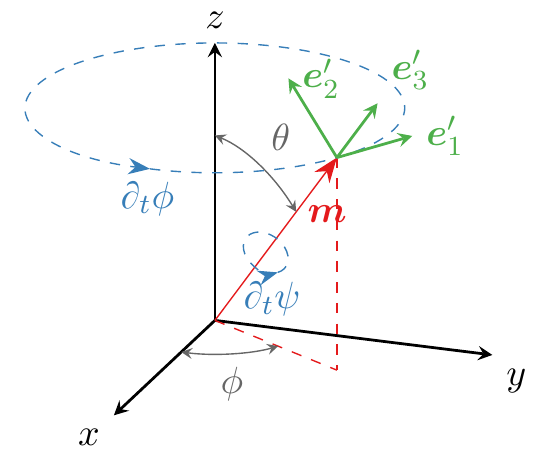}
  \caption{
    Absolute and fixed-body coordinates for the magnetization $\vec{m}$.
    The magnetization is described by the spherical coordinates $\theta$ and $\phi$.
    The fixed-body frame is constructed such that the third axis $\vec{e}_3'$ is parallel to the magnetization vector $\vec{m}$.
  }
  \label{fig:llg_angles}
\end{figure}
In Micromagnetics, magnetization dynamics are described by the Landau-Lifshitz (LL) equation that was originally proposed in \cite{landau_1935}.
This equation describes the spatially resolved motion of the magnetization in an effective field.
Due to problems with the dissipative term, an alternative formulation was derived by Gilbert \cite{gilbert_1955,gilbert_2004}.
Both formulations are completely equivalent under proper parameter transformation.
However, for the purpose of distinction, the latter is usually referred to as Landau-Lifshitz-Gilbert (LLG) equation or, alternatively, as Gilbert or implicit form of the Landau-Lifshitz equation.

The LLG can be derived by means of classical Lagrangian mechanics by the choice of an appropriate action $S$.
Due to the micromagnetic unit-sphere constraint $|\vec{m}|=1$, the magnetization field $\vec{m}$ can be described by means of spherical coordinates
\begin{equation}
  \vec{m}(\vec{x}) = \begin{pmatrix}
    \sin[\theta(\vec{x})] \cos[\phi(\vec{x})]\\
    \sin[\theta(\vec{x})] \sin[\phi(\vec{x})]\\
    \cos[\theta(\vec{x})]
  \end{pmatrix}
\end{equation}
with the polar angle $\theta$ and the azimuthal angle $\phi$.
For the sake of readability we omit the spatial dependence of fields in the following.
According to Hamilton's principle, the temporal evolution of any field is given as the path with stationary action $\delta S = 0$ with the action $S$ defined as 
\begin{equation}
  S(\phi, \theta) = \int_T \int_{\Omega_m} \mathcal{L} \dx \dt
\end{equation}
where $\mathcal{L}$ is the so-called Lagrangian which, in turn, is given by
\begin{equation}
  \mathcal{L} = T - V
\end{equation}
with $T$ being the kinetic energy density and $V$ being the potential energy density.
The potential energy density $V$ is naturally given by the free energy $E = \int V \dx$ whose contributions are introduced in Sec.~\ref{sec:energetics}.
However, the choice of the kinetic energy $T$ is not immediately clear.
Due to the unit-sphere constraint, the motion of the magnetization is restricted to rotations.
Hence, it seems reasonable to assume a kinetic energy similar to that of a rotating rigid body
\begin{equation}
  T = \frac{1}{2} \vec{\Omega} \mat{I} \vec{\Omega}
  \label{eq:llg_kinetic_energy}
\end{equation}
with $\mat{I}$ being an inertial tensor and $\vec{\Omega}$ being the angular velocity vector.
In the rigid-body picture, the magnetization in a certain point $\vec{m}(\vec{x})$ is represented by a cylindrical stick with one end fixed at the coordinate origin, see Fig.~\ref{fig:llg_angles}.
We introduce the fixed-body frame with coordinate axes $\vec{e}_1'$, $\vec{e}_2'$, $\vec{e}_3'$ as shown in Fig.~\ref{fig:llg_angles} and mark vectors with a prime whose coordinates are expressed in terms of this new basis.
In the fixed-body frame, the magnetization is trivially given as
\begin{equation}
  \vec{m} = \begin{pmatrix} 0 \\ 0 \\ 1 \end{pmatrix}'.
  \label{eq:llg_m_fixed_body}
\end{equation}
Due to the cylindrical symmetry of the rigid-body representation of the magnetization, it is clear that the inertia tensor is diagonal in the fixed-body frame and thus reads
\begin{equation}
  I = \begin{pmatrix}
    I_1 & 0   & 0 \\
    0   & I_2 & 0 \\
    0   & 0   & I_3
  \end{pmatrix}'.
\end{equation}
The angular velocity vector in the fixed-body frame is given as
\begin{equation}
  \vec{\Omega} = \begin{pmatrix}
    \pdiffs{\phi}{t} \sin(\theta) \sin(\psi) + \pdiffs{\theta}{t} \cos(\psi) \\
    \pdiffs{\phi}{t} \sin(\theta) \cos(\psi) - \pdiffs{\theta}{t} \sin(\psi) \\
    \pdiffs{\phi}{t} \cos(\theta) + \pdiffs{\psi}{t}
  \end{pmatrix}'
  \label{eq:llg_omega_fixed_body}
\end{equation}
where the spherical coordinates $\theta$ and $\phi$ are complemented by the angle $\psi$ that describes the rotation of magnetization's stick representation around its symmetry axis.
In order to derive the LLG from the general kinetic energy density \eqref{eq:llg_kinetic_energy}, two assumptions are required.
The first assumption is that of vanishing moments of inertia $I_1 = I_2 = 0$.
This assumption is reasonable since the magnetization stick has no mass in a classical sense.
Hence the rotation of a magnetic moment in an external field is expected to stop instantaneously if the external field is switched off rapidly.
With this assumption, the kinetic energy \eqref{eq:llg_kinetic_energy} reduces to
\begin{equation}
  T = \frac{1}{2} I_3 \Omega_3^2.
  \label{eq:llg_kinetic_energy_reduced}
\end{equation}
The second assumption is, that the angular momentum of the rotation around the symmetric axis $L_3$ is connected to the saturation magnetization by the relation
\begin{equation}
  \Ms = \gammae L_3 = \gamma_3 I_3 \Omega_3
  \label{eq:llg_angular_momentum}
\end{equation}
where $\gamma_3$ is the electron's gyromagnetic ratio.
This relation is reasonable since the saturation magnetization takes the place of the magnetic moment in the continuous theory of micromagnetics and the magnetic moment is generated by the spin, i.e. the angular momentum connected to the symmetry axis.
Inserting \eqref{eq:llg_omega_fixed_body} into \eqref{eq:llg_kinetic_energy_reduced} and further using the relation \eqref{eq:llg_angular_momentum} yields the following expression for the variation of the integrated kinetic energy with respect to the azimuthal angle $\phi$
\begin{align}
  \delta \left[ \int_T \int_{\Omega_m} T \dx \dt \right] \big(\{\phi, \theta\}, v_\phi \big)
  &= \diff{}{\epsilon} \int_T \int_{\Omega_m} T(\phi + \epsilon v_\phi, \theta) \dx \dt \\
  &= \int_T \int_{\Omega_m} \diff{T}{\Omega_3} \diff{}{\epsilon} \Omega_3(\phi + \epsilon v_\phi, \theta) \dx \dt \\
  &= \int_T \int_{\Omega_m} I_3 \Omega_3 \pdiffs{v_\phi}{t} \cos(\theta) \dx \dt \\
  &= \int_T \int_{\Omega_m} \frac{\Ms}{\gamma} \pdiffs{\theta}{t} \sin(\theta) v_\phi \dx \dt.
  \label{eq:llg_t_variation_phi}
\end{align}
Applying the same procedure to the variation with respect to the polar angle $\theta$ yields
\begin{equation}
  \delta \left[ \int_T \int_{\Omega_m} T \dx \dt \right] \big(\{\phi, \theta\}, v_\theta \big)
  = \int_T \int_{\Omega_m} - \frac{\Ms}{\gammae} \pdiffs{\phi}{t} \sin(\theta) v_\theta \dx \dt.
  \label{eq:llg_t_variation_theta}
\end{equation}
The angular velocity \eqref{eq:llg_omega_fixed_body} can be simplified by setting $\psi = 0$.
This angle describes the rotation of the magnetization's stick representation around its symmetry axis.
Hence this assumption does not introduce any losses of generality \cite{landau_mechanics}.
\begin{equation}
  \vec{\Omega} = \begin{pmatrix}
    \pdiffs{\theta}{t} \\
    \pdiffs{\phi}{t} \sin(\theta) \\
    \pdiffs{\phi}{t} \cos(\theta) + \pdiffs{\psi}{t}
  \end{pmatrix}'
\end{equation}
This simplification leads to the following relations for the time derivatives of magnetization coordinates and their respective variations
\begin{align}
  \pdiffs{m_1}{t} &=  \sin(\theta) \pdiffs{\phi}{t} & v_{m_1} &= \sin(\theta) v_{\phi} \\
  \pdiffs{m_2}{t} &= -\pdiffs{\theta}{t} & v_{m_2} &= -v_{\theta}.
\end{align}
Inserting into the kinetic-energy variations \eqref{eq:llg_t_variation_phi} and \eqref{eq:llg_t_variation_theta} yields
\begin{align}
  \delta \left[ \int_T \int_{\Omega_m} T \dx \dt \right] \big(\vec{m}, v_{m_1} \big)
  &= \int_T \int_{\Omega_m} \frac{\Ms}{\gammae} \pdiffs{m_2}{t} v_{m_1} \dx \dt\\
  \delta \left[ \int_T \int_{\Omega_m} T \dx \dt \right] \big(\vec{m}, v_{m_2} \big)
  &= - \int_T \int_{\Omega_m} \frac{\Ms}{\gammae} \pdiffs{m_1}{t} v_{m_2} \dx \dt
\end{align}
which, in the fixed-body frame, can be summarized in the vector valued variation 
\begin{equation}
  \delta \left[ \int_T \int_{\Omega_m} T \dx \dt \right] \big(\vec{m}, \vec{v}_m \big)
  = \int_T \int_{\Omega_m} \frac{\Ms}{\gammae} (\vec{m} \times \pdiffs{\vec{m}}{t}) \cdot \vec{v}_m \dx \dt.
  \label{eq:llg_variation_t}
\end{equation}
in the fixed-body frame where the magnetization is given as $\vec{m} = (0,0,1)$, see \eqref{eq:llg_m_fixed_body}.
In order to compute the variation of the action $\delta S$, the variation of the kinetic energy \eqref{eq:llg_variation_t} has to be complemented by the variation of the potential energy which is given as
\begin{align}
  \delta \left[ \int_T \int_{\Omega_m} V \dx \dt \right] \big(\vec{m}, \vec{v}_m \big)
  &= \delta \left[ \int_T E \dt \right] (\vec{m}, \vec{v}_m)\\
  &= \int_T \delta E\Big(\vec{m}, \vec{v}_m(t) \Big) \dt\\
  &= \int_T \left[ \int_{\Omega_m} \frac{\delta E}{\delta \vec{m}} \cdot \vec{v}_m \dx
  +  \int_{\partial \Omega_m} \vec{B} \cdot \vec{v}_m \ds \right] \dt
  \label{eq:llg_variation_v}
\end{align}
where the boundary term $\vec{B}$ is the same as introduced in Sec.~\ref{sec:static} and hence depends on the particular choice of energy contributions.
Putting together the variation of the kinetic energy \eqref{eq:llg_variation_t} and the potential energy \eqref{eq:llg_variation_v} results in the variation of the action
\begin{equation}
  \delta S \big(\vec{m}, \vec{v}_m \big)
  =
  \int_T \left[
    \int_{\Omega_m} \left(
      \frac{\Ms}{\gammae} (\vec{m} \times \pdiffs{\vec{m}}{t}) \cdot \vec{v}_m
      + \frac{\delta E}{\delta \vec{m}} \cdot \vec{v}_m \right) \dx
  +  \int_{\partial \Omega_m} \vec{B} \cdot \vec{v}_m \ds \right] \dt
  = 0
  \label{eq:llg_variation_s}
\end{equation}
which is required to vanish for arbitrary variations $\vec{v}_m$ according to Hamilton's principle.
Restricting the variations to functions vanishing on the boundary $\partial \Omega_m$, i.e. $\vec{v}_m \in V^0_m$, yields the equation
\begin{equation}
  - \frac{\Ms}{\gammae} (\vec{m} \times \pdiffs{\vec{m}}{t}) = \frac{\delta E}{\delta \vec{m}}
\end{equation}
that has to hold for any $\vec{x} \in \Omega_m$.
Cross-multiplying both sides with $\gammae / \Ms \vec{m}$ from the left results in
\begin{equation}
  \pdiffs{\vec{m}}{t} - (\pdiffs{\vec{m}}{t} \cdot \vec{m}) \vec{m}
  = \pdiffs{\vec{m}}{t} 
  = \frac{\gammae}{\Ms} \vec{m} \times \frac{\delta E}{\delta \vec{m}}
  \label{eq:llg_llg_no_dissipation}
\end{equation}
where $\pdiffs{\vec{m}}{t} \cdot \vec{m}$ vanishes due to the micromagnetic unit-sphere constraint that requires any derivative of the magnetization to be perpendicular on $\vec{m}$.
The equation of motion \eqref{eq:llg_llg_no_dissipation} describes the magnetization dynamics without energy losses.
However, realistic systems are expected to lose magnetic energy by conversion to e.g. phonons, eddy currents \cite{brown_1963}.
In the framework of Lagrangian mechanics, dissipative processes are described by a Rayleigh function $D$.
The time evolution of the magnetization $\vec{m}$ subject to the dissipative function $D$ is then given as
\begin{equation}
  \delta S(\vec{m}, \vec{v}_m) =
  - \delta \left[ \int_T \int_{\Omega_m} D \dx \dt \right] (\vec{m}, \vec{v}_m).
\end{equation}
The Rayleigh function $D$ is usually chosen to be proportional to the square of the time derivative of the variable of motion.
Choosing $D = \alpha \Ms / (2 \gammae) (\pdiffs{\vec{m}}{t})^2$ in the case of magnetization dynamics yields
\begin{align}
  \delta S(\vec{m}, \vec{v}_m)
  &= - \delta \left[ \int_T \int_{\Omega_m} \alpha \frac{\Ms}{2 \gammae} (\pdiffs{\vec{m}}{t})^2  \dx \dt \right] (\vec{m}, \vec{v}_m)\\
  &= - \int_T \int_{\Omega_m} \alpha \frac{\Ms}{\gammae} \pdiffs{\vec{m}}{t} \vec{v}_m \dx \dt
\end{align}
where $\alpha \ge 0$ is a dimensionless damping parameter.
Inserting the variation of the action \eqref{eq:llg_variation_s} and once more considering variations $\vec{v}_m \in V^0_m$ that vanish on the boundary $\partial \Omega_m$, results in the equation of motion
\begin{equation}
  \pdiffs{\vec{m}}{t} 
  = \frac{\gammae}{\Ms} \vec{m} \times \frac{\delta E}{\delta \vec{m}} + \alpha \vec{m} \times \pdiffs{\vec{m}}{t}.
\end{equation}
By introducing the effective field defined as
\begin{equation}
  \heff = - \frac{1}{\mu_0 \Ms} \vdiff{E}{\vec{m}},
  \label{eq:llg_effective_field}
\end{equation}
this equation can be turned into the well-known Gilbert form of the LLG
\begin{equation}
  \pdiffs{\vec{m}}{t} 
  = -\gamma \vec{m} \times \heff + \alpha \vec{m} \times \pdiffs{\vec{m}}{t}
  \label{eq:llg_llg_gilbert}
\end{equation}
with $\gamma = \mu_0 \gammae \approx \SI{2.2128e5}{m/As}$ being the reduced gyromagnetic ratio.
This equation of motion is completed by the boundary condition $\vec{m} \times \vec{B} = 0$ which is obtained by varying $\vec{v}_m$ on the boundary for \eqref{eq:llg_variation_s} and by considering the same cross product with $\vec{m}$ that is applied to obtain \eqref{eq:llg_llg_no_dissipation}.
Note, that this boundary condition resembles the static micromagnetic boundary condition \eqref{eq:static_boundary_condition}.
From \eqref{eq:llg_variation_s} it is clear, that this boundary condition has to hold at all times.

This semi-implicit formulation introduced by Gilbert can be transformed into an explicit form by inserting the complete right-hand side of \eqref{eq:llg_llg_gilbert} into $\pdiffs{\vec{m}}{t}$ on the right-hand side of \eqref{eq:llg_llg_gilbert}.
Applying basic vector algebra and considering $\vec{m} \cdot \pdiffs{\vec{m}}{t} = 0$ and $\vec{m} \cdot \vec{m} = 1$ yields
\begin{equation}
  \pdiffs{\vec{m}}{t} 
  = -\frac{\gamma}{1 + \alpha^2} \vec{m} \times \heff - \frac{\alpha\gamma}{1 + \alpha^2}  \vec{m} \times (\vec{m} \times \heff)
  \label{eq:llg_llg}
\end{equation}
which, apart from the definition of the parameters $\gamma$ and $\alpha$, equals the original equation introduced by Landau and Lifshitz.
While the presented derivation of the LLG is not rigorous, it should be noted that the required assumptions, namely vanishing moments of inertia $I_1 = I_2 = 0$ and the connection of the remaining moment of inertia with the saturation magnetization $\Ms = \gammae I_3 \Omega_3$, are physically reasonable.
The strict application of variational calculus not only yields the LLG but also its boundary conditions depending on the contributions to the effective field $\heff$.

A more detailed investigation on the LLG as derived from a Lagrangian is presented in \cite{wegrowe_2012} where it is also shown that the kinetic contribution to the Lagrangian \eqref{eq:llg_kinetic_energy_reduced} is equivalent to the assumption 
\begin{equation}
  T = \frac{\Ms}{\gammae} \pdiffs{\phi}{t} \cos(\theta)
  \label{eq:llg_gilbert_assumption}
\end{equation}
introduced in the original work by Gilbert \cite{gilbert_1955} and earlier by Doering \cite{doring1948tragheit}.
A full quantum mechanical description of a spin subject to exchange interaction, anisotropy and Zeeman field is given in \cite{bode_2012} where the Landau-Lifshitz-Gilbert equation is also obtained in a limit case.

\subsection{Properties of the Landau-Lifshitz-Gilbert Equation}\label{sec:llg_properties}
\begin{figure}
  \centering
  \includegraphics{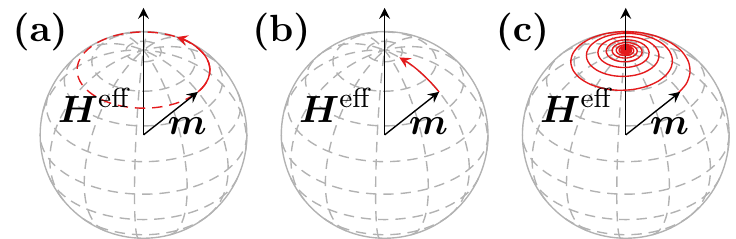}
  \caption{
    Visualization of the contributions to the magnetization dynamics as described by the Landau-Lifshitz-Gilbert equation (LLG).
    (a) Precessional motion around the effective field $\heff$.
    (b) Dissipative motion of the magnetization towards $\heff$.
    (c) Combined precessional and dissipative motion as described by the LLG.
  }
  \label{fig:llg_llg}
\end{figure}
Figure~\ref{fig:llg_llg} illustrates the damped precessional motion of the magnetization $\vec{m}$ in an effective field $\heff$ as described by the LLG.
As noted in Sec.~\ref{sec:intro}, the main assumption of micromagnetics is the constant modulus of the magnetization field $\vec{m}$.
This property is conserved by the LLG as can be seen by considering the time derivative of the squared magnetization
\begin{equation}
  \pdiffs{|\vec{m}|^2}{t} = 
  \pdiffs{(\vec{m} \cdot \vec{m})}{t} = 
  2 \pdiffs{\vec{m}}{t} \cdot \vec{m}.
\end{equation}
Inserting the right-hand side of \eqref{eq:llg_llg} yields $\pdiffs{|\vec{m}|^2}{t} = 0$ and hence also $\pdiffs{|\vec{m}|}{t} = 0$.
In classical micromagnetics, the energy connected to the magnetization, as defined by the sum of the energy contributions introduced in Sec.~\ref{sec:energetics}, may only change due to external fields varying in time or due to the energy dissipation modeled by the damping term.
In the case of an effective field that does not explicitly depend on the time $t$ the time derivative of the energy is given by
\begin{align}
  \pdiffs{E}{t}
  &= \int_{\Omega_m} \vdiff{E}{\vec{m}} \cdot \pdiffs{\vec{m}}{t} \dx \\
  &= - \mu_0 \int_{\Omega_m} \Ms \heff \cdot \pdiffs{\vec{m}}{t} \dx.
\end{align}
Inserting \eqref{eq:llg_llg} for $\pdiffs{\vec{m}}{t}$ yields
\begin{align}
  \pdiffs{E}{t}
  &= \mu_0 \int_{\Omega_m} \Ms \heff \left[ \frac{\gamma}{1 + \alpha^2} \vec{m} \times \heff + \frac{\alpha\gamma}{1 + \alpha^2} \cdot  \vec{m} \times (\vec{m} \times \heff) \right] \dx \\
  &= - \mu_0 \int_{\Omega_m} \Ms \frac{\alpha\gamma}{1 + \alpha^2} | \vec{m} \times \heff |^2 \dx \label{eq:llg_energy_loss}.
\end{align}
The value of the integral is zero in the case of an energy minimum, see Brown's condition \eqref{eq:static_brown_condition}, and positive otherwise.
Hence, for a positive damping constant $\alpha > 0$, the energy of a magnetic system is a non-increasing function in time.
In this case the LLG is said to have Lyapunov structure \cite{daquino_2005,cimrak_2007}.
In the special case of no damping $\alpha = 0$ the right-hand side of \eqref{eq:llg_energy_loss} vanishes
\begin{equation}
  \partial_t E = 0
\end{equation}
and the LLG has Hamiltonian structure, i.e. it preserves the energy.

\section{Spintronics in micromagnetics}
The term spintronics summarizes all effects caused by the interaction of electrons with solid state devices due to their spin rather then their charge.
For magnetic systems, this particularly covers the origin of spin-polarized currents and their impact on the magnetization configuration.
The term spintronics was coined in the 1980ies when the giant magnetoresistance (GMR) was discovered by Fert \cite{baibich1988giant} and Gr\"unberg \cite{binasch1989enhanced}.
Exploiting the spin of electrons, in addition to their charge, adds extra degrees of freedom and allows for the development of novel devices especially in the areas of storage and sensing technology.

\begin{figure}
  \centering
  \includegraphics{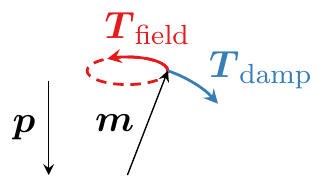}
  \caption{
    Spin-torque contributions for neglectable damping $\alpha \ll 1$ with respect to the reference polarization $\vec{p}$. The fieldlike torque $\vec{T}_\text{field}$ leads to a precessional motion of the magnetization $\vec{m}$ around $\vec{p}$.
    The dampinglike torque $\vec{T}_\text{damp}$ leads to a direct relaxation of $\vec{m}$ towards $\vec{p}$.
  }
  \label{fig:spin_torque}
\end{figure}
In the semiclassical picture of micromagnetics, the spin polarization of an electric current is described by a three dimensional vector field $\vec{p}$.
If a polarized electric current passes a magnetic region, it exerts a torque on the magnetization.
This so-called spin torque, similar to the torque generated by a magnetic field, can be split into a fieldlike contribution $T_\text{field}$ and a dampinglike contribution $T_\text{damp}$, see Fig.~\ref{fig:spin_torque}.
The fieldlike torque has the same form as the torque generated by a regular effective-field contribution, i.e. it leads to a damped precessional motion as described in Sec.~\ref{sec:llg}.
Since the Gilbert damping $\alpha$ is usually small $\alpha \ll 1$, the magnetization dynamics caused by the fieldlike torque are dominated by the precessional part.
In contrast, the dynamics caused by the dampinglike torque are dominated by the direct rotation of the magnetization towards the polarization and accompanied by a small precessional contribution.
Depending on the origin of the polarized current, the torque is either referred to as spin-transfer torque or spin-orbit torque.

\subsection{Spin-transfer torque in multilayers}\label{sec:spin_slon}
\begin{figure}
  \centering
  \includegraphics{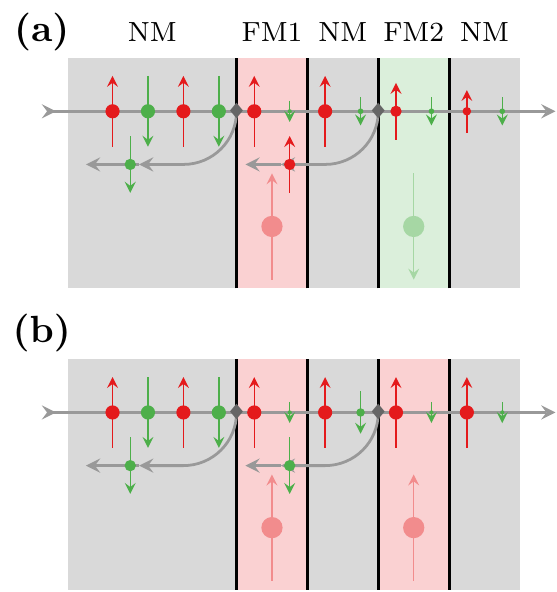}
  \caption{
    Schematic illustration of the most important scattering processes in a multilayer with magnetic layers FM1, FM2 and nonmagnetic layers NM subject to a current perpendicular to the layers.
    (a) Antiparallel magnetization configuration of FM1 and FM2.
    Electrons with opposite polarization of FM1 are scattered before entering the layer.
    Hence, FM1 acts as a spin polarizer and FM2 is subject to spin torque.
    FM1 is stabilized by electrons scattered  by FM2.
    (b) Parallel magnetization configuration of the FM1 and FM2.
    FM1 acts as spin polarizer leading to a stabilization of FM2.
    Scattered electrons from FM2 exert spin torque in FM1.
  }
  \label{fig:spin_stt}
\end{figure}
A typical device that exploits spin-transfer torque, consists of two magnetic layers separated by a nonmagnetic spacer layer and sandwiched with two nonmagnetic leads, see Fig.~\ref{fig:spin_stt}.
If passed by an electric current, the conducting electrons are subject to scattering processes depending on the spin configuration of the conducting electrons.
Even if the applied current has a net spin polarization of zero, these spin-dependent scattering processes lead to a non-vanishing spin-polarization distribution across the multilayer.
In particular, the interfaces between magnetic and non-magnetic regions act as scattering sites due to the rapid transition of magnetization.
A simplified illustration of the scattering processes for different magnetization configurations of the multilayer, i.e. antiparallel and parallel, is given in Fig.~\ref{fig:spin_stt}.

In the case of an antiparallel configuration, a scattering process takes place at the first interface that is passed by the conducting electrons, see Fig.~\ref{fig:spin_stt}(a).
Due to this scattering, the FM1 layer acts as a spin filter and the majority of the conducting electrons that reach the FM2 layer carry the polarization of FM1.
The spin-transfer torque will cause the switching of the FM2 magnetization if exceeding a critical strength.
In addition, a scattering process at the first FM2 interface will reflect electrons with the polarization of FM1 leading to a stabilization of the FM1 magnetization configuration.

For a parallel configuration, the same scattering as for the antiparallel case appears at the first interface of FM1, see Fig.~\ref{fig:spin_stt}(b).
This leads to a spin polarization parallel to the magnetization configuration of FM1 in the spacer layer layer between FM1 and FM2.
However, if the spacer layer has sufficient thickness, the spin polarization reduces due to spin-flip events.
At the first interface of FM2 the recovered electrons with antiparallel polarization to the magnetization of FM2 are scattered back to FM1.
The scattered electrons exert a torque on the magnetization of FM1 and can switch it if exceeding a critical strength.
The magnetization of FM2 on the other hand is stabilized by the electrons polarized by the FM1 layer.
Possible applications of this torque mechanism, that was first investigated in by Slonczewski \cite{slonczewski1996current}, Berger \cite{berger1996emission}, an Waintal and coworkers \cite{waintal2000role}, are the spin-transfer torque magnetoresistive random access memory (STT MRAM) \cite{huai2008spin,worledge2011spin} and spin torque oscillators (STO) \cite{houssameddine2007spin,kim2012spin}.
A comprehensive theoretical overview over spin-transfer torque is given in a work by Ralph and Stiles\cite{ralph2008spin}.

A very popular model for the description of the magnetization dynamics in spin-transfer-torque devices is the model proposed by Slonczewski \cite{slonczewski2002currents}.
This model uses the macrospin approach, where the magnetic region subject to spin torque, also referred to as free layer, is described by a single spin $\vec{m}$.
The current is assumed to be polarized by another magnetic layer, referred to as polarizing layer, whose magnetization is described by the vector $\vec{p}$.
The motion of the free-layer magnetization $\vec{m}$ is described by the extended LLG
\begin{equation}
  \pdiffs{\vec{m}}{t} = 
  -\gamma \vec{m} \times \heff
  +\alpha \vec{m} \times \pdiffs{\vec{m}}{t}
  + \vec{T}
  \label{eq:spin_llg}
\end{equation}
where the torque $\vec{T}$ consists of a dampinglike and fieldlike contribution $\vec{T} = \vec{T}_\text{damp} + \vec{T}_\text{field}$.
According to the model of Slonczewski these contributions are given by
\begin{align}
  \vec{T}_\text{damp}  &= \eta_\text{damp}(\vartheta)  \frac{j_\text{e} \gamma \hbar}{2 e \mu_0 \Ms} \vec{m} \times (\vec{m} \times \vec{p}) \label{eq:spin_t_damp}\\
  \vec{T}_\text{field} &= \eta_\text{field}(\vartheta) \frac{j_\text{e} \gamma \hbar}{2 e \mu_0 \Ms} \vec{m} \times \vec{p}
  \label{eq:spin_t_field}
\end{align}
where the dimensionless functions $\eta_\text{damp}$ and $\eta_\text{field}$ describe the angular dependence of the torque strength with $\vartheta$ being the angle between $\vec{m}$ and $\vec{p}$.
By comparing the torque contributions \eqref{eq:spin_t_damp} and \eqref{eq:spin_t_field} with the effective-field term in the LLG \eqref{eq:spin_llg}, the torque can be expressed by means of an effective field contribution $\htorque$ given by
\begin{equation}
  \htorque = - \frac{j_\text{e} \hbar}{2 e \mu_0 \Ms} \big[ \eta_\text{damp}(\vartheta) \, \vec{m} \times \vec{p} + \eta_\text{field}(\vartheta) \, \vec{p} \big].
\end{equation}
Inserting into the LLG \eqref{eq:spin_llg} and transforming the LLG into the explicit form \eqref{eq:llg_llg} yields
\begin{align}
  \pdiffs{\vec{m}}{t} =
  &- \frac{\gamma}{1 + \alpha^2} \vec{m} \times \left[
    \heff +
    \frac{j_\text{e} \hbar}{2 e \mu_0 \Ms} (\alpha \eta_\text{damp} - \eta_\text{field}) \vec{p} \right] \nonumber \\
  &- \frac{\alpha\gamma}{1 + \alpha^2}  \vec{m} \times \left(\vec{m} \times \left[
    \heff +
  \frac{j_\text{e} \hbar}{2 e \mu_0 \Ms} \left(-\frac{1}{\alpha} \eta_\text{damp} - \eta_\text{field}\right) \vec{p} 
    \right] \right)
\end{align}
where the vector identity $\vec{m} \times [\vec{m} \times (\vec{m} \times \vec{p})] = - \vec{m} \times \vec{p}$ was used.
From this formulation it is clear, that both the dampinglike torque $\vec{T}_\text{damp}$ and the fieldlike torque $\vec{T}_\text{field}$ contribute to the precessional motion as well as the dampinglike motion.
However, this intermixing highly depends on the Gilbert damping $\alpha$.
In the limit case of vanishing $\alpha$, the LLG simplifies to
\begin{equation}
  \pdiffs{\vec{m}}{t} =
   - \gamma \vec{m} \times \left[
    \heff -
   \frac{j_\text{e} \hbar}{2 e \mu_0 \Ms} \eta_\text{field} \, \vec{p} \right]
   - \gamma  \vec{m} \times \left(\vec{m} \times \left[
     \frac{j_\text{e} \hbar}{2 e \mu_0 \Ms} \eta_\text{damp} \, \vec{p} 
   \right] \right)
\end{equation}
where the fieldlike torque exclusively contributes to the precessional motion and the dampinglike torque exclusively contritbutes to the dampinglike motion.
This limit demonstrates the unique feature of the dampinglike torque to facilitate a dampinglike motion independently from the Gilbert damping $\alpha$.
In the original work of Slonczewski, the expression
\begin{equation}
  \eta(\vartheta) =
  \frac{P\Gamma}{(\Gamma + 1) + (\Gamma - 1) \cos(\vartheta)}
  \label{eq:spin_slon_orig}
\end{equation}
is derived as angular dependence of the torque for symmetric systems, i.e. systems with two identical magnetic layers.
This expression is valid for both the dampinglike and the fieldlike torque, but in general requires a different set of model parameters $P$ and $\Gamma$.
The dimensionless parameters $P$ and $\Gamma$ depend on geometry and materials of the complete system and describe the polarization strength and the angular asymmetry of the STT respectively.
A more general expression for the angular dependence is introduced in \cite{xiao2005macrospin} as
\begin{equation}
  \eta(\vartheta) =
  \frac{q^+}{A + B \cos(\vartheta)} + \frac{q^-}{A - B \cos(\vartheta)}.
  \label{eq:spin_slon_asym}
\end{equation}
This expression accounts for asymmetric devices with two different magnetic layers.
Again, the free parameters $q^+$, $q^-$, $A$, and $B$ are dimensionless and depend on the geometry and material composition of the complete stack.

The model of Slonczewski is often used to describe STT devices with one hard-magnetic layer acting as spin polarizer and one soft magnetic layer that is subject to the spin torque induced by the hard magnetic layer.
For these devices the magnetization dynamics in the hard magnetic layer, referred to as reference layer or pinned layer, are neglectable and LLG is solved for the soft magnetic layer, referred to as free layer, only \cite{apalkov2006micromagnetic}.
However, the model can also be used to describe the bidirectional coupling of the magnetization configuration in both magnetic layers \cite{rowlands2012magnetization}.

\begin{figure}
  \centering
  \includegraphics{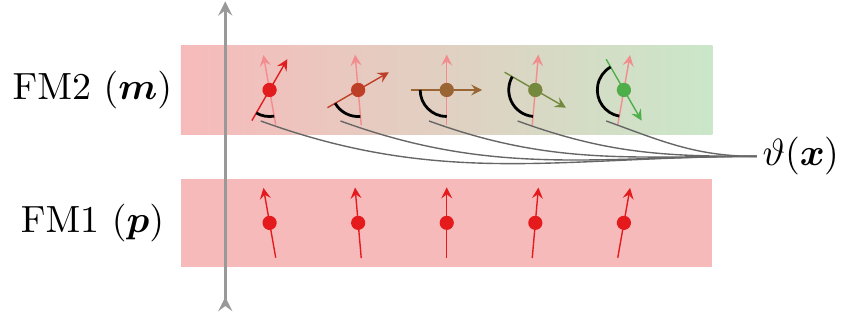}
  \caption{
    Generalization of the macrospin model of Slonczewski to laterally inhomogeneous magnetization configurations.
    The magnetization configuration of the polarizing layer FM1 $\vec{p}$ is projected onto the magnetic free layer FM2.
    The model of Slonczewski is applied locally by evaluating the angle $\vartheta(\vec{x})$ between the projected polarization $\vec{p}$ and the local magnetization $\vec{m}$.
  }
  \label{fig:spin_stt_spatial_theta}
\end{figure}
The macrospin approach as proposed in the original work of Slonczewski is accurate for small systems below the single-domain limit.
Systems of this size are dominated by the exchange interaction and hence act much like a single spin.
With growing size, other energy contributions such as the demagnetization energy gain influence leading to the generation of magnetic domains, which renders the macrospin approximation useless.
For thin film structures with large lateral dimensions but small thicknesses below the exchange length, the generalization of the macrospin model is straightforward.
In this case, the magnetization $\vec{m}$ and the polarization $\vec{p}$ in \eqref{eq:spin_t_damp} and \eqref{eq:spin_t_field} are functions of the lateral position $\vec{x}$ in the multilayer stack and the tilting angle $\vartheta$ is computed accordingly, see Fig.~\ref{fig:spin_stt_spatial_theta}. 
The torque contributions \eqref{eq:spin_t_damp} and \eqref{eq:spin_t_field} are usually applied as volume terms with the polarization $\vec{p}$ assumed to be independent of the perpendicular position in the free layer.

However, the spin-transfer torque is considered to be a surface effect rather than a volume effect.
For free-layer thicknesses below the exchange length, the treatment as volume effect does not affect the torque in a qualitative fashion, since the magnetization can be assumed constant across the free layer in this case.
Yet the strength of the torque needs to be scaled with the reciprocal free-layer thickness $1/d$ in order to account for the surface nature of the effect.

While the generalization from a macro-spin model to a spatially resolved model allows for the description of various multi layer devices, the model of Slonczewski has a number of shortcomings.
The presented generalization neglects lateral diffusion of the spins which might introduce inaccuracies for strongly inhomogeneous magnetization configurations.
Moreover, the treatment as volume term is only justified for free-layer thicknesses below the exchange length. 
Another disadvantage of this model are the free parameters introduced in \eqref{eq:spin_slon_orig} and \eqref{eq:spin_slon_asym} which depend on the geometry and material parameters of the complete system in a nontrivial fashion.
A comprehensive overview over Slonczewski-like models is given in a work by Berkov and Miltat \cite{berkov2008spin}.

\subsection{Spin-transfer torque in continuous media}\label{sec:spin_zhang}
\begin{figure}
  \centering
  \includegraphics{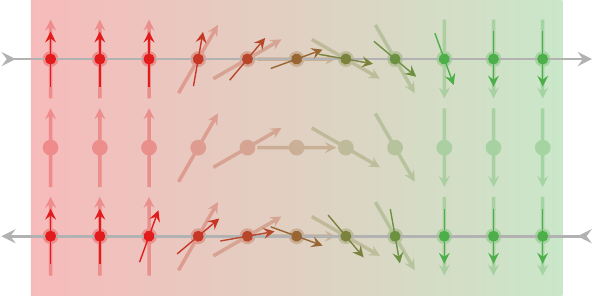}
  \caption{
    Spin-torque in magnetic domain walls as predicted by the model of Zhang and Li.
    Spin polarization is carried in the direction of electron motion and exerts a torque according to the local gradient of the magnetization.
    The magnetization is depicted with desaturated colors while the polarization of the conducting electrons, which are responsible for the spin torque, is depicted with saturated colors.
    The top row illustrates an electron motion from left to right.
    The bottom row illustrates an electron motion from right to left.
  }
  \label{fig:spin_zhangli}
\end{figure}
Spin-transfer torque can not only be exploited in magnetic multilayer structures, but also in continuous single phase magnets.
In these systems, magnetic domains take over the role of the distinct layers in multilayer stacks, i.e. they act as spin polarizer.
While the spin torque in multilayers acts on the surfaces of the magnetic layers to the spacer layer, the spin torque in single phase magnets acts in regions of high magnetization gradients, i.e. domain walls.
A simple picture for the origin of spin torque in single phase magnets is given in Fig.~\ref{fig:spin_zhangli}.
While the conducting electrons pass the magnetic region, they pick up the polarization from the local magnetization and carry this polarization in the direction of the electron flow where they exert a torque.
This mechanism can be used to move domain walls and complete domain structures with electric currents.
In contrast to field induced domain-wall motion, the spin-torque moves magnetic domain walls always in the direction of electron motion regardless of the nature of the wall, e.g. head-to-head, tail-to-tail.
This property is exploited, for example, by the magnetic racetrack memory proposed by Parkin et al. \cite{parkin_2008}.

An established model for the description of spin torque in continuous magnets is the model proposed by Zhang and Li \cite{zhang2004roles}.
In this model, the torque contribution $\vec{T}$ to the LLG is given as
\begin{align}
  \vec{T}
  =
  - b \, \vec{m} \times [\vec{m} \times (\je \cdot \vnabla) \vec{m}]
  - b \xi \vec{m} \times (\je \cdot \vnabla) \vec{m}
  \label{eq:spin_zhang_torque}
\end{align}
where $\xi$ describes the degree of nonadiabacity according to \cite{zhang2004roles} and $b$ is given as
\begin{equation}
  b = \frac{\beta \mub}{e \Ms ( 1 + \xi^2 )}
  \label{eq:spin_zhang_b}
\end{equation}
with $\beta$ being the dimensionless polarization rate of the conducting electrons, $\mub$ being the Bohr magneton, and $e$ being the elementary charge.
Instead of the coupling constant $b$, this model is often defined in terms of the spin-drift velocity $u = b j_\text{e}$ which has the dimension of a velocity.
Moreover, the letter $\beta$ is often used as degree of nonadiabacity instead of $\xi$.

The Zhang-Li model delivers reasonable results for the description of current driven domain-wall motion.
However, the description of spin-torque is purely local since the torque only depends on first derivatives of the magnetization.
This means, that the diffusion of spin polarization is completely neglected.
Consequently, the model is not suited for the description of spin torque in multilayers, since this requires the transport of spin across a nonmagnetic spacer layer.
Moreover, the lack of diffusion also introduces inaccuracies in systems with highly inhomogeneous magnetization configurations.

\subsection{Spin-diffusion}\label{sec:spin_diff}
Both, the model of Slonczewski introduced in Sec.~\ref{sec:spin_slon} and the model of Zhang and Li introduced in Sec.~\ref{sec:spin_zhang} are applicable only for specific material systems and magnetization configurations.
Also, both models neglect the diffusion of the spin polarization to some extent.
A more general approach to spin torque considers the torque generated by a vector field $\vec{s}$ referred to as spin accumulation
\begin{equation}
  \vec{T} = 
  - \frac{J}{\hbar \Ms} \vec{m} \times \vec{s}
  \label{eq:spin_diff_t}
\end{equation}
where $J$ denotes the coupling strength of the spin accumulation $\vec{s}$ and the magnetization $\vec{m}$.
The torque definition leads to the extended LLG
\begin{equation}
  \pdiffs{\vec{m}}{t} = 
  -\gamma \vec{m} \times \left (\heff + \frac{J}{\hbar \gamma \Ms} \vec{s} \right)
  +\alpha \vec{m} \times \pdiffs{\vec{m}}{t}.
  \label{eq:spin_diff_llg}
\end{equation}
The spin accumulation $\vec{s}(\vec{x})$ describes the deviation of the conducting electron's polarization compared to the equilibrium configuration at vanishing charge current $\je = 0$.
That said, by definition $\vec{s}$ is zero if no current is applied to the system.
Several variations of the spin-diffusion model have been proposed for the computation of the spin accumulation $\vec{s}$.
According to \cite{zhang2002mechanisms} the dynamics of $\vec{s}$ are given by 
\begin{equation}
  \pdiffs{\vec{s}}{t} =
  - \vnabla \cdot \js
  - \frac{\vec{s}}{\tausf}
  - J \frac{\vec{s} \times \vec{m}}{\hbar}
  \label{eq:spin_diff_dts}
\end{equation}
where $\tausf$ denotes the spin-flip relaxation time which is a material parameter.
While the LLG \eqref{eq:spin_diff_llg} is defined in the magnetic region $\Omega_m$ only, the spin accumulation $\vec{s}$ is generated also in nonmagnetic regions such as the leads or the spacer layer of an STT device and hence has to be solved in the complete sample region $\Omega$.
Consequently, \eqref{eq:spin_diff_dts} is potentially defined in composite media and thus all material parameters, such as $J$ and $\tausf$, may vary spatially.
The matrix-valued $\js$ in \eqref{eq:spin_diff_dts} denotes the spin current defined by
\begin{equation}
  \js = 2 C_0 \beta \frac{\mub}{e} \vec{m} \otimes \vnabla u - 2 D_0 \vnabla \vec{s}
  \label{eq:spin_diff_js}
\end{equation}
with $u$ being the electric potential, $\mub$ being the Bohr magneton, and $e$ being the elementary charge.
The variables $C_0$, $D_0$ and $\beta$ are material parameters.
$C_0$ is connected to the electric conductivity $\sigma$ by $\sigma = 2 C_0$ and $D_0$ denotes the material's diffusion constant.
$\beta$ is a dimensionless constant that denotes the rate of polarized conducting electrons in magnetic materials.
If the distribution of the electric potential $u$ is known, the coupled system \eqref{eq:spin_diff_llg} and \eqref{eq:spin_diff_dts} can be solved in order to simultaneously resolve the dynamics of the magnetization $\vec{m}(t)$ and the spin diffusion $\vec{s}(t)$.
However, the distribution of the electric potential $u$ might not be known upfront.
Specifically, the charge current $\je$ in the diffusion model is defined as 
\begin{equation}
  \je = - 2 C_0 \vnabla u + 2 D_0 \beta' \frac{e}{\mub} (\vnabla \vec{s})^T \vec{m}
  \label{eq:spin_diff_je}
\end{equation}
which suggests a strong coupling of the electric potential $u$ with $\vec{m}$ and $\vec{s}$.
If the charge current distribution $\je$ is known, the spin-diffusion dynamics \eqref{eq:spin_diff_dts} can be solved by inserting \eqref{eq:spin_diff_je} into \eqref{eq:spin_diff_js} via the gradient of the electric potential $\vnabla u$, which results in the spin-current definition
\begin{equation}
  \js =
  - \frac{\beta \mub}{e} \vec{m} \otimes \je
  - 2 D_0 \left(
    \vnabla \vec{s}
    - \beta \beta' \vec{m} \otimes \left[(\vnabla \vec{s})^T \vec{m}\right]
  \right).
  \label{eq:spin_diff_js_from_je}
\end{equation}
Inserting into \eqref{eq:spin_diff_dts} directly yields the spin-accumulation dynamics for a given magnetization configuration $\vec{m}$.
The resulting magnetization dynamics due to the spin torque, can be resolved by simultaneous solution of \eqref{eq:spin_diff_dts} and the LLG \eqref{eq:spin_diff_llg}.

\begin{figure}
  \centering
  \includegraphics{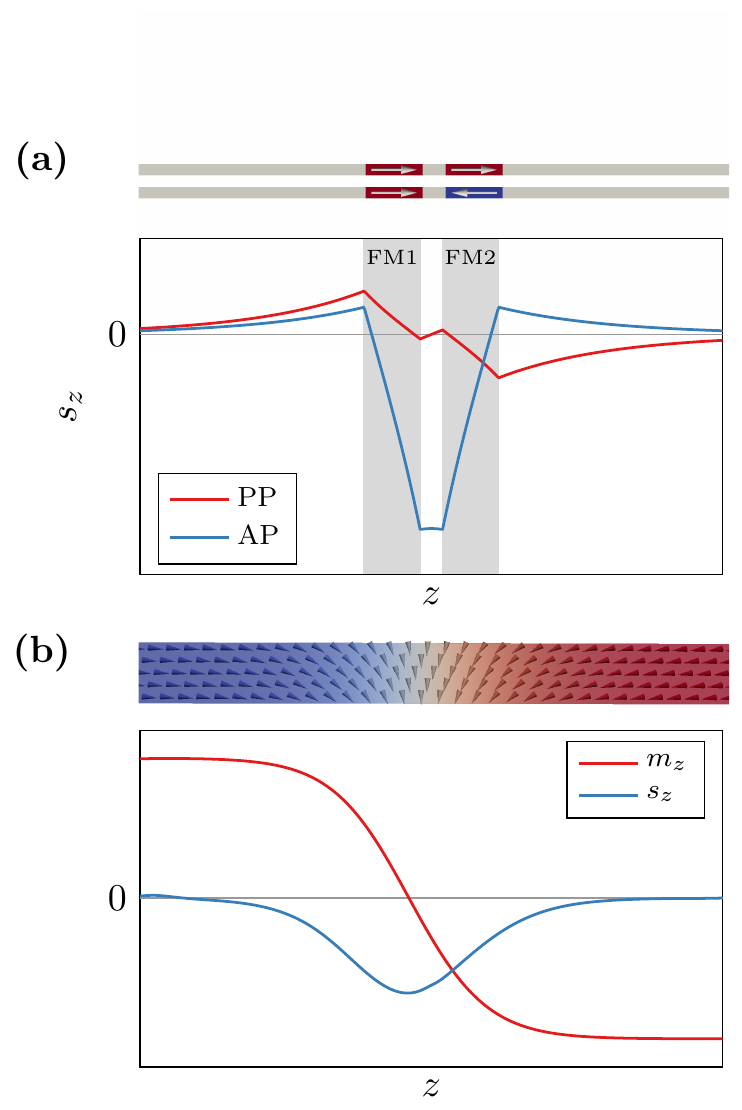}
  \caption{
    Spin accumulation $\vec{s}$ for typical magnetization configurations.
    (a) Spin accumulation in a magnetization multilayer with magnetic layers FM1 and FM2.
    $s_z$ is shown for a parallel and antiparallel magnetization configuration in $\pm z$-direction.
    (b) Spin accumulation due to a magnetic domain wall.
    The magnetization configuration $m_z$ is plotted along with the resulting spin accumulation $s_z$.
  }
  \label{fig:spin_diff_s}
\end{figure}
However, in most realistic systems, the spin accumulation relaxes two orders of magnitude faster than the magnetization configuration \cite{zhang2004roles}.
If the quantity of interest is the magnetization dynamics rather than the spin-accumulation dynamics, this difference in time scales can be exploited in order to simplify the model.
Namely, the spin-accumulation can be assumed to instantaneously relax when the magnetization changes.
In this case, the spin-accumulation does no longer explicitly depend on the time $t$, but only on the magnetization $\vec{m}$.
The defining equation for the spin accumulation $\vec{s}(\vec{m})$ is derived by setting $\pdiffs{\vec{s}}{t} = 0$ in \eqref{eq:spin_diff_dts} which results in
\begin{equation}
  \vnabla \cdot \js
  + \frac{\vec{s}}{\tausf}
  + J \frac{\vec{s} \times \vec{m}}{\hbar}
  = 0
  \quad\text{in}\quad
  \Omega.
  \label{eq:spin_diff_js_source}
\end{equation}
Inserting the definition of the spin current \eqref{eq:spin_diff_js} yields a linear partial differential equation of second order in $\vec{s}$.
Typical solutions for the spin accumulation $\vec{s}$ in STT devices as well as domain walls are depicted in Fig.~\ref{fig:spin_diff_s}.
By application of this simplified model, the treatment of the spin-accumulation $\vec{s}$ in the context of dynamical micromagnetics becomes similar to the treatment of effective-field contributions.
Instead of performing a coupled time integration on both the magnetization $\vec{m}$ and the spin accumulation $\vec{s}$ as required by \eqref{eq:spin_diff_dts}, the spin accumulation is defined by the magnetization $\vec{m}$ only.

The previous methods require the knowledge of the charge-current distribution $\je$ for the computation of the spin accumulation $\vec{s}$.
In a regular shaped sample with a homogeneous conductivity $\sigma$, the charge-current density may be assumed constant in a first-order approximation.
However, in a magnetic system subject to spin-polarized currents, Ohm's law gives only one contribution to the total conductivity which may also depend on the magnetization configuration.
In order to accurately account for magnetization dependent resistance effects, neither the electric potential $u$ nor the charge current $\je$ should be prescribed in the sample.
Instead, these entities should be solution variables like the spin accumulation $\vec{s}$.
In order to set up such a self-consistent model, the source equation \eqref{eq:spin_diff_js_source} has to be complemented by a source equation for the charge current, which is naturally given by the continuity equation
\begin{equation}
  \vnabla \cdot \je = 0
  \quad\text{in}\quad
  \Omega.
  \label{eq:spin_diff_je_source}
\end{equation}
Inserting the current definitions \eqref{eq:spin_diff_js} and \eqref{eq:spin_diff_je} in the source equations \eqref{eq:spin_diff_js_source} and \eqref{eq:spin_diff_je_source} yields a system of linear partial differential equations that can be solved for the solution pair $(\vec{s}, u)$ with the magnetization $\vec{m}$ being an input to the system.
In this self-consistent model, instead of prescribing the charge current $\je$ or potential $u$ in the complete sample, boundary conditions are used to define the potential or current inflow on specific interfaces.

\subsection{Boundary conditions for the spin-diffusion model}\label{sec:spin_diff_bc}
\begin{figure}
  \centering
  \includegraphics{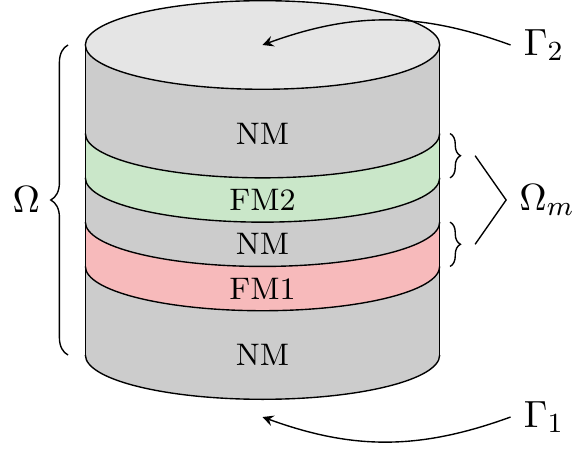}
  \caption{
    Illustration of a typical magnetic multilayer system.
    The magnetic layers FM1 and FM2 are separated by a nonmagnetic spacer layer and sandwiched by nonmagnetic leads.
    $\Omega$ denotes the volume of the complete system and $\Omega_m$ denotes the volume of magnetic layers.
    The bottom and top surface of the complete system are denoted by $\Gamma_1$ and $\Gamma_2$.
  }
  \label{fig:spin_diff_regions}
\end{figure}
Since the partial differential equations for the solution of the spin-diffusion model are of second order in both solution variables $u$ and $\vec{s}$, boundary conditions are required in order to find a unique solution.
The boundary conditions of the electric potential $u$ directly correspond to the voltage or charge current applied to the system through electric contacts.
A typical multilayer system with contact regions $\Gamma_1$ and $\Gamma_2$ is depicted in Fig.~\ref{fig:spin_diff_regions}.
By choosing either Dirichlet or Neumann conditions for $u$ on a part of the boundary, a constant electric potential or charge-current inflow can be prescribed on this part respectively.
For example, in order to simulate a constant potential $u_0$ at contact $\Gamma_1$, a Dirichlet condition is applied directly to the potential $u$
\begin{equation}
  u = u_0 \quad \text{on} \quad \Gamma_{1}.
\end{equation}
Applying a constant current inflow $j_0$ on contact $\Gamma_2$ is achieved by applying the Neumann condition
\begin{equation}
  \je \cdot \vec{n} =
  - 2 \left[
  C_0 \vnabla u + D_0 \beta' \frac{e}{\mub} \left[ (\vnabla\vec{s})^T \vec{m} \right]
  \right] \cdot \vec{n}
  =
  j_0
  \quad \text{on} \quad \Gamma_2
\end{equation}
where $\vec{n}$ denotes the outward pointing normal to $\Gamma_2$.
In order to complete the set of boundary conditions for $u$, all parts of the sample's boundary which are not used as contacts are treated with homogeneous Neumann conditions
\begin{equation}
  \je \cdot \vec{n} = 0 
  \quad \text{on} \quad \partial \Omega \setminus \Gamma_2 \cup \Gamma_2.
\end{equation}
The spin accumulation $\vec{s}$ is treated with homogeneous Neumann conditions on the complete boundary
\begin{equation}
  \vnabla \vec{s} \cdot \vec{n} = 0
  \quad \text{on} \quad \partial \Omega
  \label{eq:spin_diff_s_neumann}
\end{equation}
which is equivalent to a no-flux condition on the spin current $\js \cdot \vec{n} = 0$ for systems as depicted in Fig.~\ref{fig:spin_diff_regions}, where the contacts belong to the boundary of the nonmagnetic region $\Omega_n$.
This equivalence is obtained by multiplying~\eqref{eq:spin_diff_js_from_je} with the boundary normal $\vec{n}$ and inserting the Neumann condition which yields the boundary flux
\begin{align}
  \js \cdot \vec{n} &=
  \beta \frac{\mub}{e} \vec{m} (\je \cdot \vec{n})
  - 2 D_0 \left[
    \vnabla\vec{s} \cdot \vec{n}
    - \beta \beta' \vec{m} ((\vnabla\vec{s} \cdot \vec{n}) \cdot \vec{m})
  \right]\\
  &= \beta \frac{\mub}{e} \vec{m} (\je \cdot \vec{n}).
\end{align}
This spin-current flux is nonzero only at boundaries with both nonvanishing charge-current flux $\je \cdot \vec{n}$ and nonvanishing magnetization $\vec{m}$.

A vanishing charge-carrier flux usually implies a vanishing spin-current flux since the spin is transported by the charge carriers.
This makes the no-flux condition on the spin current a reasonable choice for parts of the boundary that do not serve as electric contacts.
For electric contacts, the no-flux condition is reasonable only if the thickness of the respective nonmagnetic regions exceeds the spin-flip relaxation length.
In this case the polarization of the current and thus the spin flux can be assumed zero at the contact, see e.g. Fig.~\ref{fig:spin_diff_s}\,(a).
If this is not the case, the contact should be treated with a Robin condition instead
\begin{equation}
  \vnabla \vec{s} \cdot \vec{n} + \frac{1}{\sqrt{2 D_0 \tausf}} \vec{s} = 0
\end{equation}
where the exponential decay of the spin accumulation in the nonmagnetic lead region is taken into account.

While the boundary conditions on the electric potential $u$ are relevant only for the self-consistent treatment of $u$ and $\vec{s}$, the boundary conditions for the spin accumulation $\vec{s}$ also hold for the simplified spin-diffusion models with prescribed electric potential or charge current respectively.

\subsection{Spin-orbit torque}
\begin{figure}
  \centering
  \includegraphics{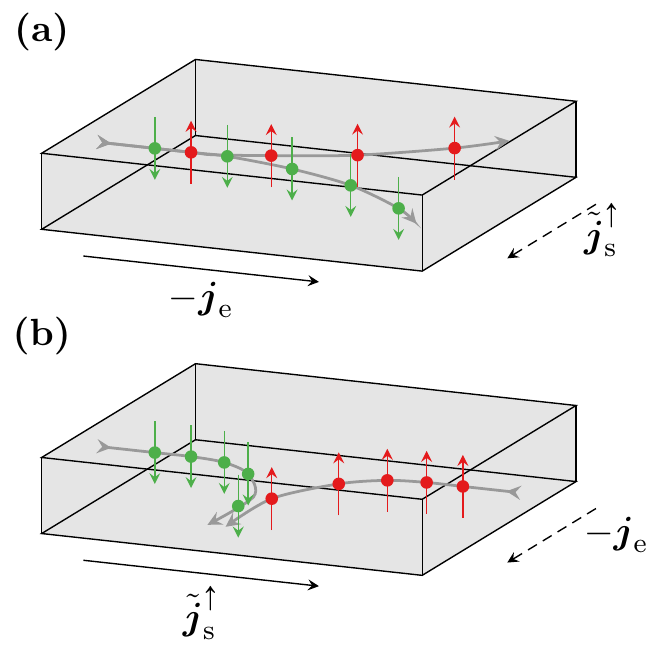}
  \caption{
    Illustration of current conversion due to the spin-Hall and inverse spin-Hall effect.
    (a) Spin-Hall effect.
    A non-polarized current is subject to spin splitting due to spin-orbit coupling.
    The result is the conversion of a charge current $\je$ into a spin current $\js$ perpendicular to $\je$.
    (b) inverse spin-Hall effect.
    A pure spin current $\js$ is considered to be constituted by two charge currents of opposite direction and spin polarization.
    The spin-orbit induced deflection of the polarized electrons leads to the creation of a charge current $\je$ perpendicular to $\js$.
  }
  \label{fig:spin_hall}
\end{figure}
Several extensions to the spin-diffusion model introduced in Sec.~\ref{sec:spin_diff} have been proposed in order to account for further spintronics effects.
An important class of effects describes the conversion of charge currents to spin currents and vice versa due to spin-orbit coupling.
The origin for this conversion is the polarization dependent deflection of the conducting electrons either due to material impurities \cite{dyakonov1971current,hirsch1999spin} or due to intrinsic asymmetries of the material \cite{murakami2003dissipationless,sinova2004universal}.
Depending on the direction of the current conversion, this spin-orbit effect is either referred to as spin-Hall effect or inverse spin-Hall effect.
The spin-Hall effect describes the conversion of charge currents into spin currents, see Fig.~\ref{fig:spin_hall}\,(a), while the conversion of spin currents into charge currents is referred to as inverse spin-Hall effect, see \ref{fig:spin_hall}\,(b).
Incorporating these effects into the spin-diffusion model is done by extending the original current definitions \eqref{eq:spin_diff_je} and \eqref{eq:spin_diff_js} according to Dyakonov \cite{dyakonov2007magnetoresistance}.
The extended current definitions $\je'$ and $\js'$ are defined in terms of the original current definitions and read
\begin{align}
  j_{\text{e},i}'  &= j_{\text{e},i} + \epsilon_{ijk} \thetash \frac{e}{\mub} j_{\text{s},jk} \label{eq:spin_diff_je_so}\\
  j_{\text{s},ij}' &= j_{\text{s},ij} - \epsilon_{ijk} \thetash \frac{\mub}{e} j_{\text{e},k} \label{eq:spin_diff_js_so}
\end{align}
where index notation was used.
Here $\epsilon_{ijk}$ is the Levi-Cevita tensor and $\thetash$ is the dimensionless spin-Hall angle.
Inserting $\je'$ and $\js'$ into the source equations \eqref{eq:spin_diff_je_source} and \eqref{eq:spin_diff_js_source} yields a self-consistent spin-diffusion model including spin-Hall effects.
Typically, the spin-Hall effects are exploited in multilayer structures with heavy-metal layers which are subject to spin-orbit coupling and neighboring magnetic layers where the spin-polarized currents interact with the magnetization configuration.
Similar to the considerations in the original spin-diffusion model, equations \eqref{eq:spin_diff_je_so} and \eqref{eq:spin_diff_js_so} together with the respective source equations are solved in the complete structure including magnetic and nonmagnetic regions, using spatially varying material parameters in order to account for the different material properties.

\subsection{Material parameters in the spin-diffusion model}
The spin-diffusion model introduces a number of material parameters to the set of parameters required by classical micromagnetics.
Depending on the exact formulation of the spin-diffusion model, different sets of parameters are used.
The spin-flip relaxation time $\tausf$ and the exchange coupling of the spin-accumulation and the magnetization $J$ are often specified in terms of the characteristic length scales $\lambdasf$ and $\lambdaj$ defined by
\begin{align}
  \lambdasf &= \sqrt{2 D_0 \tausf}\\
  \lambdaj  &= \sqrt{2 D_0 \hbar / J}.
\end{align}
Alternatively, the exchange strength $J$ may be quantified by the characteristic time $\tauj = \hbar / J$.
While classical micromagnetics is often used to describe single-phase materials, the spin-diffusion model is usually used to solve the spin accumulation in composite systems.
In order to account for such systems, that expose different material properties in different regions, all material parameters in the governing equations are scalar fields rather than constants.
It should be noted that none of the material parameters, both in classical micromagnetics as well as in the spin-diffusion model, are subject to spatial differentiation.
Hence, the material-parameter fields may comprise jumps across material interfaces without compromising the mathematical formulation of the model.

\subsection{Spin dephasing}
In addition to the spin-flip relaxation and the exchange coupling in the original spin-diffusion model \eqref{eq:spin_diff_dts}, an additional term for the description of spin-dephasing was proposed in several works \cite{petitjean2012unified,akosa2015role,haney2013current}
\begin{equation}
  \pdiffs{\vec{s}}{t} =
  - \vnabla \cdot \js
  - \frac{\vec{s}}{\tausf}
  - \frac{\vec{s} \times \vec{m}}{\tauj}
  - \frac{\vec{m} \times (\vec{s} \times \vec{m})}{\tauphi}.
  \label{eq:spin_diff_dephasing}
\end{equation}
where $\tauphi$ is the spin-dephasing time.
This extended spin-diffusion model can be solved similar to the original model either in nonequilibrium or equilibrium and either for prescribed charge current or self-consistently.

\subsection{Valet-Fert model}
\begin{figure}
  \centering
  \includegraphics{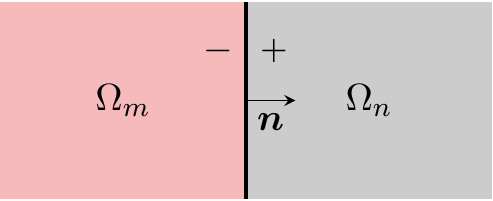}
  \caption{
    Magnetic--nonmagnetic interface in the Valet-Fert model.
    The interface normal is defined to point from the magnetic layer $\Omega_m$ to the nonmagnetic layer $\Omega_n$.
    Function evaluations at the interface are marked either by the superscript $-$ to denote evaluation in $\Omega_m$ or by the superscript $+$ to denote evaluation in $\Omega_n$.
  }
  \label{fig:spin_vf_regions}
\end{figure}
An alternative to the spin-diffusion model introduced in Sec.~\ref{sec:spin_diff} was introduced by Valet and Fert in \cite{valet1993theory}.
The originally one-dimensional model for collinear magnetization configurations was generalized to three dimensions and noncollinear configurations in \cite{niimi2012giant}.
Similar to the Zhang-Levy-Fert model introduced in Sec.~\ref{sec:spin_diff}, the Valet-Fert model defines charge and spin currents $\jevf$ and $\jsvf$ as well as an electric potential $\uvf$ and a spin potential $\svf$ which corresponds to the spin accumulation $\vec{s}$.
However, in contrast to the Zhang-Levy-Fert model, the spin current and spin potential are assumed to be collinear to the magnetization in the ferromagnetic regions $\Omega_m$
\begin{align}
  \svf &= \ssvf \vec{m} \label{eq:spin_vf_s_scalar}\\
  \jsvf &= \vec{m} \otimes \jssvf \label{eq:spin_vf_js_scalar}.
\end{align}
With these simplified assumptions, the Valet-Fert model defines the currents in the magnetic region $\Omega_m$ as
\begin{align}
  \jevf &= - \frac{\vnabla \uvf}{\rhos (1 - \betavf^2)} - \frac{\betavf \vnabla \ssvf}{2 \rhos (1 - \betavf^2)} \\
  \jssvf &= - \frac{\betavf \vnabla \phi}{\rhos (1 - \betavf^2)} - \frac{\vnabla \ssvf}{2 \rhos (1 - \betavf^2)}
\end{align}
with $\rhos$ being connected to the electric conductivity and $\betavf$ being a dimensionless polarization parameter.
In the nonmagnetic region $\Omega_n$, the magnetization $\vec{m}$ as well as the polarization parameter $\betavf$ vanishes and the currents are defined as
\begin{align}
  \jevf &= - \frac{\vnabla \uvf}{\rhos} \label{eq:spin_vf_jen}\\
  \jsvf &= - \frac{\vnabla \svf}{2 \rhos} \label{eq:spin_vf_jsn}
\end{align}
with \eqref{eq:spin_vf_jen} being Ohm's law.
The source equations for the currents are given as
\begin{align}
  \vnabla \cdot \jevf &= 0 \label{eq:spin_vf_je_source} \\
  2 \rhos \vnabla \cdot \jsvf &= - \frac{\svf}{\lambdavf^2} \label{eq:spin_vf_js_source}
\end{align}
where $\lambdavf$ is a spatially varying material parameter that denotes the characteristic spin-flip relaxation length.
In contrast to the Zhang-Levy-Fert model, the Valet-Fert model does not require the electric potential $\uvf$ and the spin potential $\svf$ to be continuous across interfaces.
Instead, these potentials are subject to a set of well-defined jump conditions.
The charge current $\je$ is continuous everywhere which includes interfaces.
The spin current $\js$ is continuous within magnetic/nonmagnetic layers.
Furthermore its longitudinal component is continuous across magnetic/nonmagnetic interfaces
\begin{equation}
  \vec{n} \cdot \jsvf^- =
  \left[ \vec{m} \cdot \jsvf^+ \right] \cdot \vec{n}
\end{equation}
where the `$-$' superscript corresponds to values in the magnetic layer and the `$+$' superscript corresponds to values in the nonmagnetic layer, see Fig.~\ref{fig:spin_vf_regions}.
Both, the charge potential $\uvf$ and the spin potential $\svf$ may have jumps at magnetic--nonmagnetic interfaces defined by
\begin{align}
  \vec{n} \cdot \jevf^+ &=
  - \frac{\uvf^+ - \uvf^-}{\rsb(1 - \gammavf^2)}
  - \gammavf \frac{\vec{m} \cdot (\svf^+ - \svf^-)}{2 \rsb(1 - \gammavf^2)} \label{eq:spin_vf_je_jump}\\
  \vec{n} \cdot \jsvf^+ &=
  -\left[
    \gammavf \frac{\uvf^+ - \uvf^-}{\rsb(1 - \gammavf^2)}
    + \frac{\vec{m} \cdot (\svf^+ - \svf^-)}{2 \rsb(1 - \gammavf^2)}
  \right] \vec{m}
  - \gup \left[
    \svf^+ - \svf^- - (\vec{m} \cdot [\svf^+ - \svf^-]) \vec{m}
  \right] \label{eq:spin_vf_js_jump}
\end{align}
where the interface properties $\rsb$, $\gammavf$ and $\gup$ denote the resistivity, the spin-flip probability and the spin-mixing conductance of the interface respectively.
While the interface resistivity $\rsb$ and the splin-flip probability $\gammavf$ have bulk counterparts in the Zhang-Levy-Fert model, namely $C_0^{-1}$ and $\tausf^{-1}$, the spin-mixing conductance $\gup$ is unique to the Valet-Fert model.
It describes the interface resistivity for electrons polarized perpendicular to the magnetization in the ferromagnetic layer and contributes significantly to the overall resistivity of the interface.

Moreover, the spin-mixing conductance $\gup$ plays an important role for the description of spin torque in the context of the Valet-Fert model.
Since the spin potential $\svf$ is collinear in the magnetic regions by definition of the model, see \eqref{eq:spin_vf_s_scalar}, it cannot exert a torque on the magnetization $\vec{m}$.
Thus, torque can only be generated at the interface between magnetic and nonmagnetic regions, where the spin potential $\svf$ can have components perpendicular to the magnetization.
In general, the spin-mixing conductance is assumed to be complex-valued with the real and imaginary part describing the strength of the dampinglike and fieldlike torque respectively.
This said, the Valet-Fert model does not predict the ratio of these torque contributions in contrast to the spin-diffusion model introduced in Sec.~\ref{sec:spin_diff}.
As for the simplified model by Slonczewski, this ratio is an input parameter to the method.

\subsection{Connecting the spintronics models}
Various micromagnetic models for the description of spin torque and other spintronics effects are described in the preceding section.
While the spin-diffusion model introduced in Sec.~\ref{sec:spin_diff} covers a multitude of spintronics effects, specialized models like the model by Slonczewski, see Sec.~\ref{sec:spin_slon}, were developed for very specific purposes.

\subsubsection{Slonczewski model}
\begin{figure}
  \centering
  \includegraphics{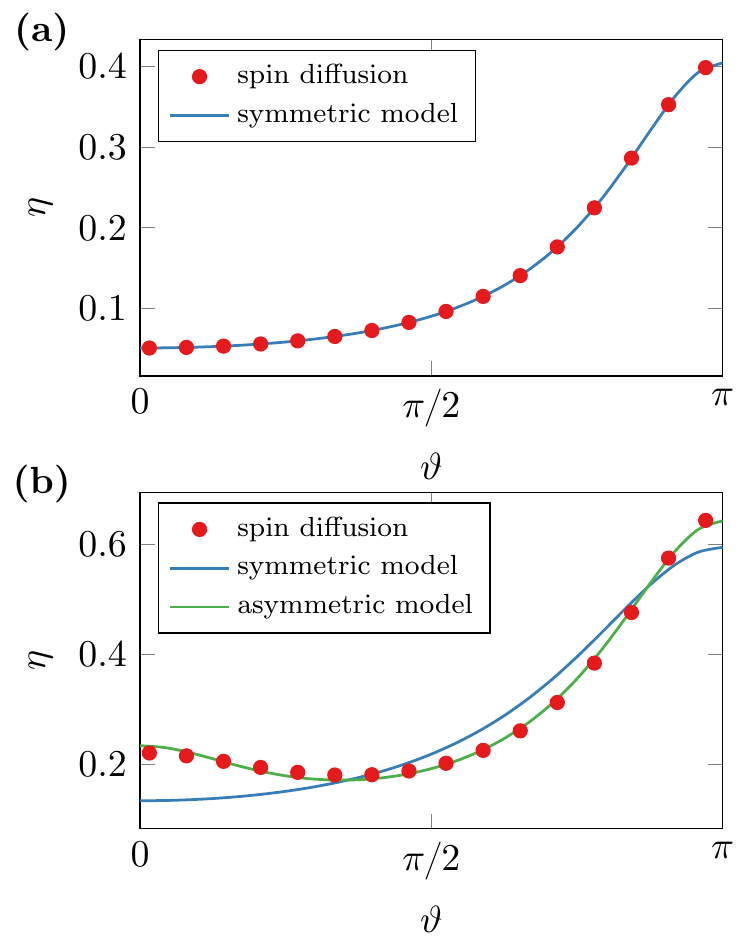}
  \caption{
    Spin-torque angular dependence $\eta$ fitted to the simulation results of the spin-diffusion model.
    (a) Angular dependence for a symmetric multilayer with two similar magnetic layers.
    The original model by Slonczewski shows a good agreement with the spin-diffusion results.
    (b) Angular dependence for an asymmetric multilayer.
    The original model by Slonczewski is insufficient, the generalized model shows a good agreement.
  }
  \label{fig:spin_diff_vs_slon}
\end{figure}
The Slonczewski model describes the spin torque in multilayers in terms of a macrospin approach.
The characteristic properties of this model are the angular dependencies $\eta_\text{damp}(\vartheta)$ and $\eta_\text{field}(\vartheta)$ of the torques defined in \eqref{eq:spin_t_damp} and \eqref{eq:spin_t_field}.
While the original angular dependency proposed by Slonczewski \eqref{eq:spin_slon_orig} is predicted to be valid for structures with two similar magnetic layers, the more general form \eqref{eq:spin_slon_asym} is expected to work also for asymmetric structures.
In order to compare the model of Slonczewski with the spin-diffusion model, the torque for a homogeneously magnetized free layer with varying tilting angle $\vartheta$ is computed with the spin-diffusion model.
The angular dependence of this torque is extracted and compared to the Slonczewski model in Fig.~\ref{fig:spin_diff_vs_slon}.
The asymmetric system used for this comparison has two magnetic layers with thicknesses $\SI{3}{nm}$ and $\SI{5}{nm}$ and typical material parameters as used in \cite{abert2018efficient}.
The symmetric system has similar material parameters, but uses the same layer thickness of $\SI{3}{nm}$ for both magnetic layers.
The symmetric structure is well fitted by the original expression for the angular expression, see Fig.~\ref{fig:spin_diff_vs_slon}\,(a).
For the asymmetric structure, i.e. a structure with different free-layer and pinned-layer thicknesses, the more general expression \eqref{eq:spin_slon_asym} is required in order obtain an accurate fit, see Fig.~\ref{fig:spin_diff_vs_slon}\,(b).
This proves agreement of the two models in the application scope of the Slonczewski model and consequently the superiority of the spin-diffusion model which is able to accurately describe further spin-transport effects and devices.

\subsubsection{Zhang-Li model}\label{sec:spin_connect_zhang}
Like the model of Slonczewski, the model of Zhang and Li introduced in Sec.~\ref{sec:spin_zhang} can be perceived as a special case of the spin-diffusion model.
Setting $D_0 = 0$ in \eqref{eq:spin_diff_js_from_je} and inserting in \eqref{eq:spin_diff_js_source} yields
\begin{align}
  - \vnabla \left( \frac{\beta \mub}{e} \vec{m} \otimes \je \right)
  + \frac{\vec{s}}{\tausf} + \frac{J}{\hbar} \vec{s} \times \vec{m}
  &= \\
  - \frac{\beta \mub}{e} (\je \cdot \vnabla) \vec{m}
  + \frac{\vec{s}}{\tausf} - \frac{J}{\hbar} \vec{m} \times \vec{s}
  &= 0.
\end{align}
Multiplying with $\vec{m}$ and $\vec{m} \times \vec{m}$ respectively and eliminating $\vec{m} \times (\vec{m} \times \vec{s})$ terms results in the torque
\begin{align}
  \vec{T}
  &= - \frac{J}{\hbar \Ms} \vec{m} \times \vec{s} \\
  &= \frac{\beta \mub}{e \Ms}
  \frac{1}{1 + \left(\frac{\hbar}{J \tausf}\right)^2}
  \left(
    \vec{m} \times [ \vec{m} \times (\je \cdot \vnabla) \vec{m} ] +
    \frac{\hbar}{J \tausf} \vec{m} \times (\je \cdot \vnabla) \vec{m}
  \right)
\end{align}
which has the form as the model of Zhang and Li \eqref{eq:spin_zhang_torque} with the degree of nonadiabacity being defined as
\begin{equation}
  \xi = \frac{\hbar}{J \tausf} = \frac{\lambdaj^2}{\lambdasf^2}.
\end{equation}
This means, that the model of Zhang and Li exactly reproduces the spin-diffusion model for vanishing diffusion $D_0$.
Neglecting the spin diffusion restricts the model of Zhang and Li to the description of local torque phenomena, i.e. torque due to local magnetization gradients such as domain walls.

\subsubsection{Valet-Fert model}
In contrast to the Slonczewski and Zhang-Li models, the Valet-Fert model is closely linked to the spin-diffusion model introduced in Sec.~\ref{sec:spin_diff}.
Within the nonmagnetic layers the models are completely similar and in the magnetic regions the equations have common terms.
A major difference of the models is the role of interfaces that have distinct properties such as the resistivity $\rsb$, the spin-flip probability $\gammavf$ and the spin-mixing conductance $\gup$ in the Valet-Fert model.
In contrast, the Zhang-Levy-Fert model introduced in Sec.~\ref{sec:spin_diff} solely relies on bulk material properties.
For the bulk properties, there is a straightforward mapping of the solution variables and material parameters.
Using \eqref{eq:spin_vf_s_scalar} and \eqref{eq:spin_vf_js_scalar} and assuming a constant magnetization in the ferromagnetic layers $\vnabla \vec{m} = 0$ yields
\begin{align}
  \jevf &= - \frac{\vnabla \phi}{\rhos (1 - \betavf^2)} - \frac{\betavf (\vnabla \svf)^T \vec{m}}{2 \rhos (1 - \betavf^2)} \label{eq:spin_vf_je}\\
  \jsvf &= - \frac{\betavf \vec{m} \otimes \vnabla \uvf}{\rhos (1 - \betavf^2)} - \frac{\vnabla \svf}{2 \rhos (1 - \betavf^2)} \label{eq:spin_vf_js}
\end{align}
for the Valet-Fert model.
With spatially resolved material parameters $\rho$ and $\betavf$, these current definitions can be used for the complete sample region $\Omega$.
Assuming the following relations, \eqref{eq:spin_vf_je} and \eqref{eq:spin_vf_js} can be identified with the definitions of the Zhang-Levy-Fert model \eqref{eq:spin_diff_je} and \eqref{eq:spin_diff_js}.
\begin{align}
  \jevf   &= \je \label{eq:spin_vf_map_je}\\
  \jsvf   &= - \frac{e}{\mub} \js \label{eq:spin_vf_map_js}\\
  \uvf    &= u \label{eq:spin_vf_map_u}\\
  \svf    &= - 2 \frac{D_0 e}{C_0 \mub} \vec{s} \label{eq:spin_vf_map_s}\\
  C_0     &= \frac{1}{2 \rhos (1 - \beta^2)}\\
  \betavf &= \beta = \beta'
\end{align}
where it should be noted that instead of the two polarization parameters $\beta$ and $\beta'$ of the Zhang-Levy-Fert model, the Valet-Fert model introduces only a single parameter $\betavf$.
Comparison of the source equations for the spin current in the different models \eqref{eq:spin_diff_js_source} and \eqref{eq:spin_vf_js_source} yields the following additional parameter mappings
\begin{align}
  J &= 0\\
  \tausf &= \frac{\lambdavf^2}{2 D_0 (1 - \betavf^2)}. \label{eq:spin_vf_map_tausf}
\end{align}
With these parameter mappings, the Zhang-Levy-Fert model can be used to solve the Valet-Fert model in both the magnetic regions $\Omega_m$ as well as the nonmagnetic regions $\Omega_n$.

While both models perfectly agree in the bulk, the essential differences of the models are the continuity conditions of the potentials.
While the Zhang-Levy-Fert model assumes continuous potentials $u$ and $\vec{s}$, the Valet-Fert model allows jumps which are defined by interface properties, namely the interface resistivity $\rsb$, the spin-flip probability $\gammavf$, and the spin-mixing conductance $\gup$.
However, these jumps can be mimicked in the Zhang-Levy-Fert model by introducing thin layers with effective material properties at the positions of the respective interfaces.
Approximation of the charge current in the Zhang-Levy-Fert model \eqref{eq:spin_diff_je} within the effective-interface layer $\Omega_e$ by means of finite differences and multiplication with the boundary normal $\vec{n}$ yields
\begin{equation}
  \vec{n} \cdot \je
  = - 2 C_0 \frac{u^+ - u^-}{d} + 2 D_0 \beta' \frac{e}{\mub} \frac{\vec{m} \cdot (\vec{s}^+ - \vec{s}^-)}{d}
\end{equation}
with $d$ being the thickness of the effective-interface layer.
Considering the potential mappings \eqref{eq:spin_vf_map_u}, \eqref{eq:spin_vf_map_s} and the current mapping \eqref{eq:spin_vf_map_je}, this translates to the following jump condition across the effective-interface layer
\begin{equation}
  \vec{n} \cdot \jevf
  = - 2 C_0 \frac{\uvf^+ - \uvf^-}{d} - \beta' C_0 \frac{\vec{m} \cdot (\svf^+ - \svf^-)}{d}
\end{equation}
with $d$ being the thickness of the layer.
Comparison with the jump condition \eqref{eq:spin_vf_je_jump} of the Valet-Fert model results in the parameter mappings
\begin{align}
  C_0    &= \frac{d}{2 \rsb (1 - \gammavf^2)} \label{eq:spin_vf_map_interface_c0}\\ 
  \beta' &= \gammavf \label{eq:spin_vf_map_interface_beta}
\end{align}
for the effective-interface layer.
Applying the same procedure to the spin current of the Zhang-Levy-Fert model \eqref{eq:spin_diff_js} yields
\begin{equation}
  \vec{n} \cdot \js
  = 2 C_0 \beta \frac{\mub}{e} \frac{\vec{m} \cdot (u^+ - u^-)}{d}
  - 2 D_0 \frac{\vec{s}^+ - \vec{s}^-}{d}
\end{equation}
which translates to
\begin{align}
  \vec{n} \cdot \jsvf
  =& - 2 C_0 \beta \frac{\vec{m} (\uvf^+ - \uvf^-)}{d}
  - C_0 \frac{\vec{\svf}^+ - \vec{\svf}^-}{d} \label{eq:spin_vf_js_jump_fd}\\
  =& - \left[ 2 C_0 \beta \frac{(\uvf^+ - \uvf^-)}{d}
  + C_0 \frac{\vec{m} \cdot (\vec{\svf}^+ - \vec{\svf}^-)}{d}
  \right] \vec{m}
  \nonumber\\
  &- \frac{C_0}{d} [
    \vec{\svf}^+ - \vec{\svf}^-
    - \vec{m} \cdot (\vec{\svf}^+ - \vec{\svf}^-) \vec{m}
  ]
\end{align}
and leads to the additional mappings
\begin{align}
  \beta &= \gammavf\\
  \gup  &= \frac{C_0}{d}.
\end{align}
According to these relations, the spin-mixing conductance $\gup$ depends implicitly on $\rsb$ and $\gammavf$ which contradicts the Valet-Fert model where $\gup$ is an independent interface property.
However, while the charge current $\je$ is approximately constant throughout the effective-interface layer $\Omega_e$ due to the continuity equation \eqref{eq:spin_diff_je_source}, the spin current $\js$ has sources in $\Omega_e$ according to \eqref{eq:spin_diff_js_source} which renders the finite-difference approximation \eqref{eq:spin_vf_js_jump_fd} inaccurate.
While an appropriate choice of $\tausf$ and $J$ in the effective-interface layer can approximately reproduce the behavior of the Valet-Fert model, the dependency of these material parameters from the spin-mixing conductance $\gup$ is nontrivial and is best resolved by a fitting procedure.

For the special case of a collinear magnetization configuration in the magnetic layers, the parameter mapping between the Zhang-Levy-Fert model and the Valet-Fert model is exact.
In this case, the spin-accumulation in both models is also collinear to the magnetization.
As a consequence, the spin-mixing-conductance term vanishes which reflects the fact, that a collinear magnetization configuration results in a vanishing spin torque.
That said, applying the bulk mappings \eqref{eq:spin_vf_map_je} -- \eqref{eq:spin_vf_map_tausf}, and adding effective-interface layers with thickness $d$ and material parameters \eqref{eq:spin_vf_map_interface_c0} -- \eqref{eq:spin_vf_map_interface_beta} in order to simulate the interface properties of the Valet-Fert model in the Zhang-Levy-Fert model, results in the same spin accumulation and electric potential for both models in the limit of small $d$.

The Valet-Fert model is very popular in the experimental community where the interface properties $\rsb$, $\lambdavf$ and $\gup$ are discussed and determined for various material systems.
However, the Zhang-Levy-Fert model provides a more general approach for the description of spin transport and spin torque.
In particular, the bidirectional description of spin torque that accounts for both the torque exerted from the current polarization on the magnetization and vice versa leads to a better representation of the physical processes.
Interface properties as defined by the Valet-Fert model can be modeled with additional thin layers.

\subsection{Beyond the spin-diffusion model}
While the spin-diffusion model introduced in Sec.~\ref{sec:spin_diff} has been shown to incorporate various models for the description of spin torque and other spintronics effects, its area of application is restricted to diffusive transport.
Some effects that are not included in the equations presented in Sec.~\ref{sec:spin_diff}, such as inplane GMR, spin pumping, and anomalous Hall effect might be added by means of additional terms to the diffusion model since they are in principle compatible with the diffusive transport assumption \cite{tserkovnyak2002spin}.
An important class of spintronics devices though is making use of magnetic tunnel junctions in order to exploit tunnel-magneto resistance (TMR) and spin torque.
The spin transport in tunnel junctions, however, is not diffusive.
Various ab initio models have been developed in order to accurately describe magnetic tunnel junctions \cite{mathon2001theory,caffrey2011prediction,butler2001spin}.
However, ab initio models are computationally challenging and do not integrate well with the semiclassical micromagnetic model.
The development of a suitable model that integrates well with the spin-diffusion model and micromagnetics is still subject to ongoing research.

\section{Discretization}
The micromagnetic model as introduced in the preceding sections defines a set of nonlinear partial differential equations in space and time, which can only be solved analytically for simple edge cases.
In general, the solution of both static and dynamic micromagnetics calls for numerical methods.
However, the development of efficient numerical methods is challenging due to the following properties of the micromagnetic equations.
\begin{enumerate}
  \item The demagnetization field, that describes the dipole--dipole interaction in the magnetic material, is a long-range interaction.
    A naive implementation of such an interaction has a computational complexity of $\mathcal{O}(\n^2)$ with $\n$ being the number of simulation cells.
    Various methods have been proposed to reduce this complexity to $\mathcal{O}(\n \log \n)$ \cite{berkov1993solving} or even $\mathcal{O}(\n)$ \cite{seberino2001concise}.
  \item The exchange interaction adds a local coupling with high stiffness due to its second order in space.
    While the competition of the long-range demagnetization field with the local exchange field is crucial for the generation of magnetic domains, it also poses high demands on the numerical time-integration methods.
  \item The nonlinear nature of most of the energy contributions leads to a complex energy landscape, which makes it difficult to efficiently seek for energy minima.
    In the context of quasistatic hysteresis computation, the nearest local energy minimum to a given magnetization configuration has to be found.
   A complex energy landscape increases the risk to miss a local minimum, and calls for a thoughtful choice of minimization algorithm.
\end{enumerate}

\begin{figure}
  \centering
  \includegraphics{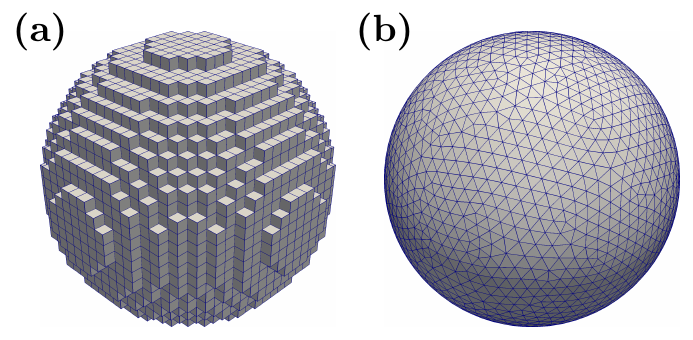}
  \caption{
    Spatial discretization of a sphere.
    (a) Regular cuboid grid with 8217 cells as required by finite-difference methods.
    (b) Tetrahedral mesh with 7149 vertices as required by finite-element methods.
  }
  \label{fig:discrete_fd_vs_fe}
\end{figure}
A large variety of tailored numerical methods for the solution of micromagnetic equations has been proposed.
Typically, these methods introduce distinct discretizations for space and time.
Among the spatial discretizations, the most popular methods applied in micromagnetics are the finite-difference method (FDM) and the finite-element method (FEM).
For both methods the magnetic region is subdivided into simulation cells resulting in a mesh of cells.
However, the requirements for the mesh differ significantly for both methods.
While the finite-difference method usually requires a regular cuboid mesh, the finite-element method typically works on irregular tetrahedral meshes, see Fig.~\ref{fig:discrete_fd_vs_fe}.
While FDM or FEM are used for spatial discretization in order to compute the effective field $\heff$ or the respective energy contributions $E$, another class of algorithms is required in order to either minimize the total energy with respect to the magnetization configuration $\vec{m}$ or to compute the time evolution of $\vec{m}$ according to the LLG.
Independent from the discretization method, the cell size has to be chosen sufficiently small in order to accurately resolve the structure of domain walls.
The characteristic length for the domain-wall width is the so-called exchange length which is defined by
\begin{equation}
  l = \sqrt{\frac{A}{K_\text{eff}}}.
  \label{eq:discrete_exchange_length}
\end{equation}
where $K_\text{eff}$ is the effective anisotropy constant that includes contributions from the crystalline anisotropy as well as the shape anisotropy which is introduced by the demagnetization field \cite{hubert_1998}.
Note, that for both energy minimization and magnetization dynamics, the effective field needs to be computed in the magnetic domain only.
In the following sections, the spatial discretization with FDM and FEM is discussed in detail.
In further sections, numerical methods for efficient integration of the LLG, energy minimization, and energy barrier calculations will be discussed.

\subsection{The finite-difference method}
The finite-difference method is a very popular numerical tool for the solution of micromagnetic equations.
While the restrictions to regular meshes renders this method inapt for certain problems involving complex geometries, this restriction allows the application of very fast algorithms.

\subsubsection{Demagnetization field}\label{sec:fd_demag}
Among the effective-field contributions introduced in Sec.~\ref{sec:energetics}, the demagnetization field holds the special role as the only long-range interaction.
Since long-range interactions are computationally costly, the value of a spatial discretization strategy is significantly influenced by its demagnetization-field algorithm.
The finite-difference method solves partial differential equations by approximation of the differential operators with finite differences.
In case of the demagnetization-field problem \eqref{eq:energetics_demag_poisson}, which has the form of Poisson's equation, this would require the approximation of the Laplacian.
However, the application of this classical finite-difference procedure is complicated by the open boundary condition \eqref{eq:energetics_demag_open_boundary}, which prevents the restriction of the computational domain to the magnetic region $\Omega_m$.
Hence, in finite-difference micromagnetics instead of discretizing Poisson's equation, the demagnetization field is usually solved by direct integration of \eqref{eq:energetics_demag_field}.
Consider a cellwise constant normalized magnetization 
\begin{equation}
  \vec{m}(\vec{x}) = \vec{m}_i
  \quad \forall \quad
  \vec{r} \in \Omega_i
  \label{eq:fd_demag_constant}
\end{equation}
with the cells $\Omega_i$ being an arbitrary partitioning of the magnetic domain $\Omega_m$
\begin{equation}
  \Omega_m = \bigcup_i \Omega_i
  \quad \text{with} \quad
  \Omega_i \cap \Omega_j = \emptyset
  \quad \text{if} \quad
  i \neq j.
  \label{eq:fd_demag_partitioning}
\end{equation}
Inserting the discretization \eqref{eq:fd_demag_constant} into the integral formulation of the demagnetization field \eqref{eq:energetics_demag_field} yields
\begin{align}
  \hdemag(\vec{r}) &= \int_\Omega \tensor{N}(\vec{x} - \vec{x}') \vec{M}'(\vec{x}) \dx' \\
                   &= \Ms \sum_j \left[ \int_{\Omega_j} \tensor{N}(\vec{x} - \vec{x}') \dx' \right] \vec{m}_j.
\end{align}
In order to compute the demagnetization field $\hdemag$ with the same discretization as the magnetization $\vec{m}$, the field is averaged over each cell $\Omega_i$ which results in
\begin{align}
  \hdemag_i
  &= \Ms \sum_j \left[ \frac{1}{V_i} \int_{\Omega_i} \int_{\Omega_j} \tensor{N}(\vec{x} - \vec{x}') \dx \dx' \right] \vec{m}_j \\
  &= \sum_j \mat{A}_{ij} \vec{m}_j
  \label{eq:fd_demag_discrete}
\end{align}
where $V_i$ is the volume of the simulation cell $i$.
Here, $\mat{A}$ denotes the linear demagnetization-field operator, which is represented by a dense $3\n \times 3\n$ matrix with $\n$ being the number of simulation cells.

\begin{figure}
  \centering
  \includegraphics{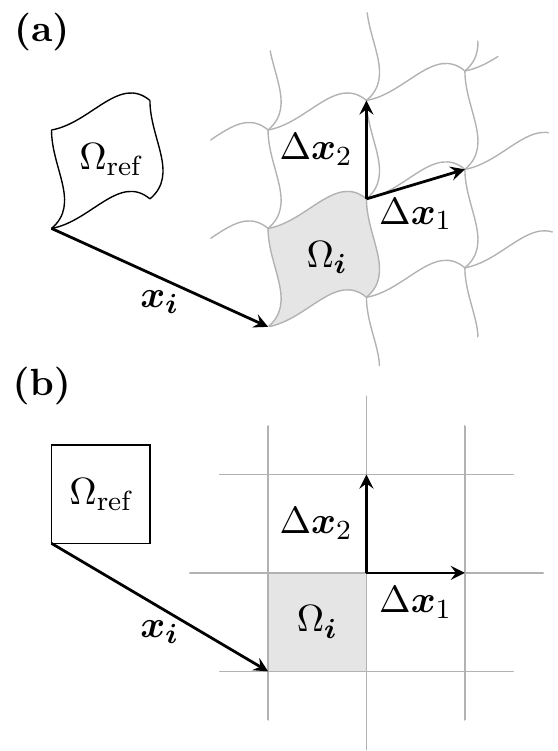}
  \caption{
    Examples of regular grids in two dimensions as required for the convolutional computation of the demagnetization field.
    (a) Irregularly shaped but periodic mesh.
    (b) Cuboid mesh as usually used for finite-difference computations.
  }
  \label{fig:fd_demag_regular_mesh}
\end{figure}
While this method could be used for the numerical demagnetization-field computation, it scales with $\mathcal{O}(\n^2)$ for both storage requirements and computational complexity which is unfeasible for large problems.
A better scaling can be accomplished by exploiting the convolutional structure of the integral equation \eqref{eq:energetics_demag_field}.
In order to preserve this structure on the discrete level, a regular spatial discretization is required, i.e. all simulation cells $\Omega_{\vec{i}}$ must be of the same shape $\Omega_\text{ref}$
\begin{equation}
  \mathbbm{1}_{\Omega_{\vec{i}}} (\vec{r}) = \mathbbm{1}_{\Omega_\text{ref}} (\vec{x} - \vec{x}_{\vec{i}})
  \label{eq:fd_demag_omega_ref}
\end{equation}
where $\mathbbm{1}_{\Omega_\text{ref}}$ denotes the indicator function of $\Omega_\text{ref}$, the multiindex $\vec{i}$ addresses the simulation cell and $\vec{x}_{\vec{i}}$ denotes the offset of the simulation cell $\Omega_{\vec{i}}$ from the reference cell $\Omega_\text{ref}$, see Fig.~\ref{fig:fd_demag_regular_mesh}.
Consequently, the offset from a simulation cell $\Omega_{\vec{i}}$ to another simulation cell $\Omega_{\vec{j}}$ is given by a multiple of the cell spacing $\Delta \vec{x}$ in every spatial dimension
\begin{equation}
  \vec{x}_{\vec{i}} - \vec{x}_{\vec{j}} 
  =
  \sum_k (i_k - j_k) \Delta \vec{x}_k.
  \label{eq:fd_demag_periodic}
\end{equation}
Using \eqref{eq:fd_demag_omega_ref} and \eqref{eq:fd_demag_periodic}, the integration in \eqref{eq:fd_demag_discrete} can be carried out over the reference cell $\Omega_\text{ref}$
\begin{align}
  \hdemag_{\vec{i}} &= \Ms \sum_{\vec{j}} \left[ \frac{1}{V_{\vec{i}}} \iint_{\Omega_\text{ref}} \tensor{N} \left(
  \sum_k (i_k - j_k) \Delta \vec{x}_k + \vec{x} - \vec{x}' \right) \dx \dx' \right] \vec{m}_{\vec{j}}.
\end{align}
which has the form of a discrete convolution since it only depends on the difference of the multiindices $\vec{i}$ and $\vec{j}$
\begin{align}
  \hdemag_{\vec{i}} &= \Ms \sum_{\vec{j}} \tensor{N}_{\vec{i} - \vec{j}} \vec{m}_{\vec{j}} \label{eq:fd_demag_discrete_convolution} \\
  \tensor{N}_{\vec{i} - \vec{j}} &= \frac{1}{V_{\vec{i}}} \iint_{\Omega_\text{ref}} \tensor{N} \left(
  \sum_k (i_k - j_k) \Delta \vec{x}_k + \vec{x} - \vec{x}' \right) \dx \dx' \label{eq:fd_demag_discrete_demag_tensor}.
\end{align}
The discrete demagnetization tensor $\tensor{\n}_{\vec{i} - \vec{j}}$ has entries for every possible cell distance which amounts to $\prod_k (2\n_k-1) \approx 8 \n$, where $\n_k$ denotes the number of cells in spatial dimension $k$.
Compared to the direct demagnetization-field operator $\mat{A}$, the demagnetization tensor reduces the storage requirements from $\mathcal{O}(\n^2)$ to $\mathcal{O}(\n)$.
The computational complexity, however, still amounts to $\mathcal{O}(\n^2)$ when implementing \eqref{eq:fd_demag_discrete_convolution} literally.

\begin{figure}
  \centering
  \includegraphics{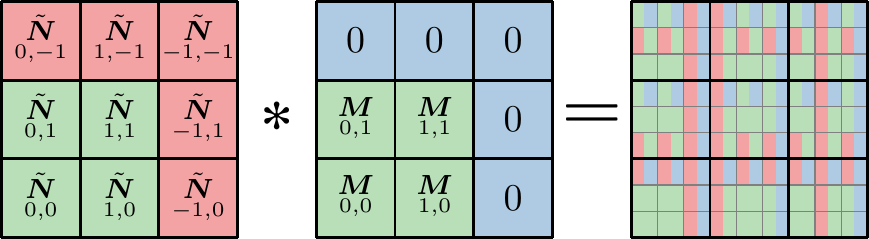}
  \caption{
    Visualization of the discrete convolution of the magnetization field $\vec{M}$ with the demagnetization-tensor field $\tensor{N}$.
    The color blocks in the result matrix represent the multiplications of the respective input values.
  }
  \label{eq:fd_demag_convolution}
\end{figure}
In order to reduce the computational complexity, the discrete convolution \eqref{eq:fd_demag_convolution} is computed in Fourier space where it reduces to a cell-wise multiplication according to the convolution theorem
\begin{equation}
  \mathcal{F}(\tensor{N} \ast \vec{m}) =
  \mathcal{F}(\tensor{N}) \mathcal{F}(\vec{m})
  \label{eq:fd_demag_convolution_theorem}
\end{equation}
where the Fourier transform is applied componentwise.
Since the cell-wise multiplication has a low complexity of $\mathcal{O}(\n)$, the overall complexity of the demagnetization-field computation is governed by the complexity of the Fourier-transform computation.
In case of the fast Fourier transform this amounts to $\mathcal{O}(\n \log \n)$.

Note that this fast-convolution algorithm requires the discrete demagnetization tensor $\tensor{N}_{\vec{i} - \vec{j}}$ to be of the same size as the discrete magnetization $\vec{m}_{\vec{i}}$ in order to perform cell-wise multiplication in Fourier space.
However, while the magnetization is discretized with $\prod_i \n_i$ cells, the discrete demagnetization tensor has a size of $\prod_i 2 \n_i - 1$.
Hence, the discrete magnetization has to be expanded in order to match the size of the demagnetization tensor.
Due to the cyclic nature of the discrete Fourier transform
\begin{equation}
  (f \ast g)_i = \sum_{j=0}^{\n-1} = f_{(i-j+\n)\%\n} \cdot g_j
\end{equation}
all entries of the demagnetization tensor $\tensor{N}_{\vec{i} - \vec{j}}$ are considered for every field evaluation $\hdemag_{\vec{i}}$.
For instance, the negative-distance entry $\tensor{N}_{-1, -1}$ is considered for the computation of the field $\hdemag$ at position $(0,0)$, although the $(0,0)$ cell has no neighbors at negative distances. 
In order to neglect these unphysical distances, the only reasonable choice for the expansion of the magnetization is by adding zero entries, which is often referred to as zero-padding, see \cite{press2007numerical}.
The complete convolution algorithm for the demagnetization-field computation is visualized in Fig.~\ref{eq:fd_demag_convolution}, where $\vec{m}$ and $\tensor{N}$ are reduced to two spatial dimensions for the sake of simplicity.
The result of the convolution algorithm is of the size $\prod_i 2 \n_i -1$ like the demagnetization tensor.
Physical meaningful values of the computed field $\hdemag$, however, are found only in the first $\prod_i \n_i$ entries.
The remaining entries are algorithmic byproducts and can be neglected.

Note, that the only requirement for the application of the fast convolution is a mesh regularity as described by \eqref{eq:fd_demag_omega_ref} and \eqref{eq:fd_demag_periodic}.
However, the evaluation of the demagnetization tensor \eqref{eq:fd_demag_discrete_demag_tensor} might be unfeasible for complicated reference cells such as illustrated in Fig.~\ref{fig:fd_demag_regular_mesh}\,(a).
For three-dimensional cuboid cells, an analytical formula for the demagnetization tensor $\tensor{N}_{\vec{i} - \vec{j}}$ was derived by Newell et al. \cite{newell1993generalization}.
According to this work the diagonal element $N_{1,1}$ of the tensor is given by
\begin{multline}
  N_{1,1}(\vec{x}, \Delta \vec{x}) =  \frac{1}{4 \pi \Delta x_1 \Delta x_2 \Delta x_3}
  \sum_{\vec{i}, \vec{j} \in \{0,1\}}
  (-1)^{\sum_x i_x + j_x} \\
  f[x_1 + (i_1 - j_1) \Delta x_1, x_2 + (i_2 - j_2) \Delta x_2, x_3 + (i_3 - j_3) \Delta x_3]
\end{multline}
where the function $f$ is defined as
\begin{align}
  f(x_1, x_2, x_3)
  &= \frac{|x_2|}{2} (x_3^2 - x_1^2) \sinh^{-1}\left( \frac{|x_2|}{\sqrt{x_1^2 + x_3^2}} \right) \nonumber \\
  &+ \frac{|x_3|}{2} (x_2^2 - x_1^2) \sinh^{-1}\left( \frac{|x_3|}{\sqrt{x_1^2 + x_2^2}} \right) \nonumber \\
  &- |x_1 x_2 x_3| \tan^{-1} \left( \frac{|x_2 x_3|}{x_1 \sqrt{x_1^2 + x_2^2 + x_3^2}} \right) \nonumber \\
  &+ \frac{1}{6} (2 x_1^2 - x_2^2 - x_3^2) \sqrt{x_1^2 + x_2^2 + x_3^2}.
\end{align}
The elements $N_{2,2}$ and $N_{3,3}$ are obtained by circular permutation of the coordinates
\begin{align}
  N_{2,2}(\vec{x}, \Delta \vec{x})
  &= N_{1,1} [ (x_2, x_3, x_1), (\Delta x_2, \Delta x_3, \Delta x_1) ] \\
  N_{3,3}(\vec{r}, \Delta \vec{r})
  &= N_{1,1} [ (x_3, x_1, x_2), (\Delta x_3, \Delta x_1, \Delta x_2) ].
\end{align}
The off-diagonal element $N_{1,2}$ is given by
\begin{multline}
  N_{1,2}(\vec{x}, \Delta \vec{x}) =  \frac{1}{4 \pi \Delta x_1 \Delta x_2 \Delta x_3}
  \sum_{\vec{i}, \vec{j} \in \{0,1\}}
  (-1)^{\sum_x i_x + j_x} \\
  g[x_1 + (i_1 - j_1) \Delta x_1, x_2 + (i_2 - j_2) \Delta x_2, x_3 + (i_3 - j_3) \Delta x_3]
\end{multline}
where the function $g$ is defined as
\begin{align}
  g(x_1, x_2, x_3)
  &= (x_1 x_2 x_3) \sinh^{-1}\left( \frac{x_3}{\sqrt{x_1^2 + x_2^2}} \right) \nonumber \\
  &+ \frac{x_2}{6} (3 x_3^2 - x_2^2) \sinh^{-1}\left( \frac{x_1}{\sqrt{x_2^2 + x_3^2}} \right) \nonumber \\
  &+ \frac{x_1}{6} (3 x_3^2 - x_1^2) \sinh^{-1}\left( \frac{x_2}{\sqrt{x_1^2 + x_3^2}} \right) \nonumber \\
  &- \frac{x_3^3}{6} \tan^{-1} \left( \frac{x_1 x_2}{x_3 \sqrt{x_1^2 + x_2^2 + x_3^2}} \right)
   - \frac{x_3 x_2^2}{2} \tan^{-1} \left( \frac{x_1 x_3}{x_2 \sqrt{x_1^2 + x_2^2 + x_3^2}} \right) \nonumber \\
  &- \frac{x_3 x_1^2}{2} \tan^{-1} \left( \frac{x_2 x_3}{x_1 \sqrt{x_1^2 + x_2^2 + x_3^2}} \right)
   - \frac{x_1 x_2 \sqrt{x_1^2 + x_2^2 + x_3^2}}{3}.
\end{align}
Again other off-diagonal elements are obtained by permutation of coordinates
\begin{align}
  N_{1,3}(\vec{x}, \Delta \vec{x})
  &= N_{1,2} [ (x_1, x_3, x_2), (\Delta x_1, \Delta x_3, \Delta x_2) ] \\
  N_{2,3}(\vec{x}, \Delta \vec{x})
  &= N_{1,2} [ (x_2, x_3, x_1), (\Delta x_2, \Delta x_3, \Delta x_1) ].
\end{align}
Like the continuous tensor $\tensor{N}(\vec{x} - \vec{x}')$ the discrete tensor $\tensor{N}_{\vec{i} - \vec{j}}$ is symmetric
\begin{equation}
  N_{ij} = N_{ji}.
\end{equation}
Hence, the above definitions of $N_{1,2}$, $N_{1,3}$ and $N_{2,3}$ can be used to obtain the remaining off-diagonal elements.
While these analytical expressions are exact, their numerical evaluation can lead to inaccuracies due to floating-point errors.
These errors especially occur for large cell distances and degenerated cells and can be avoided by numerical integration as shown by Lebecki et al. \cite{lebecki2008periodic} and Kr\"uger et al. in \cite{kruger2013fast}.

The fast-convolution algorithm can be further optimized by considering the specific properties of the demagnetization-field problem, i.e. Fourier transforms of zero values and evaluation of unneeded data can be omitted \cite{kanai2010micromagnetic} and symmetries in the demagnetization tensor can be exploited \cite{miltat2007numerical} to further speed up computations.

Instead of computing the demagnetization field directly from a convolution of the magnetization $\vec{m}$ with the demagnetization tensor $\tensor{N}$, the field can also be computed as negative gradient of the scalar potential $u$ \cite{abert2012fast,fu2016finite}.
The scalar potential $u$ itself is the result of a convolution, see \eqref{eq:energetics_demag_potential_convolution}, and can be computed by the fast-convolution algorithm.
The gradient can be approximated with finite differences.
Compared to the direct computation, the advantage of this approach is the reduction of Fourier-transform operations.
However, the additional gradient computation compromises the overall accuracy of the computation.

\subsubsection{Local field contributions}
Except for the demagnetization field, all energy contributions introduced in Sec.~\ref{sec:energetics} have either local or short-range character in the sense that they depend on derivatives of the magnetization.
Local contributions to the effective field, such as the anisotropy fields \eqref{eq:energetics_aniso_uniaxial_field} and \eqref{eq:energetics_aniso_cubic_field} are simply evaluated cellwise.
Differential operators in short-range contributions such as the exchange field \eqref{eq:static_exchange_field} are approximated with finite differences.
The second-order finite-difference approximations for the first and second spatial derivative of the discretized magnetization $\vec{m}_{\vec{i}}$ read
\begin{align}
  \left( \pdiff{\vec{m}}{x_i} \right)_{\vec{j}} &=
  \frac{\vec{m}_{\vec{j} + \vec{e}_i} - \vec{m}_{\vec{j} - \vec{e}_i}}{2 \Delta x_i} \label{eq:fd_fd_first}\\
  \left( \frac{\partial^2 \vec{m}}{\partial x_i^2} \right)_{\vec{j}} &=
  \frac{\vec{m}_{\vec{j} + \vec{e}_i} - 2 \vec{m}_{\vec{j}}+ \vec{m}_{\vec{j} - \vec{e}_i}}{\Delta x_i^2}. \label{eq:fd_fd_second}
\end{align}
which results in the following discretization of the exchange field $\hex$
\begin{equation}
  \hex_{\vec{i}} =
  \frac{2 A}{\mu_0 \Ms} \Delta \vec{m}_{\vec{i}}
  \approx
  \frac{2 A}{\mu_0 \Ms}
  \sum_k \frac{
    \vec{m}_{\vec{i} + \vec{e}_k} - 2\vec{m}_{\vec{i}} + \vec{m}_{\vec{i} - \vec{e}_k}
  }{
    \Delta x_k^2
  }.
\end{equation}
While higher-order finite-differences might lead to better convergence properties for some applications, they also increase the computational effort by involving more neighboring cells in the computation.
Since second-order approximation is usually considered sufficiently accurate, most finite-difference codes stick with this approximation.

\begin{figure}
  \centering
  \includegraphics{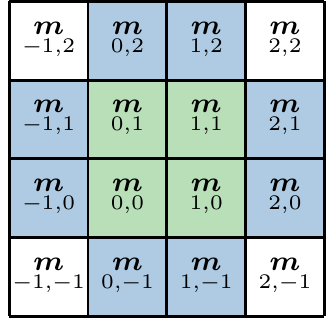}
  \caption{
    Visualization of the virtual cells, marked with a blue background color, introduced for the finite-difference implementation of boundary conditions.
  }
  \label{fig:fd_fd_boundary}
\end{figure}
The boundary condition $\vec{B} = 0$, as derived in Sec.~\ref{sec:static}, is usually implemented by introducing virtual cells surrounding the magnetic region $\Omega_m$, see Fig.~\ref{fig:fd_fd_boundary}, and computing the magnetization values $\vec{m}_{\vec{i}}$ in these cells accordingly.
In case of the exchange interaction this boundary condition is defined by $\pdiff{\vec{m}}{\vec{n}} = 0$ as derived in Sec.~\ref{sec:energetics_static_exchange}.
For the boundary at $x_1 = 0$ this leads to the following equation
\begin{equation}
  \left( \pdiff{\vec{m}}{x_i} \right)_{(0, j, k)}
  =
  \frac{\vec{m}_{(1,j,k)} - \vec{m}_{(-1,j,k)}}{2 \Delta x_i}
  =
  0
\end{equation}
which needs to hold for all $0 \leq j \leq \n_2 - 1$ and $0 \leq k \leq \n_3 - 1$ and determines the values of the  virtual cells $\vec{m}_{(-1,j,k)}$ as $\vec{m}_{(-1,j,k)} = \vec{m}_{(1,j,k)}$.
After computing the values of the virtual cells, the derivatives required for the distinct field computations can be evaluated according to \eqref{eq:fd_fd_first} and \eqref{eq:fd_fd_second} without special treatment for the boundary cells.
Note, that the boundary conditions, and thus the computation of the virtual cells, differ depending on the choice of field contributions, see Sec.~\ref{sec:static_multiple_bc}.

\subsubsection{Spintronics}
The implementation of the spin-torque effects of Slonczewski as well as Zhang and Li as introduced in Sec.~\ref{sec:spin_slon} and \ref{sec:spin_zhang} are straightforward as the torque contributions depend on the local magnetization and their derivatives only.
In order to solve the spin-diffusion model, a second-order partial differential equation in space has to be solved in order to compute the spin-accumulation $\vec{s}$ and the electric potential $u$.
The discretization of equations \eqref{eq:spin_diff_js_source} and \eqref{eq:spin_diff_je_source} with finite differences is straightforward.
However, care has to be taken in order to properly account for discontinuous material parameters when dealing with multilayer structures.
For instance, the spin-current $\js$, as defined by \eqref{eq:spin_diff_js}, may be discontinuous across material interfaces due to discontinuities of the material constants $C_0$ and $D_0$.
Hence, the discretization of the divergence in \eqref{eq:spin_diff_js_source} must be carefully chosen to account for these jumps in a distributional sense.
A possible finite-difference discretization of the time dependent spin-diffusion model with prescribed electric current \eqref{eq:spin_diff_dts} and \eqref{eq:spin_diff_js_from_je} is presented by Garcia-Cervera et al.\cite{garcia2007spin}.

\subsubsection{Existing software packages}
Various software packages implementing the finite-difference method with FFT accelerated demagnetization-field computation have been developed.
Probably the most popular open-source finite-difference micromagnetic software is OOMMF \cite{donahue1999oommf}.
OOMMF is a multi platform code running on central processing units (CPUs).
Other finite-difference CPU codes include the open-source software Fidimag \cite{fidimag} and the commercial package MicroMagus \cite{micromagus}.
A very simple CPU implementation of the finite-difference algorithms with the Python library NumPy is presented in \cite{abert2015full}.

The recent advent of general-purpose graphics-processing units (GPGPUs) allowed for the significant acceleration of scientific software.
A popular open-source package for finite-difference micromagnetics on GPGPUs is MuMax3 \cite{leliaert2018fast,vansteenkiste2014design}.
Other GPGPU codes include magnum.fd \cite{magnum.fd} and the recently developed GPGPU extension to OOMMF \cite{fu2016finite}.

\subsection{The finite-element method}
The finite-element method is a powerful numerical tool for the solution of partial differential equations.
In contrast to the finite-difference method where the differential operators are discretized directly, in finite-elements the original problem is transformed into a variational problem before discretization.
Consider Poisson's equation
\begin{equation}
  - \Delta u = f
  \quad\text{in}\quad
  \Omega.
\end{equation}
Multiplying both sides with a test function $v$ from a suitable function space $V$ and integrating the original problem turns the original problem into a variational problem.
The solution $u \in V$ is required to satisfy
\begin{equation}
  - \int_\Omega \Delta u \, v \dx = \int_\Omega f \, v \dx
  \quad\forall\quad v \in V.
\end{equation}
Applying integration by parts yields
\begin{equation}
  \int_\Omega \vnabla u \cdot \vnabla v \dx =
  \int_\Omega f \, v \dx
  + \int_{\partial \Omega} u_\text{N} \, v \ds
  \quad\forall\quad v \in V
  \label{eq:fem_poisson}
\end{equation}
where $\vnabla u \cdot \vec{n}$ in the boundary integral was replaced by $u_\text{N}$ in order to implement Neumann boundary conditions $\vnabla u \cdot \vec{n} = u_\text{N}$ for $x \in \partial \Omega$.
This formulation is referred to as weak form of the original problem as it weakens the requirements on the solution $u$.
While the original equation requires the solution to be twice differentiable, the weak form requires $u$ to be only once differentiable almost everywhere.
Dirichlet conditions are applied on $\Gamma_\text{D} \subset \partial \Omega$ by restricting the solution space to functions satisfying the Dirichlet condition 
\begin{equation}
  u(\vec{x}) = u_\text{D}
  \;\forall\;
  \vec{x} \in \Gamma_\text{D} \subset \partial \Omega
\end{equation}
and additionally restricting the test functions to $V_\text{D}$ given by
\begin{equation}
  V_\text{D} = \{v \in V : v(\vec{x}) = 0 \;\forall\; \vec{x} \in \Gamma_\text{D} \}
\end{equation}
so that the solution $u$ is not tested on the Dirichlet boundary $\Gamma_\text{D}$.
By appropriate choice of the function space $V$, the variational problem \eqref{eq:fem_poisson}, can be shown to have a unique solution \cite{braess2007finite}.

\begin{figure}
  \centering
 
  \includegraphics{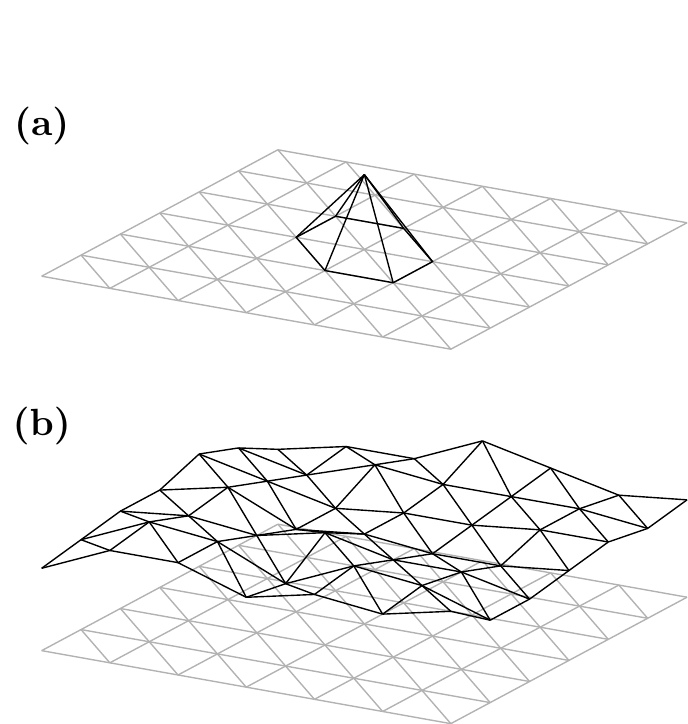}
  \caption{
    Piecewise affine, globally continuous basis functions for the finite-element method in two dimensions.
    (a) Single basis function.
    (b) Possible discrete function obtained by superposition of basis functions.
  }
  \label{fig:demag_fem}
\end{figure}
Discretization of the continuous problem \eqref{eq:fem_poisson} is achieved by choice of a discrete function space $V_h \subset V$.
While the finite-element method is very general concerning the choice of discrete function spaces \cite{braess2007finite}, the most common choice is that of piecewise affine, globally continuous functions.
A suitable function basis is constructed using a tetrahedral mesh as depicted in Fig.~\ref{fig:discrete_fd_vs_fe}\,(b).
Each mesh node $\vec{x}_i$ is associated with a basis function $\phi_i$ with node values
\begin{equation}
  \phi_i(\vec{x}_j) = \delta_{ij}
\end{equation}
which is affine within each cell, see Fig.~\ref{fig:demag_fem}.
In order to discretize the weak formulation \eqref{eq:fem_poisson}, both the solution function $u$ and the test function $v$ are expressed in terms of the basis functions
\begin{align}
  u_h &= \sum_i u_i \phi_i\\
  v_h &= \sum_i v_i \phi_i.
\end{align}
Inserting into \eqref{eq:fem_poisson} and neglecting the boundary term for the sake of simplicity yields
\begin{equation}
  \sum_{i} u_i \int_\Omega \vnabla \phi_i \cdot \vnabla \phi_j \dx =
  \int_\Omega f \, \phi_j \dx
  \quad\forall\quad j \in [1,\n].
  \label{eq:fem_poisson_discrete}
\end{equation}
Instead of testing with all possible test functions $v_h$, the test functions are reduced to the individual basis functions $\phi_j$.
Since both, the left-hand side and the right-hand side of \eqref{eq:fem_poisson} are linear in the test function $v$, the equality \eqref{eq:fem_poisson_discrete} holds also true for any test function $v_h$.
The discretized solution $u_i$ is given by a linear system of equations
\begin{equation}
  \sum_i A_{ij} u_i = b_j
\end{equation}
with 
\begin{align}
  A_{ij} &= \int_\Omega \vnabla \phi_i \cdot \vnabla \phi_j \dx\\
  b_j    &= \int_\Omega f \, \phi_j \dx
\end{align}
where the matrix $A_{ij}$, which is referred to as stiffness matrix, depends only on the geometry of the mesh.
Furthermore $A_{ij}$ is sparsely populated due to the choice of basis functions that result in nonzero contributions only for neighboring nodes.
The sparsity is an important aspect from a computational point of view, since it reduces the storage requirements from $\mathcal{O}(\n^2)$ to $\mathcal{O}(\n)$.
Furthermore this property can be exploited by the use of iterative methods for the solution of linear systems.
These methods avoid the direct inversion of the matrix and require only the computation of matrix--vector multiplications which leads to a superior scaling compared to direct methods \cite{saad2003iterative}.

The procedure of turning a partial differential equation into a variational problem by multiplication with test functions, which is referred to as Galerkin method, can be applied to a large variety of problems.

\subsubsection{Demagnetization field}\label{sec:fem_demag}
\begin{figure}
  \centering
  \includegraphics{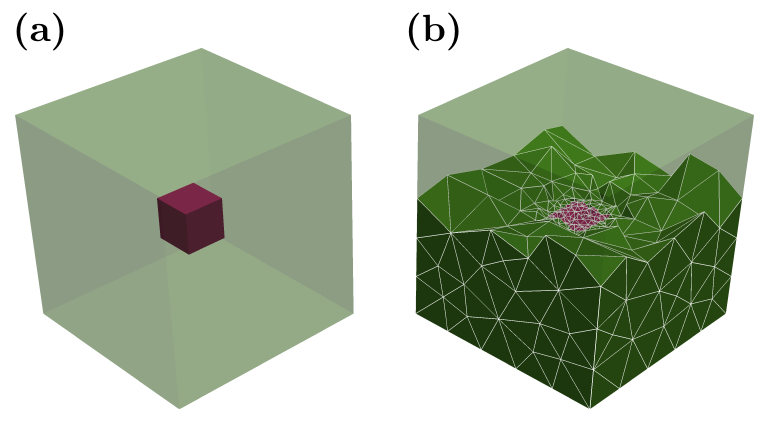}
  \caption{
    Finite-element mesh for the computation of the demagnetization field with the truncation approach.
    (a) The magnetic region is marked red and the external region, marked green, is chosen to be approximately five times larger than the magnetic region in each spatial dimension.
    (b) The magnetic region, as the region of interest, is discretized with a small mesh size.
    In the external region the mesh is coarsened towards the outer boundary in order to reduce the overall number of mesh nodes.
  }
  \label{fig:fem_demag_truncation}
\end{figure}
As shown in Sec.~\ref{sec:energetics_demag}, the demagnetization-field potential is given by Poisson's equation \eqref{eq:energetics_demag_poisson}.
However, in contrast to the example given in the preceding section, the demagnetization-field problem is subject to open boundary conditions \eqref{eq:energetics_demag_open_boundary}.
Hence, Poisson's equation has to be solved in the complete space $\mathbb{R}^3$ which is not trivially possible with the finite-element method that applies only to finite regions.

A simple approach to approximate the required boundary conditions with finite elements is the so-called truncation approach.
In order to compute the demagnetization field $\hdemag$ of a finite magnetization region $\Omega_m$, the outer region $\mathbb{R}^3 \setminus \Omega_m$ is approximated with a finite external region $\Omega_e$.
Since the magnetic scalar potential is known to decay outside of the magnetic region, the open boundary conditions can be approximated by applying homogeneous Dirichlet or Neumann conditions to the outer boundary of the problem domain $\partial \Omega$ with $\Omega = \Omega_m \cup \Omega_e$.
Starting from the problem definition \eqref{eq:energetics_demag_poisson_raw}, the magnetic scalar potential is given by the weak formulation
\begin{equation}
  \int_\Omega \vnabla \cdot (\vnabla u - \Ms \vec{m}) \, v \dx = 0.
\end{equation}
Performing integration by parts and restricting both the solution function $u$ and the test function $v$ to the function space $V_0$ with
\begin{equation}
  V_0 = \{ v \in V : v(\vec{x}) = 0 \;\forall\; \vec{x} \in \partial \Omega \}
\end{equation}
yields
\begin{equation}
  \int_\Omega \vnabla u \cdot \vnabla v \dx = 
  \int_\Omega \Ms \vec{m} \cdot \vnabla v \dx
  \quad\forall\quad
  v \in V_0
  \label{eq:fem_demag_weak}
\end{equation}
which solves the homogeneous Dirichlet problem in the domain $\Omega = \Omega_m \cup \Omega_e$.
Note, that the right-hand side of \eqref{eq:fem_demag_weak} is integrated by parts in order to loosen the continuity requirements on the magnetization $\vec{m}$.
This is essential in order to accurately account for the discontinuity of the magnetization at the boundary of the magnetization region $\partial \Omega_m$.
If the magnetization $\vec{m}$ is discretized with the usual piecewise affine, globally continuous functions, this discontinuity can be described by restricting the integration domain of the right-hand side of \eqref{eq:fem_demag_weak} to the magnetic domain $\Omega_m$.

The accuracy of the truncation solution is significantly influenced by the size of the external region $\Omega_e$. 
However, a larger size of the external region usually increases the number of mesh nodes and thus the computational effort.
Choosing $\Omega_e$ five times the size of $\Omega_m$ has been found to be a reasonable trade-off between accuracy and computational complexity \cite{chen1997review}.
In order to further reduce the computational costs, the mesh in the external region can be gradually coarsened to the outside, see Fig.~\ref{fig:fem_demag_truncation}.
This leads to a reduction of degrees of freedom and has no significant impact on the accuracy of the solution, since the scalar potential is known to smoothly decay to the outside.

\begin{figure}
  \centering
  \includegraphics{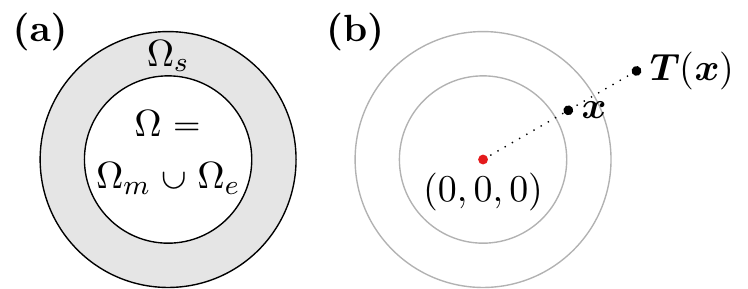}
  \caption{
    Visualization of the shell-transformation method for the demagnetization-field computation.
    (a) Definition of regions, the shell domain $\Omega_s$ that is subject to the transformation surrounds the inner domain $\Omega$ which is defined as the union of the magnetic domain $\Omega_m$ and the external domain $\Omega_e$.
    (b) The transformation in a spherical shell is performed in a radial fashion.
  }
  \label{fig:fem_demag_transform}
\end{figure}
A related approach to the truncation method is the so-called shell-transformation method.
Instead of describing the exterior space $\mathbb{R}^3 \setminus \Omega_m$ with a relatively large region $\Omega_e$, this method uses a rather small shell region $\Omega_s$ and employs a transformation $\vec{T}$ that maps the shell region onto the complete exterior space
\begin{equation}
  \vec{T}: \Omega_s \rightarrow \mathbb{R}^3 \setminus \Omega.
  \label{eq:fem_demag_trans_map}
\end{equation}
Specifically, the transformation $\vec{T}$ maps the inner boundary of the shell region onto itself and the outer outer shell boundary onto infinity
\begin{align}
  \vec{T}(\vec{x}) &= \vec{x} \;\text{for}\; \vec{x} \in \partial \Omega\\
  \vec{T}(\vec{x}) &\rightarrow \infty \;\text{for}\; \vec{x} \in \partial (\Omega \cup \Omega_s).
\end{align}
For a spherical shell, the transformation is illustrated in Fig.~\ref{fig:fem_demag_transform}.
By substitution of variables, integrals over the exterior space $\mathbb{R}^3 \setminus \Omega$ can be expressed by integrals over the shell region $\Omega_s$.
For the left-hand side of the weak formulation \eqref{eq:fem_demag_weak} this substitution yields
\begin{equation}
  \int_{\mathbb{R}^3 \setminus \Omega} \vnabla u \cdot \vnabla v \dx =
  \int_{\Omega_s} (\vnabla u)^T \mat{g} \vnabla v \dx
\end{equation}
where $\mat{g}$ is the so-called metric tensor defined by
\begin{equation}
  \mat{g} = (\mat{J}^{-1})^T \left|\det \mat{J}\right| \mat{J}^{-1}.
  \label{eq:fem_demag_trans_metric_tensor}
\end{equation}
with $J = \vnabla \vec{T}$ being the Jacobian of the transformation $\vec{T}$.
This leads to the weak formulation
\begin{equation}
  \int_\Omega \vnabla u \cdot \vnabla v \dx + 
  \int_{\Omega_s} (\vnabla u)^T \mat{g} \vnabla v \dx =
  \int_{\Omega_m} \Ms \vec{m} \cdot \vnabla v \dx
  \label{eq:fem_demag_trans_weak}
\end{equation}
for the solution of the scalar potential $u$.
Various shell geometries and transformations have been proposed to solve \eqref{eq:fem_demag_trans_weak}  \cite{imhoff1990original,brunotte1992finite,henrotte1999finite,abert2013numerical}.
Like for the truncation method, application of the shell-transformation method results in sparse linear systems. 
The advantage of the transformation method over the truncation approach is a finite representation of the infinite exterior region.
However, the metric tensor $\mat{g}$ is by definition singular at the outer shell boundary which makes the computation of the discrete entries for the system matrix difficult.
Furthermore, this singularity leads to a bad matrix condition, which has a negative effect on the numerical solution of the linear system.

In order to further reduce the degrees of freedom, it is desirable to consider only the magnetic region $\Omega_m$ for spatial discretization.
This can be achieved by a hybrid finite-element--boundary-element method proposed by Fredkin and Koehler \cite{fredkin1990hybrid}.
From \eqref{eq:energetics_demag_poisson_raw} and the divergence theorem the following jump conditions for the scalar potential at the boundary of the magnetic region $\partial \Omega_m$ apply
\begin{align}
  u^- - u^+ &= 0  \label{eq:fem_demag_bem_cont1} \\
  (\vnabla u^- - \vnabla u^+) \cdot \vec{n} &= \Ms \vec{m} \cdot \vec{n} \label{eq:fem_demag_bem_cont2}
\end{align}
where $u^-$ denotes the value in the magnetic region and $u^+$ denotes the corresponding value outside.
Consider the following splitting of the potential $u$
\begin{equation}
  u = u_1 + u_2
\end{equation}
where $u_1$ is solved by
\begin{align}
  \Delta u_1 &= - \vnabla \cdot (\Ms \vec{m}) \\
  \pdiff{u_1}{\vec{n}} &= \Ms \vec{n} \cdot \vec{m} \text{ on } \partial \Omega_m
\end{align}
in the magnetic region $\Omega_m$ and zero in the exterior region $\mathbb{R}^3 \setminus \Omega_m$.
While $u_1$ satisfies the jump condition \eqref{eq:fem_demag_bem_cont2}, it violates the continuity condition \eqref{eq:fem_demag_bem_cont1}.
Thus $u_2$ has to be chosen to fix \eqref{eq:fem_demag_bem_cont1} while preserving \eqref{eq:fem_demag_bem_cont2} and being a solution to the Laplace equation $\Delta u_2 = 0$, i.e.
\begin{align}
  u^-_2 - u^+_2 &= u^-_1 \\
  \pdiff{u^-_2}{\vec{n}} - \pdiff{u^+_2}{\vec{n}} &= 0 \\
  \Delta u_2 &= 0 \;\text{in}\; \mathbb{R}^3.
\end{align}
These requirements are fulfilled by the double-layer potential defined as
\begin{equation}
  u_2 = \int_{\partial \Omega} u_1 \pdiff{ }{\vec{n}} \frac{1}{|\vec{x} - \vec{x}'|} \dx.
  \label{eq:fem_demag_double_layer}
\end{equation}
While this integral expression can be used to evaluate $u_2$ in $\Omega_m$, this procedure has a high computational complexity of $\mathcal{O}(\n^2)$.
Hence, \eqref{eq:fem_demag_double_layer} is only evaluated at the boundary $\partial \Omega_m$ using the boundary-element method.
The result of this computation is then used as Dirichlet boundary condition for a Laplace problem that is solved with finite-elements.
This means that a computation of the scalar potential $u$ requires the finite-element solution of Poisson's equation with Neumann boundary conditions, followed by a boundary-element evaluation of the double-layer potential and the finite-element solution of a Laplace problem with Dirichlet conditions.

Compared to the pure finite-element presented in this section, the advantage of this hybrid method is that only the magnetic region $\Omega_m$ has to be discretized.
This is even true if $\Omega_m$ consists of various separated domains.
The disadvantage, however, is the loss of sparsity.
The boundary-element operator for the double-layer computation is a dense $\n_\text{B} \times \n_\text{B}$ matrix with $\n_\text{B}$ being the number of boundary nodes.
A common procedure to reduce the storage requirements of this matrix is the application of hierarchical matrices \cite{hackbusch2015hierarchical,popovic2005applications} which reduces the storage requirements as well as the computational complexity for the potential evaluation from $\mathcal{O}(\n_\text{B}^2)$ to $\mathcal{O}(\n_\text{B} \log \n_\text{B})$.
An alternative hybrid method using the single-layer potential was proposed by Garc\'ia-Cervera and Roma \cite{garcia2006adaptive}.

\begin{figure}
  \centering
  \includegraphics{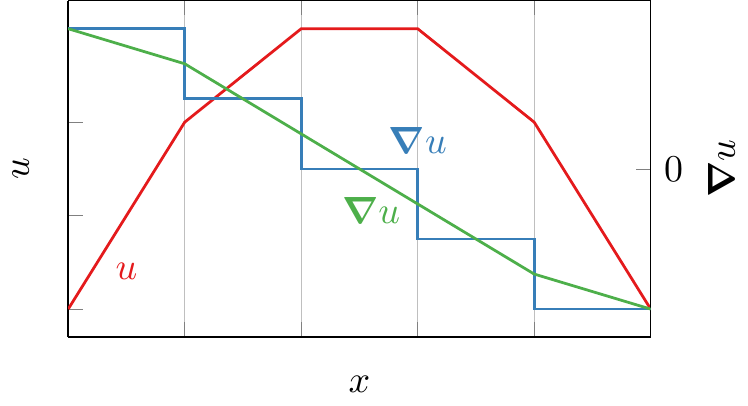}
  \caption{
    Gradient computation of a piecewise affine, globally continuous function $u$.
    The analytical calculation yields a piecewise constant gradient marked as blue.
    Projection onto the function space of piecewise affine, globally continuous functions yields the green curve.
  }
  \label{fig:fem_demag_gradient}
\end{figure}
The methods presented in this section are devoted to the calculation of the magnetic scalar potential $u$ rather then the demagnetization field $\hdemag$ which is defined as the negative gradient of the potential $\hdemag = -\vnabla u$.
Since the discrete solution of $u$ is a piecewise affine function, the gradient of $u$ can be easily computed.
However, the gradient of a piecewise affine function is a piecewise constant, discontinuous function, see Fig.~\ref{fig:fem_demag_gradient}.
Often it is desirable to compute the demagnetization field with the same discretization as the magnetization.
This is required for nodewise operations such as the nodewise evaluation of the right-hand side of the Landau-Lifshitz-Gilbert equation \eqref{eq:llg_llg_gilbert}.
The approximation of the demagnetization field as a piecewise affine, globally continuous function can be achieved by solving the weak form
\begin{equation}
  \int_{\Omega_m} \hdemag \cdot \vec{v} \dx =
  - \int_{\Omega_m} \vnabla u \cdot \vec{v} \dx
  \quad\forall\quad
  \vec{v} \in V
  \label{eq:fem_demag_projection}
\end{equation}
with $\hdemag \in V$.
This procedure is referred to as projection.
Discretization of \eqref{eq:fem_demag_projection} with the usual basis functions yields the linear system
\begin{equation}
  \sum_i H_i \int_{\Omega_m} \vec{\phi}_i \cdot \vec{\phi}_j \dx =
  - \sum_i u_i \int_{\Omega_m} \vnabla \phi_i \cdot \vec{\phi}_j \dx.
\end{equation}
The system matrix resulting from this weak form is called mass matrix $\mat{M}$
\begin{equation}
  M_{ij} = \int_{\Omega_m} \vec{\phi}_i \cdot \vec{\phi}_j \dx.
\end{equation}
Like the stiffness matrix arising from discretization of the Laplacian, the mass matrix is sparse since only basis functions of neighboring mesh nodes give nonzero contributions.
In order to avoid solving another linear system for the computation of the demagnetization field, the mass matrix can be approximated by a diagonal matrix $\mat{M}^\ast$ given by
\begin{equation}
  M^\ast_{ij}
  = \delta_{ij} \int_{\Omega_m} \vec{\phi}_i \cdot \vec{1} \dx
\end{equation}
with $\vec{1} = (1,1,1)$.
This approximation is referred to as mass lumping since all off-diagonal mass entries are lumped in the diagonal
\begin{equation}
  M^\ast_{ii} = \sum_j M_{ij}.
\end{equation}
This approximation preserves the integral of the solution variable $\hdemag$ which justifies its application
\begin{equation}
  \int_{\Omega_m} \hdemag_h \dx
  = \sum_{ij} H_i \int_{\Omega_m} \vec{\phi}_i \cdot \vec{\phi}_j \dx
  = \sum_{ij} \delta_{ij} H_i \int_{\Omega_m} \vec{\phi}_i \cdot \vec{1} \dx.
\end{equation}
Since the lumped mass matrix is diagonal, its inverse is trivially given by the reciprocal diagonal entries.
Hence, instead of solving a linear system, the projection can be expressed as a single sparse matrix--vector multiplication
\begin{align}
  \hdemags_i &= \sum_j A_{ij} u_j\\
  A_{ij} &=
  - \left[ \int_{\Omega_m} \vec{\phi}_i \cdot \vec{1} \dx \right]^{-1}
  \int_{\Omega_m} \vnabla \phi_j \cdot \vec{\phi}_i \dx.
\end{align}
The illustrative description of mass lumping is that of an weighted average with the weight being the cell sizes as described in \cite{schrefl2007numerical}.
The result of a gradient projection is depicted in Fig.~\ref{fig:fem_demag_gradient} along with the original function.

\subsubsection{Local field contributions}\label{sec:fem_other}
Local field contributions are usually computed with a similar procedure as applied for the gradient computation in Sec.~\ref{sec:fem_demag}
For the exchange field defined by \eqref{eq:static_exchange_field} this procedure yields
\begin{align}
  \int_{\Omega_m} \hex \cdot \vec{v} \dx
  &= \int_{\Omega_m} \frac{2}{\mu_0 \Ms} \vnabla \cdot (A \vnabla \vec{m}) \cdot \vec{v} \dx \label{eq:fem_other_exchange_laplace}\\
  &= - \int_{\Omega_m} A \vnabla \vec{m} : \vnabla \left( \frac{2}{\mu_0 \Ms} \vec{v} \right) \dx
     + \int_{\partial \Omega_m} \frac{2A}{\mu_0 \Ms} \pdiff{\vec{m}}{\vec{n}} \cdot \vec{v} \ds
\end{align}
where integration by parts was applied in order to avoid second spatial derivatives.
Due to the boundary condition \eqref{eq:static_exchange_bc}, the boundary integral on the right-hand side vanishes which results in
\begin{equation}
  \int_{\Omega_m} \hex \cdot \vec{v} \dx
  = - \int_{\Omega_m} 2A \vnabla \vec{m} : \vnabla \left( \frac{1}{\mu_0 \Ms} \vec{v} \right) \dx.
\end{equation}
While this weak form poses no further difficulties for single-phase magnets with a constant saturation magnetization $\Ms$, the gradient on the test function diverges at material interfaces with discontinuous $\Ms$.
In order to avoid this problem, the original problem \eqref{eq:fem_other_exchange_laplace} can be multiplied with $- \mu_0 \Ms$ before integrating by parts resulting in
\begin{equation}
  - \int_{\Omega_m} \mu_0 \Ms \hex \cdot \vec{v} \dx
  = 2 \int_{\Omega_m} A \vnabla \vec{m} : \vnabla \vec{v} \dx
  = \delta \Eex(\vec{m}, \vec{v}).
\end{equation}
with $\delta \Eex$ being the exchange differential defined in \eqref{eq:static_exchange_variation}.
This method can be used to compute any local field contribution introduced in Sec.~\ref{sec:energetics}.
In order to avoid the solution of a linear system for the retrieval of field contributions, the mass-lumping procedure introduced in Sec.~\ref{sec:fem_demag} can be applied.
For energy contributions quadratic in the magnetization $\vec{m}$, such as the exchange field, the field computation can be reduced to a single sparse matrix--vector multiplication given by
\begin{align}
  H_i &= \sum_j A_{ij} m_j \label{eq:fem_other_matrix_vector}\\
  A_{ij} &=
  - \left[ \int_{\Omega_m} \mu_0 \Ms \vec{\phi}_i \cdot \vec{1} \dx \right]^{-1}
  \delta E(\vec{\phi_j}, \vec{\phi_i}). \label{eq:fem_other_matrix}
\end{align}
Note, that this method accounts for the boundary conditions introduced by the exchange and antisymmetric exchange field in a generic fashion since it considers the differential of the energy $\delta E$ rather than the functional derivative $\vdiff{E}{\vec{m}}$.
Since the boundary conditions are embedded in the differential, they do not have to be explicitly applied.

While some effective-field contributions like the uniaxial-anisotropy field \eqref{eq:energetics_aniso_uniaxial_field} can also be computed nodewise for single-phase magnets, the nodewise computation fails for discontinuous material parameters.
Due to the integral formulation, the calculation according to \eqref{eq:fem_other_matrix_vector} and \eqref{eq:fem_other_matrix} also works in these cases.

\subsubsection{Interface contributions}\label{sec:fem_interface}
Some energy contributions introduced in Sec.~\ref{sec:energetics} are interface effects rather than bulk effects.
This means that the energy connected to these effects is the result of an interface integral.
These interface contributions usually enter the micromagnetic formalism through boundary conditions, see e.g. Sec.~\ref{sec:static_aniso}.
However, in the framework of dynamical micromagnetics, it is often desirable to be able to express all energy contribution in terms of an effective field instead of a boundary condition in order to add up the different contributions for their use in the LLG.

In order to compute a discretized effective field that accounts for interface energy contributions, we require the energy generated by the effective field to equal the interface energy contribution.
Consider the interface exchange energy given by \eqref{eq:energetics_interlayer_exchange_energy}.
Since this energy contribution is quadratic in $\vec{m}$, the equality of energies reads
\begin{equation}
  E = - \frac{1}{2} \int_{\Omega_m} \mu_0 \Ms \vec{m} \cdot \heff \dx
  = - \int_\Gamma \vec{m} \cdot \vec{p}(\vec{m}) \ds
\end{equation}
with $\vec{p} = A \vec{m}[P(\vec{x})]$.
Discretization of the field $\heff$ yields
\begin{equation}
  \sum_i - \frac{1}{2} H_i \int_{\Omega_m} \mu_0 \Ms \vec{m} \cdot \vec{\phi}_i \dx
  = - \int_\Gamma \vec{m} \cdot \vec{p} \ds
\end{equation}
which has to hold for arbitrary magnetizations $\vec{m}$.
Hence, the magnetization can be replaced with basis functions and the equality
\begin{equation}
  \sum_i - \frac{1}{2} H_i \int_{\Omega_m} \mu_0 \Ms \vec{\phi}_j \cdot \vec{\phi}_i \dx
  = - \int_\Gamma  \vec{\phi}_j \cdot \vec{p} \ds
\end{equation}
has to hold for any $\vec{\phi}_j$.
Application of mass lumping yields the system
\begin{align}
  H_i &= \sum_j A_{ij} m_j\\
  A_{ij} &=
  2 \left[ \int_{\Omega_m} \mu_0 \Ms \vec{\phi}_i \cdot \vec{1} \dx \right]^{-1}
  \int_{\Gamma} \vec{\phi}_j \cdot \vec{p}(\vec{\phi}_j) \ds\\
   &=
 - \left[ \int_{\Omega_m} \mu_0 \Ms \vec{\phi}_i \cdot \vec{1} \dx \right]^{-1}
  \delta E(\vec{\phi_j}, \vec{\phi_i})\\
\end{align}
which has the exact same form as any local bulk-field contribution as shown in Sec.~\ref{sec:fem_other}.

\subsubsection{Spintronics}
As for the finite-difference method, the simplified spin-torque models by Slonczewski as well as Zhang and Li can be incorporated in finite-element micromagnetics along the lines of the local field contributions shown in Sec.~\ref{sec:fem_other}.
In order to account for the Slonczewski spin-torque as an interface effect, the consideration of Sec.~\ref{sec:fem_interface} can be applied.

A more challenging task is the discretization of the spin-diffusion model introduced in Sec.~\ref{sec:spin_diff}.
The spin-diffusion model requires the computation of the spin accumulation $\vec{s}$ which is the solution to a partial differential equation.
This equation is turned into a weak form using the Galerkin method.
In order to solve the equilibrium spin accumulation $\vec{s}$ for a given magnetization $\vec{m}$ and a prescribed charge current $\je$ as defined by \eqref{eq:spin_diff_js_from_je} and \eqref{eq:spin_diff_js_source}, the following weak form applies
\begin{multline}
    2 \int_\Omega D_0 \vnabla \vec{s} : \vnabla \vec{v} \dx
  + 2 \int_{\Omega_m} D_0 \beta \beta' \vec{m} \otimes \left[(\vnabla \vec{s})^T \vec{m}\right] : \vnabla \vec{v} \dx\\
  + \int_\Omega \frac{\vec{s}}{\tausf} \cdot \vec{v} \dx 
  + \int_{\Omega_m} J \frac{\vec{s} \times \vec{m}}{\hbar} \cdot \vec{v} \dx\\
  =
  \int_{\Omega_m} \frac{\beta \mub}{e} \vec{m} \otimes \je : \vnabla \vec{v} \dx
  - \int_{\partial \Omega \cap \partial \Omega_m} \frac{\beta \mub}{e} (\je \cdot \vec{n}) (\vec{m} \cdot \vec{v}) \ds.
  \label{eq:fem_spin_prescribed_je}
\end{multline}
Note, that integration by parts was not only applied in order to eliminate second derivatives of the spin accumulation $\vec{s}$, but also in order to eliminate any derivative on the magnetization $\vec{m}$ or material parameters since these variables may have discontinuities in the problem domain $\Omega$.
While material parameters may have arbitrary discontinuities at material interfaces, the magnetization is continuous within the magnetic domain $\Omega_m$.
However, since the magnetization vanishes in the nonmagnetic domain $\Omega \setminus \Omega_m$, it is by definition discontinuous across magnetic--nonmagnetic interfaces.
For the discrete formulation, these properties are take into account by appropriate choices of function spaces and integration domains.
Namely, all material parameters are discretized with piecewise constant functions.
Since the magnetization is continuous within the magnetic region, it is discretized with the usual piecewise affine, globally continuous functions.
In order to account for the discontinuity at magnetic--nonmagnetic interfaces, any integral including the magnetization $\vec{m}$ is restricted to the magnetic domain $\Omega_m$ which is equivalent to a rapid drop of the magnetization to zero outside the magnetic domain.

The boundary integrals arising from partial integration of $\Delta \vec{s}$ vanish due to the homogeneous Neumann boundary condition on $\vec{s}$, see Sec.~\ref{sec:spin_diff_bc}.
Since the spin accumulation $\vec{s}$ is assumed to be continuous even across material interfaces, both $\vec{s}$ as well as the test functions are discretized with componentwise piecewise affine, globally continuous functions $\vec{s}, \vec{v} \in \vec{V}$.

For the solution of the self-consistent spin-diffusion model given by the current definitions \eqref{eq:spin_diff_je}, \eqref{eq:spin_diff_js} and their respective source equations \eqref{eq:spin_diff_je_source}, \eqref{eq:spin_diff_je_source}, the function space for both the solution and the test functions has to be extended.
The solution of the self-consistent model comprises both the spin accumulation $\vec{s}$ and the electric potential $u$.
Both the components of $\vec{s}$ and the scalar field $u$ are discretized with piecewise affine, globally continuous functions $\{\vec{s}, u\} \in \vec{V} \times V$.
Applying the Galerkin method and performing integration by parts in order to avoid diverging derivatives, yields two coupled weak formulations.
The weak formulation of \eqref{eq:spin_diff_je} and \eqref{eq:spin_diff_je_source} reads
\begin{equation}
  \int_\Omega 2 C_0 \vnabla u \cdot \vnabla v \dx
  - \int_{\Omega_m} 2 \beta' D_0 \frac{e}{\mub} (\vnabla\vec{s})^T \vec{m}
  \cdot \vnabla v \dx
  =
  - \int_{\Gamma_\text{N}} (\je^0 \cdot \vec{n}) v \ds
  \quad\forall\quad
  v \in V
  \label{eq:fem_spin_self_consistent_u}
\end{equation}
and the weak formulation of \eqref{eq:spin_diff_js} and \eqref{eq:spin_diff_js_source} is given by
\begin{multline}
  \int_{\Omega_m} 2 \beta C_0 \frac{\mub}{e} \vec{m} \otimes \vnabla u : \vnabla \vec{v} \dx
  - \int_{\partial \Omega_m \cap \Gamma_\text{D}} 2 \beta C_0 \frac{\mub}{e} (\vnabla u \cdot \vec{n})(\vec{m} \cdot \vec{v}) \ds
  \\
  - \int_\Omega 2 D_0 \vnabla \vec{s} : \vnabla\vec{v} \dx
  - \int_\Omega \frac{\vec{s} \cdot \vec{v}}{\tausf} \dx
  - \int_{\Omega_m} J \frac{(\vec{s} \times \vec{m}) \cdot \vec{v}}{\hbar} \dx \\
  = 
  - \int_{\partial \Omega_m \cap \Gamma_\text{N}} 
  \beta \frac{\mub}{e} (\je^0 \cdot \vec{n}) (\vec{m} \cdot \vec{v}) \ds.
  \quad\forall\quad
  \vec{v} \in \vec{V}.
  \label{eq:fem_spin_self_consistent_s}
\end{multline}
Here $\Gamma_\text{N} \in \partial \Omega$ denotes all external interfaces that act as contacts with prescribed charge-current inflow $\je^0 \cdot \vec{n}$ and $\Gamma_\text{D} \in \partial \Omega$ denotes contacts with prescribed electric potential $u_0$.
While the current inflow is implemented as natural Neumann boundary condition, the prescribed potential is implemented as Dirichlet boundary condition, see Sec.~\ref{sec:spin_diff_bc}.
In addition to the boundary conditions on the potential $u$, homogeneous Neumann conditions are applied to the spin accumulation $\vec{s}$.
Discretization of \eqref{eq:fem_spin_self_consistent_u} and \eqref{eq:fem_spin_self_consistent_s} yields a single sparse system of the size $4\n \times 4\n$ with $\n$ being the number of mesh nodes.

Incorporating spin-orbit interactions given by \eqref{eq:spin_diff_je_so} and \eqref{eq:spin_diff_js_so} or spin dephasing given by \eqref{eq:spin_diff_dephasing} is straightforward and can be done along the lines of the weak forms \eqref{eq:fem_spin_prescribed_je} -- \eqref{eq:fem_spin_self_consistent_s}.

\subsubsection{Existing software packages}
The implementation of finite-element solvers is a challenging task, since it involves the nontrivial generation of tetrahedral meshes, the numerical computation of integrals for the system-matrix assembly and the solution of large linear systems.
Various software packages and libraries have been developed in order to solve one or more of these tasks.
Popular open-source packages for the mesh generation are Gmsh \cite{geuzaine2009gmsh} and NetGen \cite{schoberl1997netgen}.
With ONELAB \cite{geuzaine2013onelab} and NGSolve \cite{schoberl2014c}, these mesh generators also act as full stack finite-element libraries.
Other open-source libraries for the formulation and solution of finite-element problems are Escript \cite{gross2005escript} and MFEM \cite{mfem}.
A very comprehensive and fast, yet easy to use finite-element library is FEniCS \cite{alnaes2015fenics} which, by default, uses PETSc \cite{balay2017petsc} as linear-algebra backend.
Libraries for the boundary-element method as required by the hybrid demagnetization-field method introduced in Sec.~\ref{sec:fem_demag} include BEM++ \cite{smigaj2015solving} and H2Lib \cite{h2lib}.
Both libraries are open source and provide routines for the assembly of boundary-element matrices and their compression with hierarchical matrices.

Besides these multi purpose libraries, a number of specialized micromagnetic finite-element packages are available.
The open-source library FinMag \cite{finmag} and the closed-source library magnum.fe \cite{abert2013magnum} are micromagnetic simulators based on FEniCS.
While FinMag concentrates on classical micromagnetics, magnum.fe also implements the spin-diffusion model \cite{abert2015three,ruggeri2016coupling,abert2016self}.
Another closed-source finite-element code that solves the micromagnetic equations coupled to the spin-diffusion model is FEELLGOOD \cite{alouges2012convergent,sturma2015geometry}.
Other finite-element codes include the open-source packages Magpar \cite{magpar} and NMag \cite{fischbacher2007systematic} as well as the closed-source package FEMME \cite{femme}.
The closed-source packages Tetramag \cite{kakay2010speedup} and Fastmag \cite{chang2011fastmag} make use of graphics processing units (GPUs) to speed up computations.

\subsection{Other spatial discretization methods}
While the finite-difference method and the finite-element are by far the most common discretization methods used in the micromagnetic community, other methods have been proposed to solve parts of the micromagnetic model.
Especially the computation of the demagnetization field is an ongoing matter of research.

A method that aims to combine the speed of the FFT based convolution with the flexibility of irregular meshes is the non-uniform FFT \cite{kritsikis2014beyond,exl2014non}.
Both the FFT accelerated convolution as well as the finite-element demagnetization-field computation exhibit at least a computational complexity of $\mathcal{O}(\n \log \n)$.
A well known method for the interaction of particle clouds which scales with $\n$ is the fast multipole method which has been shown to be also applicable to the continuous demagnetization-field problem \cite{apalkov2003fast,palmesi2017highly}.
Another class of methods that scales below $\mathcal{O}(\n \log \n)$ employs low-rank tensor approximations \cite{exl2012fast,exl2014fft}.

\subsection{Time integration}\label{sec:int}
Numerical integration of the Landau-Lifshitz-Gilbert equation (LLG) \eqref{eq:llg_llg_gilbert} poses several challenges on the applied method.
One of these challenges is the high stiffness that is introduced by the exchange interaction \cite{suess2002time}.
Another challenge is the micromagnetic unit-sphere constraint $|\vec{m}| = 1$ that is required to be preserved by a time-integration scheme.
Several methods, that address these difficulties, have been proposed in order to efficiently integrate the LLG.

Most numerical time-integration methods used in micromagnetics completely separate the spatial discretization from the time discretization.
For these methods, both the magnetization $\vec{m}$ and its time derivative $\pdiffs{\vec{m}}{t}$ are represented by vectors and the differential equation in space and time is transformed into $\n$ coupled ordinary differential equation
\begin{equation}
  \pdiff{m_i}{t} = f_i(t, \vec{m}).
\end{equation}
In order to evaluate the time derivative $f_i$, the effective-field contributions are computed according to the methods introduced in the preceding sections, and the right-hand side of the LLG \eqref{eq:llg_llg_gilbert} or \eqref{eq:llg_llg} is evaluated cellwise/nodewise.

Due to its first order in time, the LLG is an initial value problem.
We denote the initial value of the magnetization at time $t_0$ as
\begin{equation}
  \vec{m}(t_0) = \vec{m}^0.
\end{equation}
A discrete time-integration scheme approximates the magnetization dynamics as a series of magnetization snapshots at times $t_i$ that we denote by
\begin{equation}
  \vec{m}^i \approx \vec{m}(t_i) 
  \quad\text{with}\quad
  t_i = t_0 + i \Delta t.
\end{equation}
In the following, the most common time integrators in micromagnetics will be discussed in detail.
A more comprehensive overview of methods can be found in \cite{cimrak_2007}.

\subsubsection{Explicit Runge-Kutta methods}\label{sec:int_rk}
A reliable and efficient class of numerical integrators for initial value problems are the explicit Runge-Kutta methods.
According to the most general form of the Runge-Kutta method, the magnetization configuration $\vec{m}^i$ is obtained from a known magnetization configuration $\vec{m}^{i-1}$ by
\begin{align}
  \vec{m}^i &= \vec{m}^{i-1} + \Delta t \sum_j b_j \vec{k}_j \label{eq:int_rk}\\
  \vec{k}_j &= \pdiffs{\vec{m}}{t}\left[t_{i-1} + c_j \Delta t, \vec{m}^{i-1} + \Delta t \left(\sum_{k<j} a_{jk} \vec{k}_k\right)\right] \label{eq:int_rk_k}
\end{align}
where a set of coefficients $b_j$, $c_j$ and $a_{ij}$ defines a particular Runge-Kutta method.
The auxiliary results $\vec{k}_j$ are computed one after another.
Due to the restriction $j < k$ in the summation on the right-hand side, each $\vec{k}_j$ depends only on previously computed $\vec{k}_k$ and the previous magnetization $\vec{m}^{i-1}$ which makes this method explicit.
Explicit methods are usually computationally cheap, since they are linear in the solution variable $\vec{m}^{i}$.
However, explicit methods lack stability which might be a disadvantage especially in the case of stiff problems.

The simplest Runge-Kutta method is the explicit Euler method which is obtained by setting $b_1 = 1$ and $c_1 = 0$
\begin{equation}
  \vec{m}^i = \vec{m}^{i-1} + \Delta t \, \pdiffs{\vec{m}}{t}(t_{i-1}, \vec{m}^{i-1}).
  \label{eq:int_euler}
\end{equation}
This first-order method is not a good choice in terms of accuracy and efficiency.
However, it is well suited to investigate the preservation of the unit-sphere constraint.
Inserting the LLG \eqref{eq:llg_llg_gilbert} into \eqref{eq:int_euler} results in
\begin{equation}
  \vec{m}^i = \vec{m}^{i-1} + \Delta t \, \vec{m}^{i-1} \times \big[ - \gamma \heff(\vec{m}^{i-1}) + \alpha \pdiffs{\vec{m}}{t}(\vec{m}^{i-1}) \big].
\end{equation}
Multiplying with $\vec{m}^i + \vec{m}^{i-1}$ and reinserting \eqref{eq:int_euler} yields
\begin{equation}
  |\vec{m}^i|^2 = |\vec{m}^{i-1}|^2 + \Big| \Delta t \, \vec{m} \times \big[ - \gamma \heff(\vec{m}^{i-1}) + \alpha \pdiffs{\vec{m}}{t}(\vec{m}^{i-1}) \big] \Big|^2.
\end{equation}
and thus $|\vec{m}^i| \leq |\vec{m}^{i-1}|$ where the preservation of norm $|\vec{m}^1| = |\vec{m}^{i-1}|$ only holds for a vanishing right-hand side of the LLG $\pdiffs{\vec{m}}{t} = 0$.
In order to enforce the norm preservation, the magnetization is usually renormalized after each Runge-Kutta step, i.e. the magnetization $\vec{m}^i$ is replaced by ${\vec{m}^i}'$ given by
\begin{equation}
  {\vec{m}^i}' = \frac{\vec{m}^i}{|\vec{m}^i|}.
\end{equation}
This renormalization is also performed for higher order methods.
While these methods reduce the violation of the unit-sphere constraint, they do not guarantee its preservation.
Higher-order schemes are obtained by the choice of appropriate parameters $b_i$, $c_j$ and $a_{ij}$.
The classical Runge-Kutta method requires the computation of four auxiliary results $\vec{k}_j$ and is of fourth order.
Since the evaluation of each $\vec{k}_j$ comes at the price of an effective-field evaluation, the computation of a single integration step is more expensive for higher-order methods than for lower-order methods.
However, usually this disadvantage is more than compensated by the size of the time step which may by significantly larger for higher-order methods without compromising accuracy.

While the classical fourth-order Runge-Kutta method is very efficient, the choice of time step is difficult and may even change in the course of integration.
This problem can be solved by application of Runge-Kutta methods with adaptive stepsize control.
These methods derive different-order approximations from a shared pool of auxiliary results $\vec{k}_j$ and estimate the integration error by comparison of these approximations.
Prominent candidates which have proven valuable for micromagnetics are the Runge-Kutta-Fehlberg method \cite{fehlberg1969low} and the Dormand-Prince method \cite{dormand1980family} that both use fourth/fifth-order approximations for the stepsize control.

The usage of explicit methods is usually not advised for stiff problems because of their lack of numerical stability \cite{burden2010numerical}.
In micromagnetics, a strong exchange coupling can lead to a very high stiffness of the differential equation.
However, since the time-step size is coupled to the size of the smallest mesh cell, the use of a regular grid can mitigate this problem \cite{suess2002time}.
Hence, Runge-Kutta methods are the standard choice in finite-difference micromagnetics \cite{miltat2007numerical}.

\subsubsection{Implicit midpoint scheme}
An implicit integration scheme that specifically accounts for the micromagnetic unit-sphere constraint is the implicit midpoint rule \cite{daquino_2005}.
According to this method, the magnetization at $\vec{m}_i$ is obtained from the previous magnetization snapshot $\vec{m}_{i-1}$ by
\begin{equation}
  \vec{m}^i = \vec{m}^{i-1} + \Delta t \, \pdiffs{\vec{m}}{t} \left[t_{i-1} + \frac{\Delta t}{2}, \frac{\vec{m}^i + \vec{m}^{i-1}}{2} \right].
\end{equation}
Inserting the time derivative of $\vec{m}$ according to the LLG \eqref{eq:llg_llg_gilbert} and setting $\pdiffs{\vec{m}}{t} = (\vec{m}^i - \vec{m}^{i-1}) / \Delta t$ on the right-hand side yields
\begin{align}
  \vec{m}^i = \vec{m}^{i-1} + \Delta t
  \frac{\vec{m}^i + \vec{m}^{i-1}}{2} \times \left(
    - \gamma \heff\left[t_{i-1} + \frac{\Delta t}{2}, \frac{\vec{m}^i + \vec{m}^{i-1}}{2} \right]
    + \alpha \frac{\vec{m}^i - \vec{m}^{i-1}}{\Delta t}
  \right).
  \label{eq:int_midpoint}
\end{align}
Multiplying both sides with $\vec{m}^i + \vec{m}^{i-1}$ immediately yields $|\vec{m}^i| = |\vec{m}^{i-1}|$.
Hence, the midpoint rule exactly preserves the magnetization norm for arbitrary time-step sizes.
Moreover, it can be shown by a similar procedure, that the midpoint scheme preserves the energy of a magnetic system with quadratic energy contributions in $\vec{m}$ for vanishing damping $\alpha$ \cite{daquino_2005}.
However, the implicit nature of this method comes at the price of nonlinearity in the solution variable $\vec{m}^i$.
Since the nonlinear system \eqref{eq:int_midpoint} cannot be solved analytically, it requires the application of an iterative procedure such as Newton's method for the computation of $\vec{m}^i$.
Each Newton iteration requires the evaluation of the effective field, which results in a high computational effort.

\subsubsection{Tangent-plane integration}
Another class of integrators especially suited for the application in the framework of finite elements was introduced by Alouges et al. \cite{alouges2008new}.
This method relies on an alternative formulation of the LLG which is obtained by cross-multiplying the LLG \eqref{eq:llg_llg_gilbert} with $\vec{m}$
\begin{equation}
  \alpha \pdiffs{\vec{m}}{t} + \vec{m} \times \pdiffs{\vec{m}}{t} =
  \gamma \heff - \gamma (\vec{m} \cdot \heff) \vec{m}.
  \label{eq:int_tangent_llg}
\end{equation}
This formulation is equivalent to the original LLG since all terms in \eqref{eq:llg_llg_gilbert} are perpendicular to $\vec{m}$.
In order to solve for the time derivative $\pdiffs{\vec{m}}{t}$, \eqref{eq:int_tangent_llg} is reformulated in a weak form.
However, instead of seeking for the solution to $\vec{w} = \pdiffs{\vec{m}}{t}$ in the complete solution space $V: \mathbb{R}^3 \rightarrow \mathbb{R}^3$, the solution space is restricted to the tangent space of the magnetization $V_T = \{\vec{v}: \vec{v} \cdot \vec{m} = 0\}$.
This also allows for the restriction of the test space to the same space $V_T$ which simplifies the weak formulation of \eqref{eq:int_tangent_llg} to
\begin{equation}
  \int_{\Omega_m} (\alpha \vec{w} + \vec{m} \times \vec{w}) \cdot \vec{v} \dx =
  \int_{\Omega_m} \gamma \heff(\vec{m}) \cdot \vec{v} \dx
  \quad\forall\quad \vec{v} \in V_T.
\end{equation}
Instead of the original LLG, the right-hand side of this form is of the same order in the magnetization $\vec{m}$ as the effective field $\heff$.
This feature can be exploited in order to construct an implicit integration scheme for field terms linear in $\vec{m}$ without losing the linearity of the weak form in $\vec{w}$.
The time derivative $\vec{w}$ at $t_{i-1}$ is obtained by setting
\begin{equation}
  \vec{m} = \vec{m}^{i-1} + \theta \Delta t \vec{w}
\end{equation}
with $0 \leq \theta \leq 1$ where $\theta = 0$ leads to an explicit scheme and $\theta=1$ leads to an implicit scheme.
Employing the considerations concerning discontinuous material parameters in Sec.~\ref{sec:fem_other} yields the weak form
\begin{equation}
  \int_{\Omega_m} \mu_0 \Ms (\alpha \vec{w} + \vec{m}^{i-1} \times \vec{w}) \cdot \vec{v} \dx =
  \gamma \delta E(\vec{m}^{i-1} + \theta \Delta t \vec{w}, \vec{v})
\end{equation}
Considering only the exchange field, the weak formulation reads
\begin{equation}
  \int_{\Omega_m} \mu_0 \Ms (\alpha \vec{w} + \vec{m}^{i-1} \times \vec{w}) \cdot \vec{v} \dx =
  2 \gamma \int_{\Omega_m} A \vnabla (\vec{m}^{i-1} + \theta \Delta t \vec{w}): \vnabla \vec{v} \dx.
\end{equation}
Each field contribution can be treated with an individual $\theta$.
While the exchange field adds a high measure of stiffness to the problem which calls for an implict integration scheme, other field terms may well be treated explicitly \cite{goldenits2012effective}.
This usually applies to the demagnetization field.
Despite its linearity in $\vec{m}$ the demagnetization field can not be treated implicitly in the same manner as the exchange field, since it is the solution to another partial differential equation, see Sec.~\ref{sec:fem_demag}.
After computing the time derivative $\vec{w} = \pdiffs{\vec{m}}{t}$, the actual time step is performed by computing
\begin{equation}
  \vec{m}^i = \frac{\vec{m}^{i-1} + \Delta t \vec{w}}{|\vec{m}^{i-1} + \Delta t \vec{w}|}
\end{equation}
where the right-hand side is evaluated nodewise.
Various techniques have been proposed in order to implement the tangent-plane function space $V_T$ within a finite-element formulation.
A possible solution is the application of Lagrange multipliers \cite{goldenits2012effective,abert2013magnum} that enforce the orthogonality condition on $\vec{w}$.
An alternative approach uses a local mapping of two-dimensional vector fields $\mathbb{R}^3 \rightarrow \mathbb{R}^2$ onto the tangent plane $V_T$ \cite{ruggerithesis}.

A combination of this integration scheme with a time marching scheme for the spin accumulation, yields a method for the coupled solution of micromagnetics with the dynamic spin-diffusion equation \eqref{eq:spin_diff_dts} \cite{abert2014spin,abert2015three}.

Variants of this method include higher-order methods \cite{alouges2012convergent,kritsikis2014beyond}.
However, by increasing the order of the method, the linearity in the solution variable $\vec{w}$ is lost, which leads to a higher computational effort.

\subsubsection{Backward differentiation formula}\label{sec:int_bdf}
All previously introduced methods are so-called single-step methods, that require the knowledge of a single magnetization snapshot $\vec{m}^{i-1}$ in order to compute the subsequent magnetization $\vec{m}^i$.
In contrast, the backward differentiation formula (BDF) is a multi-step method, i.e. the magnetization $\vec{m}^i$ is computed from a series of preceding snapshots $\vec{m}^{i-j}$ with $0 \leq j \leq s$ and $s$ being the order of the method.
In the most general form, the BDF method is given by
\begin{equation}
  \sum_{j=0}^s a_i \vec{m}^{i-j} +
  \Delta t \beta \, \pdiffs{\vec{m}}{t}(t_i, \vec{m}^{i-1})
  = 0.
\end{equation}
where $a_i$ and $\beta$ can be chosen such that the method is of order $s$.
Normalizing the parameter $a_i$ and $\beta$ so that $a_0 = -1$ yields
\begin{equation}
  \vec{G}(\vec{m}^i) = 
  \vec{m}^i - \Delta t \beta \, \pdiffs{\vec{m}}{t} (t_i, \vec{m}^i) - \sum_{j=1}^s a_i \vec{m}^{i-j} = 0
\end{equation}
which is a nonlinear equation in $\vec{m}^i$ that can be solved with Newton's method.
Starting from an initial configuration $\vec{m}^{i,0}$ that is usually approximated by extrapolation of previous snapshots $\vec{m}^{i-j}$, the Newton iteration reads
\begin{equation}
  \vdiff{\vec{G}}{\vec{m}^i} (\vec{m}^{i,k}) (\vec{m}^{i,k+1} - \vec{m}^{i,k}) = - \vec{G}(\vec{m}^{i,k}).
  \label{eq:int_bdf_newton}
\end{equation}
where the variational derivative $\vdiff{\vec{G}}{\vec{m}^i}$ is given by
\begin{equation}
  \vdiff{\vec{G}}{\vec{m}^i} = \mat{1} - \Delta t \beta \vdiff{\vec{w}}{\vec{m}}
  \quad\text{with}\quad
  \vec{w} = \pdiffs{\vec{m}}{t}.
  \label{eq:int_bdf_jacobian}
\end{equation}
In order to avoid the numerically expensive inversion of $\vdiff{\vec{G}}{\vec{m}}$, the discretized version of the linear equation \eqref{eq:int_bdf_newton} is usually solved iteratively for $\Delta \vec{m}^{i,k} = \vec{m}^{i,k+1} - \vec{m}^{i,k}$.
However, in order to reduce the number of required iterations, it is advised to employ a preconditioning procedure to \eqref{eq:int_bdf_newton}.
Instead of solving \eqref{eq:int_bdf_newton} directly, the preconditioned system
\begin{equation}
  \vdiff{\vec{G}}{\vec{m}^i} (\vec{m}^{i,k}) \mat{P}^{-1} \mat{P} \Delta \vec{m}^{i,k} = - \vec{G}(\vec{m}^{i,k})
\end{equation}
is solved for $\mat{P} \Delta \vec{m}^{i,k}$.
The preconditioner $\mat{P}$ is chosen to approximate the original problem while being easily invertible.
A good choice for this purpose is a simplification of \eqref{eq:int_bdf_jacobian} where the linearization of the LLG $\vdiff{\vec{w}}{\vec{m}}$ only includes local and linear effective-field contributions like the exchange field.
Following this procedure, the computation of a single time step is computationally expensive since for every Newton iteration, the preconditioning system has to be solved.
However, this method has proven to be an superior choice for a large range of problems in the framework of finite-element micromagnetics \cite{suess2002time}.

\subsubsection{Existing software packages}
The implementation of explicit integration schemes, such as the Runge-Kutta scheme introduced in Sec.~\ref{sec:int_rk} is straightforward and does not necessarily require the use of external libraries.
However, implicit methods call for the solution of nonlinear systems.
Linear algebra libraries such as PETSc \cite{balay2017petsc} provide useful functionality for the implementation of suitable Newton methods.
The open-source library SUNDIALS \cite{hindmarsh2005sundials} includes a very efficient implementation of an adaptive-time-step BDF integration scheme as introduced in Sec.~\ref{sec:int_bdf}

\subsection{Energy minimization and barrier computation}
While dynamical micromagnetic simulations are useful to gain insight into the mechanism of fast magnetization processes like magnetization switching, they are not feasible for investigations in the \si{MHz} regime and below.
A typical example for quasi static micromagnetics is the computation of hysteresis curves.
Experimental measurements of hysteresis curves are performed with field sweeps that are orders of magnitude slowers than the response time of magnetic systems which is in the \si{GHz} regime.
Hence, the hysteresis curve can be computed by energy minimization instead of dynamical simulation.
This is done by increasing the external field stepwise end computing the new energy minimum for each step.
In order to compute hysteresis properties, the minimization algorithm is required to converge into the nearest local energetic minimum starting from a given magnetization configuration.
A global minimization would yield unique magnetization configurations for a given external field and hence be unsuited for hysteresis computations.

Numerical minimization is usually performed iteratively based on gradient evaluations.
These gradient methods are particularly useful for hysteresis computation since they start from a given magnetization configuration and converge to local minima.
The simplest gradient based minimization algorithm is the steepest-descent method with a single iteration given by
\begin{equation}
  \vec{m}^i = \vec{m}^{i-1} - \tau \vdiff{E}{\vec{m}}
  \label{eq:minimize_steepest_descent}
\end{equation}
with $\tau$ being the step size.
In order to account for the micromagnetic unit-sphere constraint $|\vec{m}| = 1$, the descent direction is usually projected onto the tangent space of the magnetization
\begin{equation}
  \vdiff{E}{\vec{m}}_{\perp_{\vec{m}}} =
  \vdiff{E}{\vec{m}} - \left( \vdiff{E}{\vec{m}} \cdot \vec{m} \right) \vec{m} =
  - \vec{m} \times \left( \vec{m} \times \vdiff{E}{\vec{m}} \right).
\end{equation}
Inserting into \eqref{eq:minimize_steepest_descent} and considering the definition of the effective field \eqref{eq:llg_effective_field} yields
\begin{equation}
  \vec{m}^i = \vec{m}^{i-1} - \tau \mu_0 \Ms \vec{m}^{i-1} \times \big[ \vec{m}^{i-1} \times \heff(\vec{m}^{i-1}) \big]
  \label{eq:minimize_steepest_descent_project}
\end{equation}
which is equivalent to an integration step of the LLG-damping-term with an explicit Euler method.
While this method still violates the unit-sphere constraint as shown in Sec.~\ref{sec:int_rk}, it represents a significant improvement compared to the original steepest-descent method \eqref{eq:minimize_steepest_descent}.
Instead of using an explicit Euler method, any of the integration methods introduced in Sec.~\ref{sec:int} may be used to progress the steepest descent.

However, time-integration methods are optimized to accurately solve for the complete magnetization trajectory, while the only measure of interest for a minimization problem is the final magnetization configuration.
Various tailored approaches for micromagnetic energy minimization including optimized steepest descend methods and variants of the conjugate gradient method haven been proposed in order to achieve high performance and reduce the risk to miss local minima \cite{fischbacher2017nonlinear,exl2014labonte,scholz2003scalable,hertel2001micromagnetic}.

\begin{figure}
  \centering
  \includegraphics{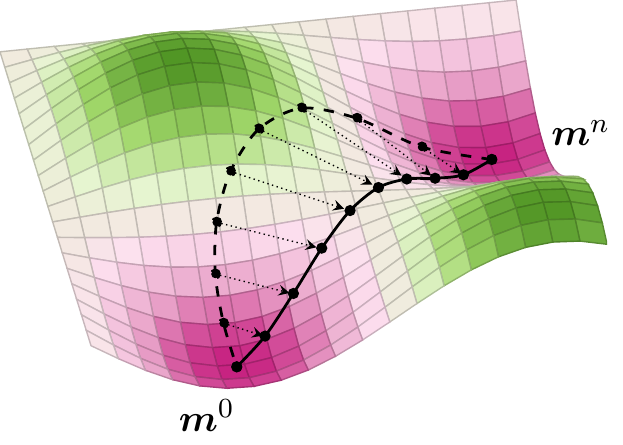}
  \caption{
    Illustration of the string method for the computation of minimum energy paths between two magnetization configurations $\vec{m}^0$ and $\vec{m}^{\n}$ in two dimensions.
    An initial transition path is discretized with a finite number of magnetization images and evolved towards the minimum.
  }
  \label{fig:barrier_string}
\end{figure}
Another class of minimization methods addresses the computation of energy barriers between two given local minima.
Energy barriers are an important measure for the stability of magnetic devices.
In magnetic storage devices, binary information is usually stored by putting the system in one of two possible energy minima, e.g. the up or down configuration in an anisotropic magnetic layer.
The energy barrier between those minima defines the characteristic lifetime of the information by the N\'eel-Arrhenius equation
\begin{equation}
  \tau_N = \tau_0 \exp \left( \frac{E}{k_\text{B} T} \right)
\end{equation}
with $E$ being the energy barrier, $T$ being the temperature and $\tau_0$ being the attempt time.
A robust yet simple method for the numerical calculation of the energy barrier between two magnetic states is the string method that yields the minimum energy path between two states \cite{weinan2007simplified}.
The string method is an iterative method the evolves an initial magnetization path $\vec{m}(\varphi)$ towards a minimum energy path defined by
\begin{equation}
  \Big( \vnabla E[\vec{m}(\varphi)] \Big)_\perp = 0
\end{equation}
where $\varphi$ denotes the position on the path and the $\perp$ subscript denotes magnetization variations that are orthogonal to the path $\vec{m}(\varphi)$.
The magnetization path is discretized by a finite number of magnetization images
\begin{equation}
  \vec{m}(\varphi) \rightarrow \vec{m}^i
  \quad\text{with}\quad
  i \in \{0,1,\dots,\n\}.
\end{equation}
As illustrated in Fig.~\ref{fig:barrier_string}, these images are evolved individually.
For a single string-method step, each magnetization image is relaxed towards an energetic minimum by a certain amount.
Using the projected steepest descent method \eqref{eq:minimize_steepest_descent_project}, the updated images are given as
\begin{equation}
  \vec{m}^{i,j} = \vec{m}^{i,j-1} - \zeta \vec{m}^{i,j-1} \times \big[ \vec{m}^{i,j-1} \times \heff(\vec{m}^{i,j-1}) \big]
  \label{eq:barrier_string_step}
\end{equation}
with $\zeta$ being the step size of the descent method.
Evolving each magnetization image towards the nearest local minimum would eventually relax every image into either the first minimum $\vec{m}^0$ or the second minimum $\vec{m}^\n$.
In this case all information about the transition path, and thus also the barrier information, would be lost.
In order to prevent the images to separate on the transition path, each evolution step \eqref{eq:barrier_string_step} is followed by a reparametrization of the path.
The position $\varphi_i$ of each magnetization image $\vec{m}^i$ on the transition path is determined by the pairwise distance norm
\begin{equation}
  \varphi_0 = 0 ,\quad
  \varphi_i = \varphi_{i-1} + \|\vec{m}^i - \vec{m}^{i-1}\|
  \quad\text{with}\quad
  \varphi \in \{1,2,\dots,\n\}.
\end{equation}
After determining the position $\phi_i$, the magnetization images $\vec{m}^i$ are interpolated onto the regular grid
\begin{equation}
  \varphi_i' = \frac{i \varphi_\n}{\n}
\end{equation}
by cubic spline interpolation.
Various variants of this method have been proposed in order to optimize the accuracy and convergence properties.
An effective improvement is the use of an energy weighted norm for the reparametrization in order to increase the image density in the barrier regime.
A popular alternative to the string method is the nudged elastic band method that introduces a spring force between the images in order to preserve a homogeneous discretization of the transition path \cite{dittrich2002path}.

\section{Applications}
This section is dedicated to applications of micromagnetic simulations.
While the first application deals with a classical micromagnetic problem and discusses the performance differences of finite-difference and finite-element tools, the remaining examples focus on the simulation of spintronics effects.

\subsection{Standard problem \#4}\label{sec:app_sp4}
\begin{figure}
  \centering
  \includegraphics{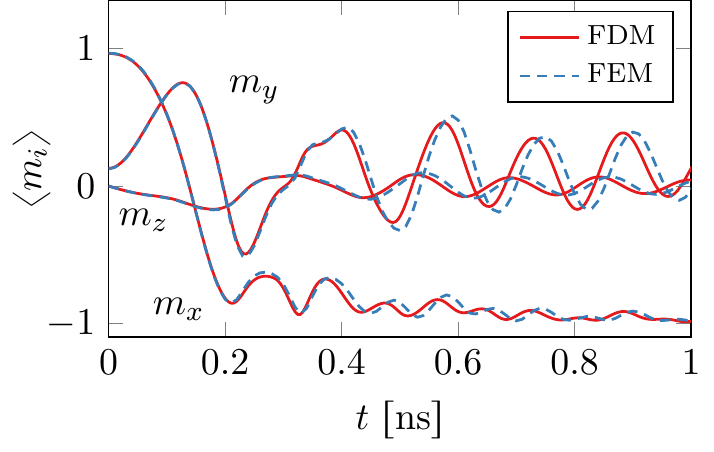}
  \caption{
    Time evolution of the average magnetization components for the switching process of a permalloy thin film according to the first part of the \textmu MAG standard problem \#4 computed with a finite-difference solver (FDM) and a finite-element solver (FEM).
  }
  \label{fig:app_sp4_m}
\end{figure}

A well known dynamical micromagnetic problem is the standard problem \#4 \cite{mumag4} that was developed by the \textmu MAG group with the aim to serve as a benchmark problem for micromagnetic simulation tools.
The problem considers a cuboid shaped magnetic system with dimensions $\SI{500x125x3}{nm}$ and material parameters similar to permalloy $\Ms = \SI{8e5}{A/m}$, $A = \SI{1.3d-11}{J/m}$, $K_\text{u1} = 0$ and $\alpha = 0.02$.
The system is prepared in a so-called s-state e.g. by relaxing the initial magnetization $\vec{m}_0 = \big( \cos[0.1], \sin[0.1], 0 \big)$ into an energetic minimum.
After relaxation, an external field with a magnitude of \SI{25}{mT} is applied directed \SI{170}{\degree} counterclockwise from the positive $x$-axis.
This field results in the switching of the magnetization in the thin film.
The dynamics of the switching process should be resolved by numerical integration of the LLG.
An alternative field which is defined in the original problem specification in not considered in this work.

Figure~\ref{fig:app_sp4_m} shows the time evolution of the averaged magnetization components computed with the finite-difference solver magnum.fd \cite{magnum.fd} and the finite-element solver magnum.fe \cite{abert2013magnum} that exhibit good agreement.
The finite-difference solver employs the FFT-accelerated demagnetization-field method presented in Sec.~\ref{sec:fd_demag} and an adaptive Runge-Kutta-Fehlberg method of 4/5 order for the time integration, see Sec.~\ref{sec:int_rk}.
The finite-element method employs the hybrid FEM/BEM method presented in Sec.~\ref{sec:fem_demag} and an adaptive preconditioned BDF scheme for time integration, see Sec.~\ref{sec:int_bdf}.
For the finite-difference method we choose a discretization of $200 \times 50 \times 1 = 10000$ cells and for the finite-element method we use a regular tetrahedral grid based on a $100 \times 24 \times 2$ cuboid grid which leads to 7878 mesh nodes.
Both grids are chosen considering the exchange length of permalloy in order obtain accurate results.

\begin{figure}
  \centering
  \includegraphics{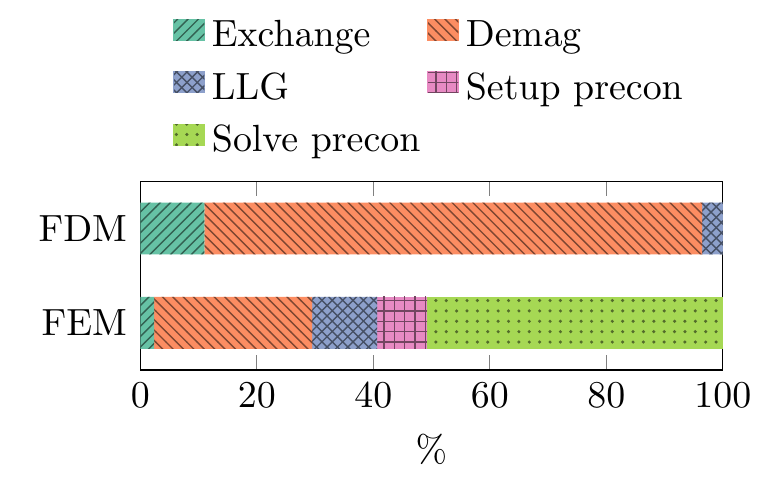}
  \caption{
    Relative time consumption for different parts of the solution process.
    Comparison of a finite-difference solver (FDM) with a finite-element solver (FEM).
  }
  \label{fig:app_sp4_timing1}
\end{figure}
Due to the different algorithms for demagnetization-field computation and time integration, the finite-difference solver and the finite-element solver show significantly deviating computation times for the problem, see Fig.~\ref{fig:app_sp4_timing1}.
While the finite-element solver spends more than half of the computation time on the assembly and solution of the preconditioner of the time integration scheme, the finite-difference solver is completely dominated by the demagnetization-field computation.

\begin{figure}
  \centering
  \includegraphics{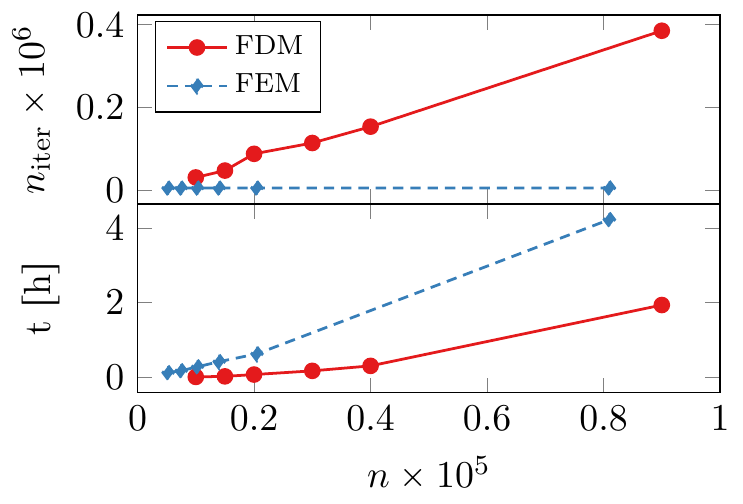}
  \caption{
    Performance comparison of a finite-difference solver (FDM) with a finite-element solver (FEM).
    The $x$-axes denotes the number of cells in the case of FDM and the number of mesh nodes in the case of FEM.
    The upper plot shows the number of required right-hand-side evaluations for the computation of the first \SI{2}{ns} of the standard problem \#4.
    The lower plot shows the overall computation time for single core computations on an Intel Core i7 system.
  }
  \label{fig:app_sp4_timing2}
\end{figure}
Another interesting comparison of the two methods is illustrated in Fig.~\ref{fig:app_sp4_timing2} where the required number of LLG right-hand-side evaluations and the total simulation time is compared for different problem sizes.
Starting from the above discretization, we perform mesh refinement on both the finite-difference grid and the finite-element mesh, which leads to an increased stiffness of the problem.
For the implicit time integration scheme used in the finite-element code, the number of right-hand-side evaluations remains almost the same for a given integration accuracy.
In contrast, the number of right-hand-side evaluations of the finite-difference solver increase faster than linear with the number of simulation cells.
However, despite the much larger number of right-hand-side evaluations of the finite-difference method, it still beats the finite-element method with respect to the overall simulation time.
One of the reasons for the superiority of the finite-difference method for this example, is the thin shape of the problem domain.
While this reduces the demagnetization-field computation to a two-dimensional convolution in the case of the finite-difference solver, the demagnetization-field algorithm of the finite-element code is dominated by the boundary-element method that has inferior scaling properties.

\subsection{Standard problem \#5}
\begin{figure}
  \centering
  \includegraphics{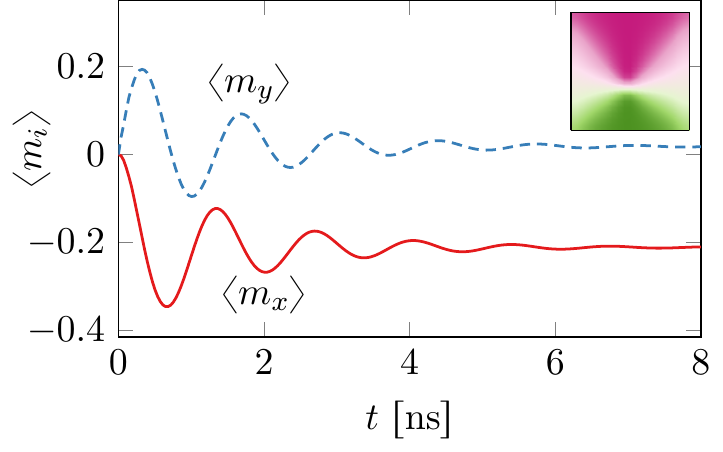}
  \caption{
    Time evolution of the averaged magnetization components for the current induced motion of a magnetic vortex according to the \textmu MAG standard problem \#5.
    The inset shows the new equilibrium vortex configuration.
  }
  \label{fig:app_sp5}
\end{figure}
The first and only \textmu MAG standard problem dealing with spintronics effects is the standard problem \#5 \cite{mumag5} which is derived from a work of Najafi et al. \cite{najafi2009proposal}.
It describes the precessional motion of a magnetic vortex core due to spin-torque.
A cuboid magnetic thin film with dimensions \SI{100x100x10}{nm} and the material parameters of permalloy, see Sec.~\ref{sec:app_sp4}, is initialized in a magnetic vortex state.
The application of a homogeneous charge current leads to the precessional motion of the vortex core which eventually relaxes in a shifted equilibrium position.
The problem definition suggests the application of the model of Zhang and Li as introduced in Sec.~\ref{sec:spin_zhang} for the calculation of the magnetization dynamics.
The current $\je$ is driven through the sample in $x$-direction.
Setting the product of the coupling constant and the current density $b j_\text{e} = \SI{72.17}{m/s}$ and choosing the degree of nonadiabaticity $\xi= 0.05$ and the damping $\alpha = 0.1$ yields the damped precessional motion depicted in Fig~\ref{fig:app_sp5}.

\begin{figure}
  \centering
  \includegraphics{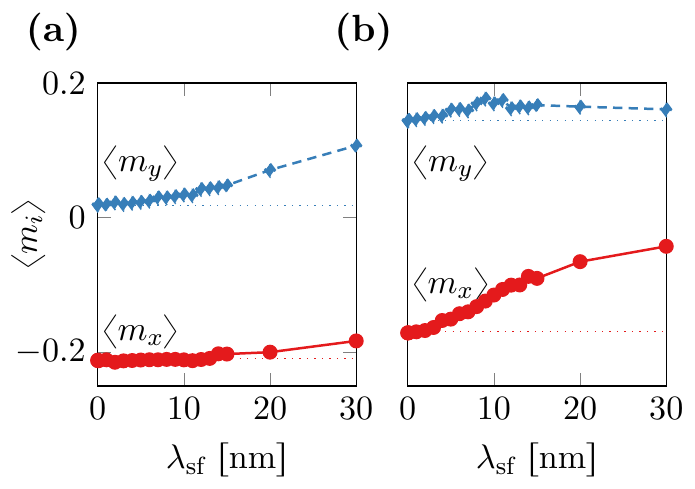}
  \caption{
    Averaged magnetization components for equilibrium vortex configurations according to the \textmu MAG standard problem \#5 computed with the spin-diffusion model for different diffusion lengths $\lambdasf$.
    (a) Results for $\xi = 0.05$.
    (b) Results for $\xi = 0.5$.
  }
  \label{fig:app_sp5_diffusion}
\end{figure}
As shown in Sec.~\ref{sec:spin_connect_zhang}, the spin-diffusion model is equivalent to the model of Zhang and Li in the case of a vanishing diffusion constant $D_0$.
The equality of the two models can also be achieved with a finite $D_0$ by considering the limit of vanshing $\lambdasf$ and $\lambdaj$ while preserving the ratio
\begin{equation}
  \frac{\lambdaj^2}{\lambdasf^2} = \xi = 0.05.
  \label{eq:app_sp5_xi}
\end{equation}
In this limit, the terms linear in $D_0$ become neglectable and the resulting torque is the same as for vanishing $D_0$.
In order to investigate the influence of diffusion effects on the magnetization dynamics we consider a finite diffusion constant of $D_0 = \SI{e-3}{A/m}$ which is a reasonable choice for magnetic materials \cite{shpiro2003self}.
Furthermore we choose $\beta' = 0.8$ and $\beta = 0.9$ which yields a current density of $j_\text{e} = \SI{1.15e12}{A/m^2}$ according to the required definition of $b j_\text{e}$.
We perform dynamic simulations for different $\lambdasf$ while always choosing $\lambdaj$ to fulfill \eqref{eq:app_sp5_xi}.

The resulting equilibrium magnetizations are summarized in Fig.~\ref{fig:app_sp5_diffusion}\,(a).
While the simulations with small diffusion lengths $\lambdasf$ and $\lambdaj$ show a perfect agreement with the results obtained from the model of Zhang and Li, there are significant deviations for larger diffusion lengths.
These deviations are most significant in the $x$-shift of the vortex core which is characterized by the $y$-component of the averaged magnetization.
The same comparison for a different degree of nonadiabacity $\xi = 0.5$ yields significant deviations in the $y$-shift of the vortex core, see Fig.~\ref{fig:app_sp5_diffusion}\,(b).
These simulations demonstrate that the model of Zhang and Li is not accurate in the presence of spin diffusion.
While the simplifications of the Zhang-Li model make it very attractive for computational micromagnetics, the results obtained from this model have to be evaluated with care.
Simulation results for materials with large diffusion lengths might be useful for qualitative investigations.
However, for an accurate description, the solution of the full diffusion model should be considered.

\subsection{Spin-torque oscillator}
\begin{figure}
  \centering
  \includegraphics{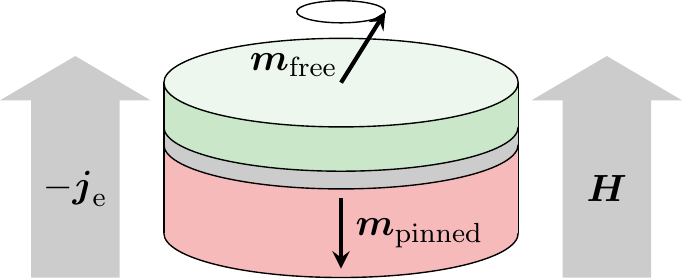}
  \caption{
    Schematic illustration of a field-stabilized spin-torque oscillator.
    The pinned layer acts as spin polarizer and the spin-torque compensates the damping contribution from the external field.
  }
  \label{fig:app_oscillator}
\end{figure}
Another spin-torque driven device with various potential applications is the spin-torque oscillator (STO) \cite{houssameddine2007spin,kim2012spin}.
A simple STO consists of two magnetic layers separated by a nonmagnetic layer.
One of the magnetic layers is designed to be very hard magnetic in order to act as stable spin polarizing layer.
The other magnetic layer, referred to as free layer, is soft magnetic and usually stabilized by an external field.
By applying a current perpendicular to the layer system, the free layer is subject to spin torque which drives its magnetization out of the external-field direction.
By suitable choice of current strength, the free-layer magnetization performs a precessional motion due to the external field while the damping contribution of the external field is exactly compensated by the spin torque, see Fig.~\ref{fig:app_oscillator}.
Numerous variants of STOs have been proposed and simulated with the spin-torque model of Slonczewski \cite{zhu2006bias,pribiag2007magnetic,firastrau2008modeling,rowlands2012magnetization}.
The model of Slonczewski is in principle perfectly suited for the simulation of STOs.
However, the input parameters to this model are the angular dependencies $\eta_\text{damp}$ and $\eta_\text{field}$ that cannot be trivially derived from the geometry and material parameters of the system.

\begin{figure}
  \centering
  \includegraphics{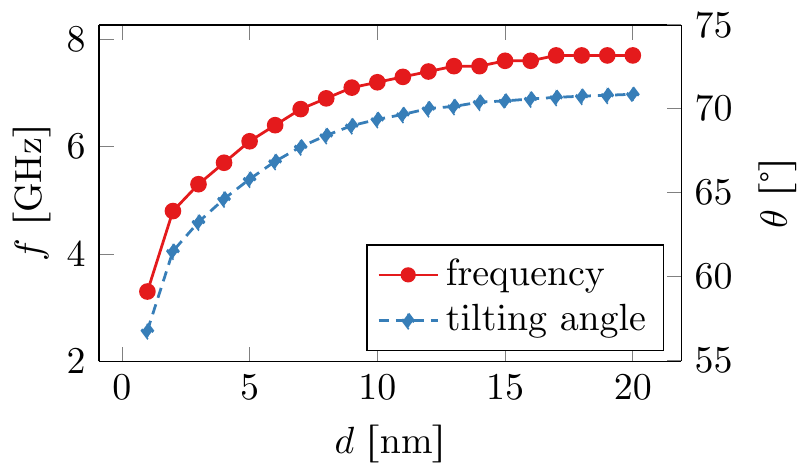}
  \caption{
    Oscillation frequency and tilting angle of the free-layer magnetization for varying pinned-layer thicknesses.
  }
  \label{fig:app_oscillator_plot}
\end{figure}
The spin-diffusion model renders very useful for the investigation of geometry dependent properties of such devices.
We simulate a cylindrical system with a radius of $R = \SI{30}{nm}$, a free-layer thickness of $d_\text{free} = \SI{3}{nm}$ and a spacer-layer thickness of $d_\text{spacer} = \SI{1.5}{nm}$.
The material parameters of the magnetic layers are chosen as $\alpha = 0.1$, $A = \SI{2.8e-11}{J/m}$, $D_0 = \SI{e-3}{A/m}$, $\beta = 0.8$, $\beta' = 0.9$, $\lambdasf = \SI{10}{nm}$, $\lambdaj = \SI{2.24}{nm}$.
For the free layer we further choose $\mu_0 \Ms = \SI{1}{T}$ and $K_\text{u1} = 0$ and for the spin-polarizing layer we chose $\mu_0 \Ms = \SI{1.24}{T}$ and $K_\text{u1} = \SI{e6}{J/m^3}$ with a perpendicular anisotropy axis.
The nonmagnetic spacer layer is simulated with material parameters $D_0 = \SI{5e-3}{A/m}$ and $\lambdasf = \SI{10}{nm}$.
We set the external field $\mu_0 H = \SI{0.6}{T}$ in $z$-direction and a current density $j_\text{e} = \SI{4e11}{A/m^2}$ in negative $z$-direction as shown in Fig.~\ref{fig:app_oscillator}.
This system is simulated with varying polarization-layer thicknesses with the spin-diffusion model with prescribed current density.
The resulting oscillation frequencies and tilting angles are shown in Fig.~\ref{fig:app_oscillator_plot}.
The qualitative dependence of the frequency and tilting angle from the thickness of the polarizing layer does not come as a surprise, since a thicker polarizing layer will obviously lead to higher spin polarization of the electrons in the free layer.
However, the spin-diffusion model does not only account for geometry changes, but also for the changes of material parameters in distinct layers.
In this respect it outperforms the simple model of Slonczewski that, on the other hand, has a much lower computational complexity which makes it a fast alternative to the spin-diffusion model for many problem settings.

\subsection{Spin-orbit torque MRAM}
\begin{figure}
  \centering
  \includegraphics{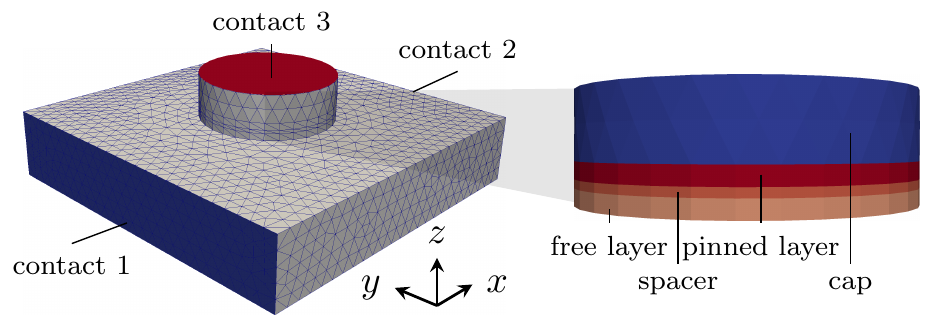}
  \caption{
    Geometry and domains of a model system consisting of a magnetic multilayer structure on top of a heavy-metal strip.
  }
  \label{fig:app_sot_domains}
\end{figure}
In a final numerical experiment, we demonstrate the capabilities of the self-consistent spin-diffusion model with spin-orbit interactions.
We aim to simulate the write and read process of a perpendicular spin-orbit torque magnetoresistive random-access memory (SOT MRAM) \cite{liu2012spin,cubukcu2014spin}.
Consider a circular magnetic multilayer with a radius of $R = \SI{10}{nm}$ consisting of 4 layers with thicknesses \SIlist{1; 0.5; 1; 4}{nm} from bottom to top, see Fig.~\ref{fig:app_sot_domains}.
From these 4 layers, only the bottom layer (free layer) and the third layer from below (pinned layer) are magnetic while all layers are conducting.
The circular stack is centered on top of a rectangular conducting underlayer with dimensions \SI{50x50x10}{nm} and the complete structure is meshed with prescribed mesh size of \SI{2}{nm}.
We define the two faces of the underlayer lying in the $yz$-plane as contact 1 and contact 2 respectively and the top interface of the circular stack as contact 3.
Material parameters in the conducting underlayer are chosen as $D_0 = \SI{e-3}{m^2/s}$, $C_0 = \SI{6e6}{A/Vm}$, $\tausf = \SI{2e-15}{s}$, $\theta = 0.3$
which is typical for heavy metals such as Ta that give rise to the spin Hall effect.
For the magnetic free layer we choose material parameters typical for perpendicular MRAM, namely $\Ms = \SI{0.796e6}{A/m}$, $\alpha = 0.02$, $A = \SI{16e-12}{J/m}$, $K_\text{u1} = \SI{0.4e6}{J/m^3}$, $D_0 = \SI{e-3}{m^2/s}$, $C_0 = \SI{e6}{A/Vm}$, $\beta = 0.9$, $\beta' = 0.8$, $\tausf = \SI{5e-14}{s}$, $J = \SI{2.1e-17}{J}$, and $\theta = 0$ with the anisotropy axis pointing in $z$-direction.
For the pinned layer, we choose the same parameters, but with a higher anisotropy $K_\text{u1} = \SI{e6}{J/m^3}$.
The nonmagnetic spacer and cap layers are simulated with parameters similar to Ag $D_0 = \SI{5e-3}{m^2/s}$, $C_0 = \SI{6e6}{A/Vm}$, $\tausf = \SI{1.225e-13}{s}$, and $\theta = 0.0$.

For the simulation of the write process, the magnetic layers are initialized in positive $z$-direction.
By applying an in-plane electric current in the underlayer, the spin Hall effect gives rise to a spin current in the $z$-direction.
This leads to spin-accumulation in the multilayer stack and thus to spin torque.
The torque on the free layer is expected to be much larger than on the pinned layer due its direct neighborhood with the underlayer.
Moreover the pinned layer has a much higher $K_\text{u1}$ than the free layer.
Hence the free layer is expected to switch easily while the pinned layer is expected to preserve its magnetization configuration even at high currents.

\begin{figure}
  \centering
  \includegraphics{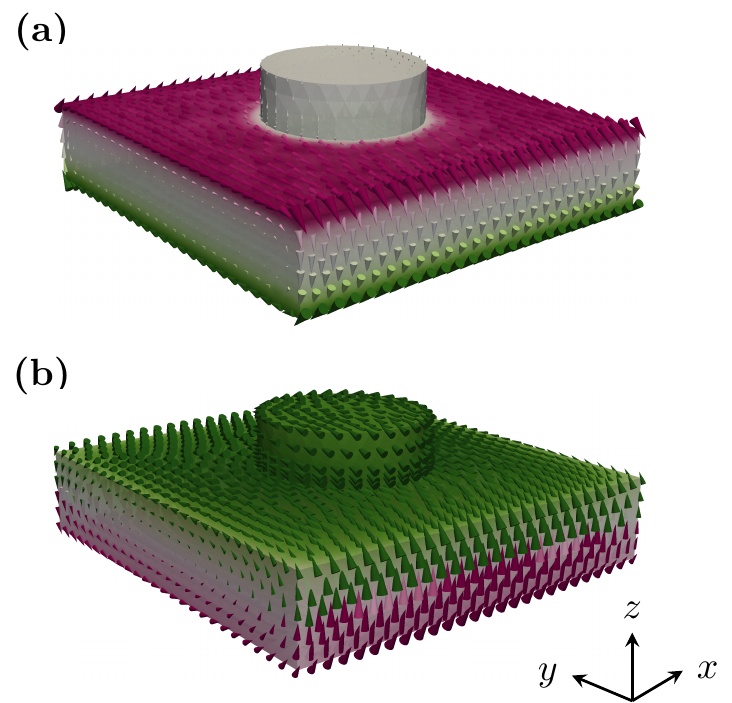}
  \caption{
    Effective-field contributions computed for an electric current in $x$-direction and both free-layer and pinned-layer magnetization pointing in $z$-direction
    (a) spin accumulation
    (b) Oersted field
  }
  \label{fig:fields}
\end{figure}
In a first numerical experiment, we determine the current dependent effective-field contributions, namely the spin accumulation $\vec{s}$ and the Oersted field $\vec{H}_\text{c}$, for the initial magnetization configuration.
We prescribe a constant electric potential at contact 1 ($u = 0$) and a constant current outflow at contact 2 ($j_\text{e} = \SI{e12}{A/m^2}$).
All remaining interfaces are treated with homogeneous Neumann conditions ($\partial\je / \partial\vec{n} = 0$).
The resulting fields are plotted in Fig.~\ref{fig:fields}.
Both the spin accumulation and the Oersted field exhibit a curl-like structure but with opposite sign.
In contrast to the Oersted field, the spin accumulation is much smaller in the stack compared to the underlayer, 
This is due to the interplay of the spin accumulation with the magnetization in the magnetic stack layers.

In the next experiment we perform time integration in order to resolve the switching of the free layer.
In order to enable the spin-orbit torque driven switching we apply an additional external field $\hzee = (-31830, 0, 0) \si{A/m}$.
Since an electric current in $x$-direction generates a spin current with polarization $y$ in the $z$-direction, this additional field is required for the perpendicular switching of the magnetization \cite{cubukcu2014spin}.
Otherwise the spin torque would draw the magnetization towards the $xy$-plane and the magnetization would return to its initial configuration once the current is turned off.

  \begin{figure}
    \centering
    \includegraphics{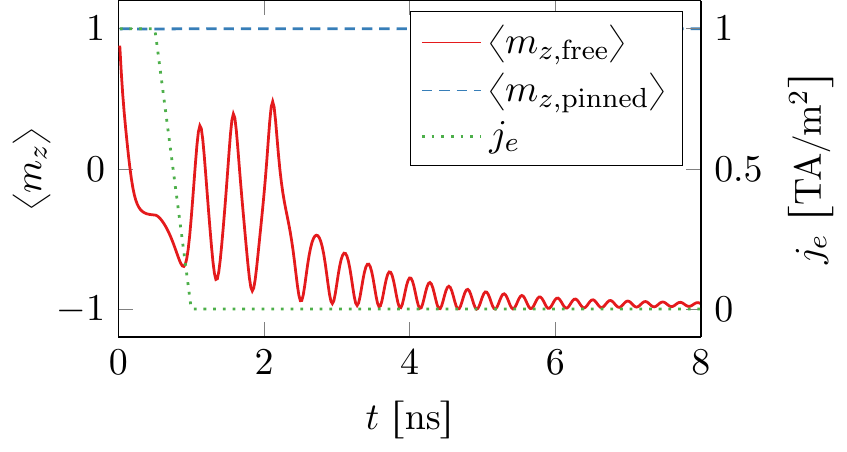}
    \caption{
      Time evolution of the averaged magnetization component $\langle m_z \rangle$ in the free layer and the pinned layer during switching due to spin-orbit torque generated by a current pulse in in the heavy-metal layer.
    }
    \label{fig:switch}
  \end{figure}
  In a first step, we perform time integration of the LLG \eqref{eq:llg_llg_gilbert} including the external field $\hzee$, the exchange field $\hex$, the demagnetization field $\hdemag$, and the anisotropy field $\haniso$ starting from the initial configuration in order to find an energetic minimum for the system without electric current.
After relaxation, we apply a constant current pulse of \SI{e12}{A/m^2} for the first \SI{0.5}{ns} that linearly decays to $0$ within another \SI{0.5}{ns}.
This pulse is applied in the same fashion as for the field computations, i.e. $u = 0$ at contact 1 and $j_\text{e}$ set as Neumann boundary condition on contact 2.
In addition to the effective-field contributions considered for the relaxation process, the spin torque due the spin accumulation as well as the Oersted field are included in the simulation.
The resulting magnetization dynamics are shown in Fig.~\ref{fig:switch}.
While the pinned-layer magnetization remains completely fixed in $z$-direction, the free-layer magnetization performs a fast switch during the current pulse and then relaxes into the $-z$-direction.

As a final experiment, we simulate the read process of the SOT MRAM.
In order to read the magnetization of the free layer, the magnetization dependent resistance of the magnetic multilayer is exploited.
When applying a current through the multilayer stack, the resistance of the structure changes either due to giant magnetoresistance (GMR) in the case of a conducting spacer or due to the tunnel magnetoresistance (TMR) in the case of of an insulating spacer.
In both cases, the multilayer is expected to have a lower resistance in case of parallel alignment of the free layer and the pinned layer and a high resistance in case of antiparallel alignment.

\begin{figure}
  \centering
  \includegraphics{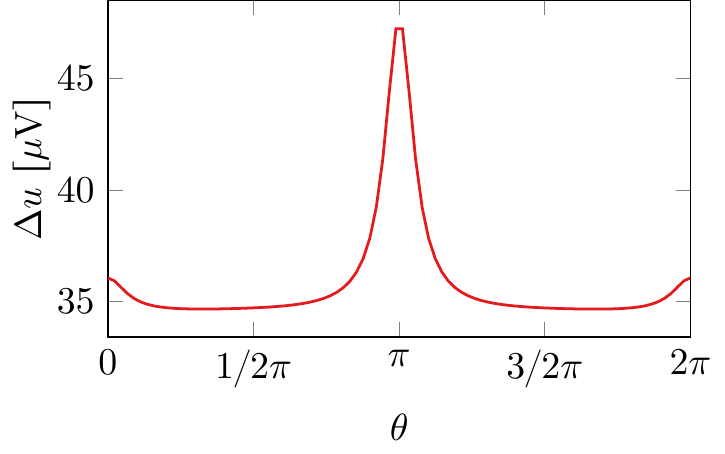}
  \caption{
    Potential difference between contact 3 and contact 1/2 for various tilting angles of the free-layer and the pinned-layer magnetization at constant current.
  }
  \label{fig:potential}
\end{figure}
For the readout we apply constant potential boundary conditions $u = 0$ at both contact 1 and contact 2 and set a constant current outflow $j_\text{e} = \SI{10}{A/m^2}$ at contact 3.
The pinned-layer magnetization is homogeneously set in $z$-direction while the the free-layer magnetization is homogeneously set to $\vec{m}_\text{free} = (0, \sin \theta, \cos \theta)$.
The coupled system for the spin accumulation and the electric potential is solved for various angles and the potential difference between contact 3 and contact 1/2 is evaluated.
Fig.~\ref{fig:potential} shows the simulation results.
As expected, the potential, and thus the resistance, of the stack is higher for antiparallel configurations.
The difference of the resistance for parallel $R_\text{p}$ and antiparallel $R_\text{a}$ configurations is significant $(R_\text{a} - R_\text{p}) / R_\text{a} \approx 23\%$.
This proves that the presented model includes all crucial physical effects for the self-consistent simulation of both the write process and the read process of an SOT MRAM device.

\section{Conclusion}
By bridging the gap between experiment and simple analytical models, computational micromagnetics has proven to be an invaluable tool for the development of magnetic devices.
With the rise of spintronics, many extensions to the classical micromagnetic model have been proposed in order to account for the bidirectional coupling of spin-polarized currents and the magnetization configuration.
Among the existing models, the spin-diffusion model by Zhang, Levy and Fert is one of the most complete approaches.
However, besides other shortcomings, the spin-diffusion model is only valid for diffusive transport which renders this model useless for the accurate description of tunnel barriers.
Since the spin-transport through tunnel barriers is dominated by quantummechnical effects, existing models mostly rely on ab initio techniques.
The efficient integration of such methods with the micromagnetic model is a challenging yet important task for future research.

The micromagnetic spintronics models introduced in this work already cover a lot of applications.
However, detailed knowledge of the applied model and its limitations is crucial for the successful application of micromagnetic simulations.
While a complex model might be required in order to understand certain details of magnetization dynamics, for other applications a simpler model might provide sufficient detail and allow to quickly perform a large number of simulations.
Moreover, the choice of discretization method has a significant impact on the accuracy and computation time and should be chosen carefully depending on the problem at hand.

\section{Acknowledgements}
The author would like to thank Prof. Dieter Suess  and Prof. Thomas Schrefl for endless discussions and valuable input concerning this article.


\begin{thebibliography}{100}

\bibitem{schrefl2007numerical}
T.~Schrefl, G.~Hrkac, S.~Bance, D.~Suess, O.~Ertl, and J.~Fidler, ``Numerical
  methods in micromagnetics (finite element method),'' {\em Handbook of
  magnetism and advanced magnetic materials}, 2007.

\bibitem{huai2008spin}
Y.~Huai, ``Spin-transfer torque {MRAM (STT-MRAM)}: {C}hallenges and
  prospects,'' {\em AAPPS bulletin}, vol.~18, no.~6, pp.~33--40, 2008.

\bibitem{parkin2004giant}
S.~S. Parkin, C.~Kaiser, A.~Panchula, P.~M. Rice, B.~Hughes, M.~Samant, and
  S.-H. Yang, ``Giant tunnelling magnetoresistance at room temperature with
  {MgO (100)} tunnel barriers,'' {\em Nature materials}, vol.~3, no.~12,
  p.~862, 2004.

\bibitem{granig2006integrated}
W.~Granig, C.~Kolle, D.~Hammerschmidt, B.~Schaffer, R.~Borgschulze, C.~Reidl,
  and J.~Zimmer, ``Integrated gigant magnetic resistance based angle sensor,''
  in {\em Proc. IEEE Sensors}, pp.~542--545, 2006.

\bibitem{brown_1963}
W.~F. {Brown Jr.}, {\em Micromagnetics}.
\newblock New York: Interscience Publisher, 1963.

\bibitem{doring1948tragheit}
W.~D{\"o}ring, ``{\"U}ber die tr{\"a}gheit der w{\"a}nde zwischen wei{\ss}schen
  bezirken,'' {\em Zeitschrift f{\"u}r Naturforschung A}, vol.~3, no.~7,
  pp.~373--379, 1948.

\bibitem{kronmuller2007general}
H.~Kronm{\"u}ller, ``General micromagnetic theory,'' {\em Handbook of Magnetism
  and Advanced Magnetic Materials}, 2007.

\bibitem{miltat2007numerical}
J.~E. Miltat and M.~J. Donahue, ``Numerical micromagnetics: Finite difference
  methods,'' {\em Handbook of magnetism and advanced magnetic materials}, 2007.

\bibitem{Leliaert_2018}
J.~Leliaert, M.~Dvornik, J.~Mulkers, J.~D. Clercq, M.~V. Milo{\v{s}}evi{\'{c}},
  and B.~V. Waeyenberge, ``Fast micromagnetic simulations on
  {GPU}{\textemdash}recent advances made with
  {\textdollar}{\textbackslash}mathsf$\lbrace$mumax$\rbrace${\^{}}3{\textdollar},''
  {\em Journal of Physics D: Applied Physics}, vol.~51, p.~123002, Feb 2018.

\bibitem{jackson_1999}
J.~D. Jackson, {\em Classical electrodynamics}.
\newblock John Wiley \& Sons, 2012.

\bibitem{griffiths_1994}
D.~J. Griffiths, {\em Introduction to Quantum Mechanics}.
\newblock New Jersey: Prentice Hall, 1994.

\bibitem{doering_1966}
W.~D\"oring, ``Mikromagnetismus,'' in {\em {H}andbuch der {P}hysik}
  (S.~Fl\"ugge, ed.), vol.~18/2, pp.~314--437, Springer, Berlin Heidelberg,
  1966.

\bibitem{hubert_1998}
A.~Hubert and R.~Sch\"afer, {\em Magnetic Domains}.
\newblock Berlin: Springer, 1998.

\bibitem{dzyaloshinsky1958thermodynamic}
I.~Dzyaloshinsky, ``A thermodynamic theory of “weak” ferromagnetism of
  antiferromagnetics,'' {\em Journal of Physics and Chemistry of Solids},
  vol.~4, no.~4, pp.~241--255, 1958.

\bibitem{moriya1960anisotropic}
T.~Moriya, ``Anisotropic superexchange interaction and weak ferromagnetism,''
  {\em Physical Review}, vol.~120, no.~1, p.~91, 1960.

\bibitem{yu2010real}
X.~Yu, Y.~Onose, N.~Kanazawa, J.~Park, J.~Han, Y.~Matsui, N.~Nagaosa, and
  Y.~Tokura, ``Real-space observation of a two-dimensional skyrmion crystal,''
  {\em Nature}, vol.~465, no.~7300, p.~901, 2010.

\bibitem{yu2012magnetic}
X.~Yu, M.~Mostovoy, Y.~Tokunaga, W.~Zhang, K.~Kimoto, Y.~Matsui, Y.~Kaneko,
  N.~Nagaosa, and Y.~Tokura, ``Magnetic stripes and skyrmions with helicity
  reversals,'' {\em Proceedings of the National Academy of Sciences}, vol.~109,
  no.~23, pp.~8856--8860, 2012.

\bibitem{bogdanov2001chiral}
A.~Bogdanov and U.~R{\"o}{\ss}ler, ``Chiral symmetry breaking in magnetic thin
  films and multilayers,'' {\em Physical review letters}, vol.~87, no.~3,
  p.~037203, 2001.

\bibitem{cortes2013influence}
D.~Cort{\'e}s-Ortu{\~n}o and P.~Landeros, ``Influence of the
  {D}zyaloshinskii-{M}oriya interaction on the spin-wave spectra of thin
  films,'' {\em Journal of Physics: Condensed Matter}, vol.~25, no.~15,
  p.~156001, 2013.

\bibitem{ruderman1954indirect}
M.~A. Ruderman and C.~Kittel, ``Indirect exchange coupling of nuclear magnetic
  moments by conduction electrons,'' {\em Physical Review}, vol.~96, no.~1,
  p.~99, 1954.

\bibitem{kasuya1956theory}
T.~Kasuya, ``A theory of metallic ferro-and antiferromagnetism on {Z}ener's
  model,'' {\em Progress of theoretical physics}, vol.~16, no.~1, pp.~45--57,
  1956.

\bibitem{yosida1957magnetic}
K.~Yosida, ``Magnetic properties of {Cu-Mn} alloys,'' {\em Physical Review},
  vol.~106, no.~5, p.~893, 1957.

\bibitem{fabian1996include}
K.~Fabian and F.~Heider, ``How to include magnetostriction in micromagnetic
  models of titanomagnetite grains,'' {\em Geophysical research letters},
  vol.~23, no.~20, pp.~2839--2842, 1996.

\bibitem{shu2004micromagnetic}
Y.~Shu, M.~Lin, and K.~Wu, ``Micromagnetic modeling of magnetostrictive
  materials under intrinsic stress,'' {\em Mechanics of Materials}, vol.~36,
  no.~10, pp.~975--997, 2004.

\bibitem{torres2003micromagnetic}
L.~Torres, L.~Lopez-Diaz, E.~Martinez, and O.~Alejos, ``Micromagnetic dynamic
  computations including eddy currents,'' {\em IEEE transactions on magnetics},
  vol.~39, no.~5, pp.~2498--2500, 2003.

\bibitem{hrkac2005three}
G.~Hrkac, M.~Kirschner, F.~Dorfbauer, D.~Suess, O.~Ertl, J.~Fidler, and
  T.~Schrefl, ``Three-dimensional micromagnetic finite element simulations
  including eddy currents,'' {\em Journal of applied physics}, vol.~97, no.~10,
  p.~10E311, 2005.

\bibitem{hertel2014hybrid}
R.~Hertel and A.~K{\'a}kay, ``Hybrid finite-element/boundary-element method to
  calculate {O}ersted fields,'' {\em Journal of Magnetism and Magnetic
  Materials}, vol.~369, pp.~189--196, 2014.

\bibitem{scholz2003scalable}
W.~Scholz, J.~Fidler, T.~Schrefl, D.~Suess, H.~Forster, V.~Tsiantos, {\em
  et~al.}, ``Scalable parallel micromagnetic solvers for magnetic
  nanostructures,'' {\em Computational Materials Science}, vol.~28, no.~2,
  pp.~366--383, 2003.

\bibitem{berkov2002fast}
D.~V. Berkov, ``Fast switching of magnetic nanoparticles: Simulation of thermal
  noise effects using the {L}angevin dynamics,'' {\em IEEE transactions on
  magnetics}, vol.~38, no.~5, pp.~2489--2495, 2002.

\bibitem{chubykalo2002langevin}
O.~Chubykalo, J.~Hannay, M.~Wongsam, R.~Chantrell, and J.~Gonzalez,
  ``{L}angevin dynamic simulation of spin waves in a micromagnetic model,''
  {\em Physical Review B}, vol.~65, no.~18, p.~184428, 2002.

\bibitem{garanin1997fokker}
D.~A. Garanin, ``{F}okker-{P}lanck and {L}andau-{L}ifshitz-{B}loch equations
  for classical ferromagnets,'' {\em Physical Review B}, vol.~55, no.~5,
  p.~3050, 1997.

\bibitem{atxitia2007micromagnetic}
U.~Atxitia, O.~Chubykalo-Fesenko, N.~Kazantseva, D.~Hinzke, U.~Nowak, and R.~W.
  Chantrell, ``Micromagnetic modeling of laser-induced magnetization dynamics
  using the {L}andau-{L}ifshitz-{B}loch equation,'' {\em Applied Physics
  Letters}, vol.~91, no.~23, p.~232507, 2007.

\bibitem{evans2012stochastic}
R.~F.~L. Evans, D.~Hinzke, U.~Atxitia, U.~Nowak, R.~W. Chantrell, and
  O.~Chubykalo-Fesenko, ``Stochastic form of the {L}andau-{L}ifshitz-{B}loch
  equation,'' {\em Physical Review B}, vol.~85, no.~1, p.~014433, 2012.

\bibitem{landau_1935}
L.~D. Landau and E.~M. Lifshitz, ``On the theory of the dispersion of magnetic
  permeability in ferromagnetic bodies,'' {\em Physikalische Zeitschrift der
  Sowjetunion}, vol.~8, pp.~153--169, 1935.

\bibitem{gilbert_1955}
T.~L. Gilbert, ``A {L}agrangian formulation of the gyromagnetic equation of the
  magnetic field,'' {\em Physical Review}, vol.~100, p.~1243, 1955.

\bibitem{gilbert_2004}
T.~L. Gilbert, ``A phenomenological theory of damping in ferromagnetic
  materials,'' {\em IEEE Transactions on Magnetics}, vol.~40, no.~6,
  pp.~3443--3449, 2004.

\bibitem{landau_mechanics}
L.~D. Landau and E.~M. Lifshitz, ``Mechanics,'' in {\em Course of Theoretical
  Physics}, Oxford: Pergamon Press, 1969.

\bibitem{wegrowe_2012}
J.-E. Wegrowe and M.-C. Ciornei, ``Magnetization dynamics, gyromagnetic
  relation, and inertial effects,'' {\em American Journal of Physics}, vol.~80,
  no.~7, pp.~607--611, 2012.

\bibitem{bode_2012}
N.~Bode, L.~Arrachea, G.~S. Lozano, T.~S. Nunner, and F.~von Oppen,
  ``Current-induced switching in transport through anisotropic magnetic
  molecules,'' {\em Physical Review B}, vol.~85, p.~115440, Mar 2012.

\bibitem{daquino_2005}
M.~d’Aquino, C.~Serpico, and G.~Miano, ``Geometrical integration of
  {L}andau-{L}ifshitz-{G}ilbert equation based on the mid-point rule,'' {\em
  Journal of Computational Physics}, vol.~209, no.~2, pp.~730--753, 2005.

\bibitem{cimrak_2007}
I.~Cimr{\'a}k, ``A survey on the numerics and computations for the
  {L}andau-{L}ifshitz equation of micromagnetism,'' {\em Archives of
  Computational Methods in Engineering}, vol.~15, no.~3, pp.~1--37, 2007.

\bibitem{baibich1988giant}
M.~N. Baibich, J.~M. Broto, A.~Fert, F.~N. Van~Dau, F.~Petroff, P.~Etienne,
  G.~Creuzet, A.~Friederich, and J.~Chazelas, ``Giant magnetoresistance of
  {(001) Fe/(001) Cr} magnetic superlattices,'' {\em Physical review letters},
  vol.~61, no.~21, p.~2472, 1988.

\bibitem{binasch1989enhanced}
G.~Binasch, P.~Gr{\"u}nberg, F.~Saurenbach, and W.~Zinn, ``Enhanced
  magnetoresistance in layered magnetic structures with antiferromagnetic
  interlayer exchange,'' {\em Physical review B}, vol.~39, no.~7, p.~4828,
  1989.

\bibitem{slonczewski1996current}
J.~C. Slonczewski, ``Current-driven excitation of magnetic multilayers,'' {\em
  Journal of Magnetism and Magnetic Materials}, vol.~159, no.~1-2, pp.~L1--L7,
  1996.

\bibitem{berger1996emission}
L.~Berger, ``Emission of spin waves by a magnetic multilayer traversed by a
  current,'' {\em Physical review B}, vol.~54, no.~13, p.~9353, 1996.

\bibitem{waintal2000role}
X.~Waintal, E.~B. Myers, P.~W. Brouwer, and D.~Ralph, ``Role of spin-dependent
  interface scattering in generating current-induced torques in magnetic
  multilayers,'' {\em Physical Review B}, vol.~62, no.~18, p.~12317, 2000.

\bibitem{worledge2011spin}
D.~Worledge, G.~Hu, D.~W. Abraham, J.~Sun, P.~Trouilloud, J.~Nowak, S.~Brown,
  M.~Gaidis, E.~O’sullivan, and R.~Robertazzi, ``Spin torque switching of
  perpendicular {Ta∣CoFeB∣MgO}-based magnetic tunnel junctions,'' {\em
  Applied Physics Letters}, vol.~98, no.~2, p.~022501, 2011.

\bibitem{houssameddine2007spin}
D.~Houssameddine, U.~Ebels, B.~Dela{\"e}t, B.~Rodmacq, I.~Firastrau,
  F.~Ponthenier, M.~Brunet, C.~Thirion, J.-P. Michel, L.~Prejbeanu-Buda, {\em
  et~al.}, ``Spin-torque oscillator using a perpendicular polarizer and a
  planar free layer,'' {\em Nature materials}, vol.~6, no.~6, p.~447, 2007.

\bibitem{kim2012spin}
J.-V. Kim, ``Spin-torque oscillators,'' in {\em Solid State Physics}, vol.~63,
  pp.~217--294, Elsevier, 2012.

\bibitem{ralph2008spin}
D.~C. Ralph and M.~D. Stiles, ``Spin transfer torques,'' {\em Journal of
  Magnetism and Magnetic Materials}, vol.~320, no.~7, pp.~1190--1216, 2008.

\bibitem{slonczewski2002currents}
J.~Slonczewski, ``Currents and torques in metallic magnetic multilayers,'' {\em
  Journal of Magnetism and Magnetic Materials}, vol.~247, no.~3, pp.~324--338,
  2002.

\bibitem{xiao2005macrospin}
J.~Xiao, A.~Zangwill, and M.~D. Stiles, ``Macrospin models of spin transfer
  dynamics,'' {\em Physical Review B}, vol.~72, no.~1, p.~014446, 2005.

\bibitem{apalkov2006micromagnetic}
D.~Apalkov, M.~Pakala, and Y.~Huai, ``Micromagnetic simulation of spin transfer
  torque switching by nanosecond current pulses,'' {\em Journal of applied
  physics}, vol.~99, no.~8, p.~08B907, 2006.

\bibitem{rowlands2012magnetization}
G.~E. Rowlands and I.~N. Krivorotov, ``Magnetization dynamics in a dual
  free-layer spin-torque nano-oscillator,'' {\em Physical Review B}, vol.~86,
  no.~9, p.~094425, 2012.

\bibitem{berkov2008spin}
D.~V. Berkov and J.~Miltat, ``Spin-torque driven magnetization dynamics:
  Micromagnetic modeling,'' {\em Journal of Magnetism and Magnetic Materials},
  vol.~320, no.~7, pp.~1238--1259, 2008.

\bibitem{parkin_2008}
S.~S. Parkin, M.~Hayashi, and L.~Thomas, ``Magnetic domain-wall racetrack
  memory,'' {\em Science}, vol.~320, no.~5873, pp.~190--194, 2008.

\bibitem{zhang2004roles}
S.~Zhang and Z.~Li, ``Roles of nonequilibrium conduction electrons on the
  magnetization dynamics of ferromagnets,'' {\em Physical Review Letters},
  vol.~93, no.~12, p.~127204, 2004.

\bibitem{zhang2002mechanisms}
S.~Zhang, P.~Levy, and A.~Fert, ``Mechanisms of spin-polarized current-driven
  magnetization switching,'' {\em Physical review letters}, vol.~88, no.~23,
  p.~236601, 2002.

\bibitem{dyakonov1971current}
M.~Dyakonov and V.~Perel, ``Current-induced spin orientation of electrons in
  semiconductors,'' {\em Physics Letters A}, vol.~35, no.~6, pp.~459--460,
  1971.

\bibitem{hirsch1999spin}
J.~Hirsch, ``Spin {H}all effect,'' {\em Physical Review Letters}, vol.~83,
  no.~9, p.~1834, 1999.

\bibitem{murakami2003dissipationless}
S.~Murakami, N.~Nagaosa, and S.-C. Zhang, ``Dissipationless quantum spin
  current at room temperature,'' {\em Science}, vol.~301, no.~5638,
  pp.~1348--1351, 2003.

\bibitem{sinova2004universal}
J.~Sinova, D.~Culcer, Q.~Niu, N.~Sinitsyn, T.~Jungwirth, and A.~MacDonald,
  ``Universal intrinsic spin {H}all effect,'' {\em Physical review letters},
  vol.~92, no.~12, p.~126603, 2004.

\bibitem{dyakonov2007magnetoresistance}
M.~Dyakonov, ``Magnetoresistance due to edge spin accumulation,'' {\em Physical
  review letters}, vol.~99, no.~12, p.~126601, 2007.

\bibitem{petitjean2012unified}
C.~Petitjean, D.~Luc, and X.~Waintal, ``Unified drift-diffusion theory for
  transverse spin currents in spin valves, domain walls, and other textured
  magnets,'' {\em Physical review letters}, vol.~109, no.~11, p.~117204, 2012.

\bibitem{akosa2015role}
C.~A. Akosa, W.-S. Kim, A.~Bisig, M.~Kl{\"a}ui, K.-J. Lee, and A.~Manchon,
  ``Role of spin diffusion in current-induced domain wall motion for disordered
  ferromagnets,'' {\em Physical Review B}, vol.~91, no.~9, p.~094411, 2015.

\bibitem{haney2013current}
P.~M. Haney, H.-W. Lee, K.-J. Lee, A.~Manchon, and M.~D. Stiles, ``Current
  induced torques and interfacial spin-orbit coupling: Semiclassical
  modeling,'' {\em Physical Review B}, vol.~87, no.~17, p.~174411, 2013.

\bibitem{valet1993theory}
T.~Valet and A.~Fert, ``Theory of the perpendicular magnetoresistance in
  magnetic multilayers,'' {\em Physical Review B}, vol.~48, no.~10, p.~7099,
  1993.

\bibitem{niimi2012giant}
Y.~Niimi, Y.~Kawanishi, D.~Wei, C.~Deranlot, H.~Yang, M.~Chshiev, T.~Valet,
  A.~Fert, and Y.~Otani, ``Giant spin {H}all effect induced by skew scattering
  from bismuth impurities inside thin film {CuBi} alloys,'' {\em Physical
  review letters}, vol.~109, no.~15, p.~156602, 2012.

\bibitem{abert2018efficient}
C.~Abert, F.~Bruckner, C.~Vogler, and D.~Suess, ``Efficient micromagnetic
  modelling of spin-transfer torque and spin-orbit torque,'' {\em AIP
  Advances}, vol.~8, no.~5, p.~056008, 2018.

\bibitem{tserkovnyak2002spin}
Y.~Tserkovnyak, A.~Brataas, and G.~E. Bauer, ``Spin pumping and magnetization
  dynamics in metallic multilayers,'' {\em Physical Review B}, vol.~66, no.~22,
  p.~224403, 2002.

\bibitem{mathon2001theory}
J.~Mathon and A.~Umerski, ``Theory of tunneling magnetoresistance of an
  epitaxial {Fe/MgO/Fe (001)} junction,'' {\em Physical Review B}, vol.~63,
  no.~22, p.~220403, 2001.

\bibitem{caffrey2011prediction}
N.~M. Caffrey, T.~Archer, I.~Rungger, and S.~Sanvito, ``Prediction of large
  bias-dependent magnetoresistance in all-oxide magnetic tunnel junctions with
  a ferroelectric barrier,'' {\em Physical Review B}, vol.~83, no.~12,
  p.~125409, 2011.

\bibitem{butler2001spin}
W.~Butler, X.-G. Zhang, T.~Schulthess, and J.~MacLaren, ``Spin-dependent
  tunneling conductance of {Fe|MgO|Fe} sandwiches,'' {\em Physical Review B},
  vol.~63, no.~5, p.~054416, 2001.

\bibitem{berkov1993solving}
D.~Berkov, K.~Ramst{\"o}ck, and A.~Hubert, ``Solving micromagnetic problems.
  towards an optimal numerical method,'' {\em physica status solidi (a)},
  vol.~137, no.~1, pp.~207--225, 1993.

\bibitem{seberino2001concise}
C.~Seberino and H.~N. Bertram, ``Concise, efficient three-dimensional fast
  multipole method for micromagnetics,'' {\em IEEE transactions on magnetics},
  vol.~37, no.~3, pp.~1078--1086, 2001.

\bibitem{press2007numerical}
W.~H. Press, S.~A. Teukolsky, W.~T. Vetterling, and B.~P. Flannery, {\em
  Numerical recipes 3rd edition: The art of scientific computing}.
\newblock Cambridge university press, 2007.

\bibitem{newell1993generalization}
A.~J. Newell, W.~Williams, and D.~J. Dunlop, ``A generalization of the
  demagnetizing tensor for nonuniform magnetization,'' {\em Journal of
  Geophysical Research: Solid Earth}, vol.~98, no.~B6, pp.~9551--9555, 1993.

\bibitem{lebecki2008periodic}
K.~M. Lebecki, M.~J. Donahue, and M.~W. Gutowski, ``Periodic boundary
  conditions for demagnetization interactions in micromagnetic simulations,''
  {\em Journal of Physics D: Applied Physics}, vol.~41, no.~17, p.~175005,
  2008.

\bibitem{kruger2013fast}
B.~Kr{\"u}ger, G.~Selke, A.~Drews, and D.~Pfannkuche, ``Fast and accurate
  calculation of the demagnetization tensor for systems with periodic boundary
  conditions,'' {\em IEEE Transactions on Magnetics}, vol.~49, no.~8,
  pp.~4749--4755, 2013.

\bibitem{kanai2010micromagnetic}
Y.~Kanai, K.~Koyama, M.~Ueki, T.~Tsukamoto, K.~Yoshida, S.~J. Greaves, and
  H.~Muraoka, ``Micromagnetic analysis of shielded write heads using symmetric
  multiprocessing systems,'' {\em IEEE Transactions on Magnetics}, vol.~46,
  no.~8, pp.~3337--3340, 2010.

\bibitem{abert2012fast}
C.~Abert, G.~Selke, B.~Kr{\"u}ger, and A.~Drews, ``A fast finite-difference
  method for micromagnetics using the magnetic scalar potential,'' {\em IEEE
  Transactions on Magnetics}, vol.~48, no.~3, pp.~1105--1109, 2012.

\bibitem{fu2016finite}
S.~Fu, W.~Cui, M.~Hu, R.~Chang, M.~J. Donahue, and V.~Lomakin,
  ``Finite-difference micromagnetic solvers with the object-oriented
  micromagnetic framework on graphics processing units,'' {\em IEEE
  Transactions on Magnetics}, vol.~52, no.~4, pp.~1--9, 2016.

\bibitem{garcia2007spin}
C.~J. Garc{\'\i}a-Cervera and X.-P. Wang, ``Spin-polarized currents in
  ferromagnetic multilayers,'' {\em Journal of computational physics},
  vol.~224, no.~2, pp.~699--711, 2007.

\bibitem{donahue1999oommf}
M.~J. Donahue, ``{OOMMF} user's guide, version 1.0,'' tech. rep., 1999.

\bibitem{fidimag}
D.~Cort{\'e}s-Ortu{\~n}o, W.~Wang, R.~Pepper, M.-A. Bisotti, T.~Kluyver,
  M.~Vousden, and H.~Fangohr, ``Fidimag v2.0.''
  \url{https://github.com/computationalmodelling/fidimag}.

\bibitem{micromagus}
D.~Berkov and N.~Gorn, ``{MicroMagus}--package for micromagnetic simulations.''
  \url{http://www.micromagus.de}, 2007.

\bibitem{abert2015full}
C.~Abert, F.~Bruckner, C.~Vogler, R.~Windl, R.~Thanhoffer, and D.~Suess, ``A
  full-fledged micromagnetic code in fewer than 70 lines of {NumPy},'' {\em
  Journal of Magnetism and Magnetic Materials}, vol.~387, pp.~13--18, 2015.

\bibitem{leliaert2018fast}
J.~Leliaert, M.~Dvornik, J.~Mulkers, J.~De~Clercq, M.~Milo{\v{s}}evi{\'c}, and
  B.~Van~Waeyenberge, ``Fast micromagnetic simulations on {GPU}—recent
  advances made with {MuMax3},'' {\em Journal of Physics D: Applied Physics},
  vol.~51, no.~12, p.~123002, 2018.

\bibitem{vansteenkiste2014design}
A.~Vansteenkiste, J.~Leliaert, M.~Dvornik, M.~Helsen, F.~Garcia-Sanchez, and
  B.~Van~Waeyenberge, ``The design and verification of {MuMax3},'' {\em AIP
  advances}, vol.~4, no.~10, p.~107133, 2014.

\bibitem{magnum.fd}
G.~Selke, B.~Kr\"uger, A.~Drews, C.~Abert, and T.~Gerhardt, ``magnum.fd.''
  \url{https://github.com/micromagnetics/magnum.fd}, 2014.

\bibitem{braess2007finite}
D.~Braess, {\em Finite elements: Theory, fast solvers, and applications in
  solid mechanics}.
\newblock Cambridge University Press, 2007.

\bibitem{saad2003iterative}
Y.~Saad, {\em Iterative methods for sparse linear systems}, vol.~82.
\newblock siam, 2003.

\bibitem{chen1997review}
Q.~Chen and A.~Konrad, ``A review of finite element open boundary techniques
  for static and quasi-static electromagnetic field problems,'' {\em IEEE
  Transactions on Magnetics}, vol.~33, no.~1, pp.~663--676, 1997.

\bibitem{imhoff1990original}
J.~Imhoff, G.~Meunier, X.~Brunotte, and J.~Sabonnadiere, ``An original solution
  for unbounded electromagnetic {2D}- and {3D}-problems throughout the finite
  element method,'' {\em IEEE Transactions on Magnetics}, vol.~26, no.~5,
  pp.~1659--1661, 1990.

\bibitem{brunotte1992finite}
X.~Brunotte, G.~Meunier, and J.-F. Imhoff, ``Finite element modeling of
  unbounded problems using transformations: a rigorous, powerful and easy
  solution,'' {\em IEEE Transactions on Magnetics}, vol.~28, no.~2,
  pp.~1663--1666, 1992.

\bibitem{henrotte1999finite}
F.~Henrotte, B.~Meys, H.~Hedia, P.~Dular, and W.~Legros, ``Finite element
  modelling with transformation techniques,'' {\em IEEE transactions on
  magnetics}, vol.~35, no.~3, pp.~1434--1437, 1999.

\bibitem{abert2013numerical}
C.~Abert, L.~Exl, G.~Selke, A.~Drews, and T.~Schrefl, ``Numerical methods for
  the stray-field calculation: A comparison of recently developed algorithms,''
  {\em Journal of Magnetism and Magnetic Materials}, vol.~326, pp.~176--185,
  2013.

\bibitem{fredkin1990hybrid}
D.~Fredkin and T.~Koehler, ``Hybrid method for computing demagnetizing
  fields,'' {\em IEEE Transactions on Magnetics}, vol.~26, no.~2, pp.~415--417,
  1990.

\bibitem{hackbusch2015hierarchical}
W.~Hackbusch, {\em Hierarchical matrices: algorithms and analysis}, vol.~49.
\newblock Springer, 2015.

\bibitem{popovic2005applications}
N.~Popovi{\'c} and D.~Praetorius, ``Applications of {H}-matrix techniques in
  micromagnetics,'' {\em Computing}, vol.~74, no.~3, pp.~177--204, 2005.

\bibitem{garcia2006adaptive}
C.~J. Garcia-Cervera and A.~M. Roma, ``Adaptive mesh refinement for
  micromagnetics simulations,'' {\em IEEE transactions on magnetics}, vol.~42,
  no.~6, pp.~1648--1654, 2006.

\bibitem{geuzaine2009gmsh}
C.~Geuzaine and J.-F. Remacle, ``Gmsh: A {3-D} finite element mesh generator
  with built-in pre-and post-processing facilities,'' {\em International
  journal for numerical methods in engineering}, vol.~79, no.~11,
  pp.~1309--1331, 2009.

\bibitem{schoberl1997netgen}
J.~Sch{\"o}berl, ``{NETGEN} an advancing front {2D}/{3D}-mesh generator based
  on abstract rules,'' {\em Computing and visualization in science}, vol.~1,
  no.~1, pp.~41--52, 1997.

\bibitem{geuzaine2013onelab}
C.~Geuzaine, F.~Henrotte, J.-F. Remacle, E.~Marchandise, and R.~Sabariego,
  ``Onelab: Open numerical engineering laboratory,'' in {\em 11e Colloque
  National en Calcul des Structures}, 2013.

\bibitem{schoberl2014c}
J.~Sch{\"o}berl, ``C++ 11 implementation of finite elements in ngsolve,'' {\em
  Institute for Analysis and Scientific Computing, Vienna University of
  Technology}, 2014.

\bibitem{gross2005escript}
L.~Gross, P.~Cochrane, M.~Davies, H.~Muhlhaus, and J.~Smillie, ``Escript:
  Numerical modelling with {Python},'' in {\em Australian Partnership for
  Advanced Computing (APAC) Conferene}, vol.~1, pp.~31--31, APAC, 2005.

\bibitem{mfem}
R.~Anderson, A.~Barker, J.~Bramwell, J.~Camier, J.~Ceverny, J.~Dahm,
  Y.~Dudouit, V.~Dobrev, A.~Fisher, T.~Kolev, D.~Medina, M.~Stowell, , and
  V.~Tomov, ``{MFEM}: A modular finite element library,'' {\em in preparation},
  2018.

\bibitem{alnaes2015fenics}
M.~S. Aln{\ae}s, J.~Blechta, J.~Hake, A.~Johansson, B.~Kehlet, A.~Logg,
  C.~Richardson, J.~Ring, M.~E. Rognes, and G.~N. Wells, ``The {FEniCS} project
  version 1.5,'' {\em Archive of Numerical Software}, vol.~3, no.~100,
  pp.~9--23, 2015.

\bibitem{balay2017petsc}
S.~Balay, S.~Abhyankar, M.~Adams, J.~Brown, P.~Brune, K.~Buschelman, L.~Dalcin,
  V.~Eijkhout, W.~Gropp, D.~Kaushik, {\em et~al.}, ``Petsc users manual
  revision 3.8,'' tech. rep., Argonne National Lab.(ANL), Argonne, IL (United
  States), 2017.

\bibitem{smigaj2015solving}
W.~{\'S}migaj, T.~Betcke, S.~Arridge, J.~Phillips, and M.~Schweiger, ``Solving
  boundary integral problems with bem++,'' {\em ACM Transactions on
  Mathematical Software (TOMS)}, vol.~41, no.~2, p.~6, 2015.

\bibitem{h2lib}
N.~Albrecht, C.~B\"orst, D.~Boysen, S.~Christophersen, and S.~B\"orm,
  ``{H2Lib}.'' \url{http://www.h2lib.org}, 2016.

\bibitem{finmag}
M.-A. Bisotti, M.~Beg, W.~Wang, M.~Albert, D.~Chernyshenko,
  D.~Cort{\'e}s-Ortu{\~n}o, R.~A. Pepper, M.~Vousden, R.~Carey, H.~Fuchs,
  A.~Johansen, G.~Balaban, L.~B.~T. Kluyver, , and H.~Fangohr, ``{FinMag}.''
  \url{https://github.com/fangohr/finmag}, 2018.

\bibitem{abert2013magnum}
C.~Abert, L.~Exl, F.~Bruckner, A.~Drews, and D.~Suess, ``magnum.fe: A
  micromagnetic finite-element simulation code based on {FEniCS},'' {\em
  Journal of Magnetism and Magnetic Materials}, vol.~345, pp.~29--35, 2013.

\bibitem{abert2015three}
C.~Abert, M.~Ruggeri, F.~Bruckner, C.~Vogler, G.~Hrkac, D.~Praetorius, and
  D.~Suess, ``A three-dimensional spin-diffusion model for micromagnetics,''
  {\em Scientific reports}, vol.~5, p.~14855, 2015.

\bibitem{ruggeri2016coupling}
M.~Ruggeri, C.~Abert, G.~Hrkac, D.~Suess, and D.~Praetorius, ``Coupling of
  dynamical micromagnetism and a stationary spin drift-diffusion equation: A
  step towards a fully self-consistent spintronics framework,'' {\em Physica B:
  Condensed Matter}, vol.~486, pp.~88--91, 2016.

\bibitem{abert2016self}
C.~Abert, M.~Ruggeri, F.~Bruckner, C.~Vogler, A.~Manchon, D.~Praetorius, and
  D.~Suess, ``A self-consistent spin-diffusion model for micromagnetics,'' {\em
  Scientific reports}, vol.~6, no.~1, p.~16, 2016.

\bibitem{alouges2012convergent}
F.~Alouges, E.~Kritsikis, and J.-C. Toussaint, ``A convergent finite element
  approximation for {L}andau-{L}ifschitz-{G}ilbert equation,'' {\em Physica B:
  Condensed Matter}, vol.~407, no.~9, pp.~1345--1349, 2012.

\bibitem{sturma2015geometry}
M.~Sturma, J.-C. Toussaint, and D.~Gusakova, ``Geometry effects on
  magnetization dynamics in circular cross-section wires,'' {\em Journal of
  Applied Physics}, vol.~117, no.~24, p.~243901, 2015.

\bibitem{magpar}
W.~Scholz, ``{MagPar}.'' \url{http://www.magpar.net/}, 2010.

\bibitem{fischbacher2007systematic}
T.~Fischbacher, M.~Franchin, G.~Bordignon, and H.~Fangohr, ``A systematic
  approach to multiphysics extensions of finite-element-based micromagnetic
  simulations: {Nmag},'' {\em IEEE Transactions on Magnetics}, vol.~43, no.~6,
  pp.~2896--2898, 2007.

\bibitem{femme}
D.~Suess and T.~Schrefl, ``{FEMME}.''
  \url{http://suessco.com/simulations/solutions/femme-software/}, 2018.

\bibitem{kakay2010speedup}
A.~Kakay, E.~Westphal, and R.~Hertel, ``Speedup of {FEM} micromagnetic
  simulations with graphical processing units,'' {\em IEEE transactions on
  magnetics}, vol.~46, no.~6, pp.~2303--2306, 2010.

\bibitem{chang2011fastmag}
R.~Chang, S.~Li, M.~Lubarda, B.~Livshitz, and V.~Lomakin, ``{FastMag}: Fast
  micromagnetic simulator for complex magnetic structures,'' {\em Journal of
  Applied Physics}, vol.~109, no.~7, p.~07D358, 2011.

\bibitem{kritsikis2014beyond}
E.~Kritsikis, A.~Vaysset, L.~Buda-Prejbeanu, F.~Alouges, and J.-C. Toussaint,
  ``Beyond first-order finite element schemes in micromagnetics,'' {\em Journal
  of Computational Physics}, vol.~256, pp.~357--366, 2014.

\bibitem{exl2014non}
L.~Exl and T.~Schrefl, ``Non-uniform {FFT} for the finite element computation
  of the micromagnetic scalar potential,'' {\em Journal of Computational
  Physics}, vol.~270, pp.~490--505, 2014.

\bibitem{apalkov2003fast}
D.~Apalkov and P.~Visscher, ``Fast multipole method for micromagnetic
  simulation of periodic systems,'' {\em IEEE transactions on magnetics},
  vol.~39, no.~6, pp.~3478--3480, 2003.

\bibitem{palmesi2017highly}
P.~Palmesi, L.~Exl, F.~Bruckner, C.~Abert, and D.~Suess, ``Highly parallel
  demagnetization field calculation using the fast multipole method on
  tetrahedral meshes with continuous sources,'' {\em Journal of Magnetism and
  Magnetic Materials}, vol.~442, pp.~409--416, 2017.

\bibitem{exl2012fast}
L.~Exl, W.~Auzinger, S.~Bance, M.~Gusenbauer, F.~Reichel, and T.~Schrefl,
  ``Fast stray field computation on tensor grids,'' {\em Journal of
  computational physics}, vol.~231, no.~7, pp.~2840--2850, 2012.

\bibitem{exl2014fft}
L.~Exl, C.~Abert, N.~J. Mauser, T.~Schrefl, H.~P. Stimming, and D.~Suess,
  ``{FFT}-based {K}ronecker product approximation to micromagnetic long-range
  interactions,'' {\em Mathematical Models and Methods in Applied Sciences},
  vol.~24, no.~09, pp.~1877--1901, 2014.

\bibitem{suess2002time}
D.~Suess, V.~Tsiantos, T.~Schrefl, J.~Fidler, W.~Scholz, H.~Forster,
  R.~Dittrich, and J.~Miles, ``Time resolved micromagnetics using a
  preconditioned time integration method,'' {\em Journal of Magnetism and
  Magnetic Materials}, vol.~248, no.~2, pp.~298--311, 2002.

\bibitem{fehlberg1969low}
E.~Fehlberg, ``Low-order classical {R}unge-{K}utta formulas with stepsize
  control and their application to some heat transfer problems,'' {\em NASA
  Technical Report}, vol.~315, 1969.

\bibitem{dormand1980family}
J.~R. Dormand and P.~J. Prince, ``A family of embedded {R}unge-{K}utta
  formulae,'' {\em Journal of computational and applied mathematics}, vol.~6,
  no.~1, pp.~19--26, 1980.

\bibitem{burden2010numerical}
R.~L. Burden and J.~D. Faires, {\em Numerical analysis}.
\newblock Cengage Learning, 2010.

\bibitem{alouges2008new}
F.~Alouges, ``A new finite element scheme for landau-lifchitz equations,'' {\em
  Discrete Contin. Dyn. Syst. Ser. S}, vol.~1, no.~2, pp.~187--196, 2008.

\bibitem{goldenits2012effective}
P.~Goldenits, G.~Hrkac, D.~Praetorius, and D.~Suess, ``An effective integrator
  for the {L}andau-{L}ifshitz-{G}ilbert equation,'' in {\em Proceedings of
  {M}athmod 2012 Conference}, 2012.

\bibitem{ruggerithesis}
M.~Ruggeri, {\em Coupling and numerical integration of the
  {L}andau-{L}ifshitz-{G}ilbert equation}.
\newblock PhD thesis, TU Wien, 2016.

\bibitem{abert2014spin}
C.~Abert, G.~Hrkac, M.~Page, D.~Praetorius, M.~Ruggeri, and D.~Suess,
  ``Spin-polarized transport in ferromagnetic multilayers: An unconditionally
  convergent {FEM} integrator,'' {\em Computers \& Mathematics with
  Applications}, vol.~68, no.~6, pp.~639--654, 2014.

\bibitem{hindmarsh2005sundials}
A.~C. Hindmarsh, P.~N. Brown, K.~E. Grant, S.~L. Lee, R.~Serban, D.~E.
  Shumaker, and C.~S. Woodward, ``Sundials: Suite of nonlinear and
  differential/algebraic equation solvers,'' {\em ACM Transactions on
  Mathematical Software (TOMS)}, vol.~31, no.~3, pp.~363--396, 2005.

\bibitem{fischbacher2017nonlinear}
J.~Fischbacher, A.~Kovacs, H.~Oezelt, T.~Schrefl, L.~Exl, J.~Fidler, D.~Suess,
  N.~Sakuma, M.~Yano, A.~Kato, {\em et~al.}, ``Nonlinear conjugate gradient
  methods in micromagnetics,'' {\em AIP Advances}, vol.~7, no.~4, p.~045310,
  2017.

\bibitem{exl2014labonte}
L.~Exl, S.~Bance, F.~Reichel, T.~Schrefl, H.~Peter~Stimming, and N.~J. Mauser,
  ``{LaBonte's} method revisited: An effective steepest descent method for
  micromagnetic energy minimization,'' {\em Journal of Applied Physics},
  vol.~115, no.~17, p.~17D118, 2014.

\bibitem{hertel2001micromagnetic}
R.~Hertel, ``Micromagnetic simulations of magnetostatically coupled nickel
  nanowires,'' {\em Journal of Applied Physics}, vol.~90, no.~11,
  pp.~5752--5758, 2001.

\bibitem{weinan2007simplified}
W.~E, W.~Ren, and E.~Vanden-Eijnden, ``Simplified and improved string method
  for computing the minimum energy paths in barrier-crossing events,'' {\em
  Journal of Chemical Physics}, vol.~126, no.~16, p.~164103, 2007.

\bibitem{dittrich2002path}
R.~Dittrich, T.~Schrefl, D.~Suess, W.~Scholz, H.~Forster, and J.~Fidler, ``A
  path method for finding energy barriers and minimum energy paths in complex
  micromagnetic systems,'' {\em Journal of Magnetism and Magnetic Materials},
  vol.~250, pp.~12--19, 2002.

\bibitem{mumag4}
``\textmu{MAG} standard problem \#4.''
  \url{https://www.ctcms.nist.gov/~rdm/std4/spec4.html}.

\bibitem{mumag5}
``\textmu{MAG} standard problem \#5.''
  \url{https://www.ctcms.nist.gov/~rdm/std5/spec5.xhtml}.

\bibitem{najafi2009proposal}
M.~Najafi, B.~Kr{\"u}ger, S.~Bohlens, M.~Franchin, H.~Fangohr, A.~Vanhaverbeke,
  R.~Allenspach, M.~Bolte, U.~Merkt, D.~Pfannkuche, {\em et~al.}, ``Proposal
  for a standard problem for micromagnetic simulations including spin-transfer
  torque,'' {\em Journal of Applied Physics}, vol.~105, no.~11, p.~113914,
  2009.

\bibitem{shpiro2003self}
A.~Shpiro, P.~M. Levy, and S.~Zhang, ``Self-consistent treatment of
  nonequilibrium spin torques in magnetic multilayers,'' {\em Physical Review
  B}, vol.~67, no.~10, p.~104430, 2003.

\bibitem{zhu2006bias}
X.~Zhu and J.-G. Zhu, ``Bias-field-free microwave oscillator driven by
  perpendicularly polarized spin current,'' {\em IEEE Transactions on
  Magnetics}, vol.~42, no.~10, pp.~2670--2672, 2006.

\bibitem{pribiag2007magnetic}
V.~Pribiag, I.~Krivorotov, G.~Fuchs, P.~Braganca, O.~Ozatay, J.~Sankey,
  D.~Ralph, and R.~Buhrman, ``Magnetic vortex oscillator driven by dc
  spin-polarized current,'' {\em Nature Physics}, vol.~3, no.~7, p.~498, 2007.

\bibitem{firastrau2008modeling}
I.~Firastrau, D.~Gusakova, D.~Houssameddine, U.~Ebels, M.-C. Cyrille,
  B.~Delaet, B.~Dieny, O.~Redon, J.-C. Toussaint, and L.~Buda-Prejbeanu,
  ``Modeling of the perpendicular polarizer-planar free layer spin torque
  oscillator: Micromagnetic simulations,'' {\em Physical Review B}, vol.~78,
  no.~2, p.~024437, 2008.

\bibitem{liu2012spin}
L.~Liu, C.-F. Pai, Y.~Li, H.~Tseng, D.~Ralph, and R.~Buhrman, ``Spin-torque
  switching with the giant spin {H}all effect of tantalum,'' {\em Science},
  vol.~336, no.~6081, pp.~555--558, 2012.

\bibitem{cubukcu2014spin}
M.~Cubukcu, O.~Boulle, M.~Drouard, K.~Garello, C.~Onur~Avci, I.~Mihai~Miron,
  J.~Langer, B.~Ocker, P.~Gambardella, and G.~Gaudin, ``Spin-orbit torque
  magnetization switching of a three-terminal perpendicular magnetic tunnel
  junction,'' {\em Applied Physics Letters}, vol.~104, no.~4, p.~042406, 2014.

\end{thebibliography}
\end{document}